\documentclass[11pt,a4paper]{article}
\pdfoutput=1 
             
\usepackage{graphicx}
             
\usepackage[utf8x]{inputenc}	
\usepackage[english]{babel}
\usepackage{paralist}
\usepackage{amsmath}
\numberwithin{equation}{section}
\usepackage{amssymb}
\usepackage{amsthm}
\usepackage{MnSymbol}
\usepackage{mathtools}		
\usepackage{dsfont}		
\usepackage{stmaryrd}		
\usepackage{cite}		
\usepackage[svgnames]{xcolor}
\usepackage{enumerate}
\usepackage{setspace}
\usepackage{booktabs}
\usepackage{multirow}
\usepackage{tikz}
\usetikzlibrary{shapes}
\usetikzlibrary{positioning}
\usetikzlibrary{chains}
\usetikzlibrary{arrows,fit,decorations.pathreplacing, shapes.misc}
\usetikzlibrary{decorations.markings}
\tikzstyle{every picture}+=[remember picture]
\tikzstyle{na} = [baseline=-.5ex]
\usepackage{todonotes}
\usepackage{pgfplots}
\usepackage[font=small,labelfont=bf]{caption}
\usepackage[font=small,labelfont=bf]{subcaption}
\usepackage[pdftex,breaklinks,colorlinks=true,urlcolor=RoyalBlue,linkcolor=blue,citecolor=blue]{hyperref}

\newcommand{\PE}{\mathrm{PE}}
\newcommand{\diff}{\mathrm{d}}

\newcommand{\rd}{\mathrm{d}}
\newcommand{\ri}{\mathrm{i}}
\newcommand{\re}{\mathrm{e}}

\newcommand{\Ncal}{\mathcal{N}}

\newcommand{\R}{\mathbb{R}}
\newcommand{\T}{\mathbb{T}}
\newcommand{\Z}{\mathbb{Z}}

\newcommand{\surm}{\mathrm{SU}}
\newcommand{\urm}{\mathrm{U}}

\newcommand{\sorm}{\mathrm{SO}}

\newcommand{\sprm}{\mathrm{Sp}}
\newcommand{\sormL}{\mathfrak{so}}

\def\ns#1{
	\node[circle, draw, fill=white] at (#1){};
	\node[cross out, draw] at (#1){};
}
\def\SevenB#1{
	\node[circle, draw, fill=white] at (#1){};
}

\def\ThreeB#1{
	\node[circle, draw, fill=black,inner sep=1] at (#1){};
}
\catcode`@=11
\allowdisplaybreaks

\begin{document}
\begin{titlepage}
\setcounter{page}{0}
\begin{flushright}KIAS-Q21006\vspace{10mm}\end{flushright}
\begin{center}

{\Large\bf 
Elliptic Quantum Curves of 6d SO(N) theories
}

\vspace{15mm}

{\large Jin Chen${}^{1}$,}\  
{\large Babak Haghighat${}^{1}$,}\
{\large Hee-Cheol Kim${}^{2,3}$,}\
{\large Kimyeong Lee${}^{4}$,}\ 
{\large Marcus Sperling${}^{1}$,}\  and \ 
{\large Xin Wang${}^{5}$} 
\\[5mm]
\noindent ${}^{1}${\em Yau Mathematical Sciences Center, Tsinghua University}\\
{\em Haidian District, Beijing, 100084, China}\\
{Email: {\tt jinchen@mail.tsinghua.edu.cn},} \\  
{\tt babakhaghighat@tsinghua.edu.cn,} \\
{{\tt msperling@mail.tsinghua.edu.cn }}
\\[5mm]
\noindent ${}^{2}${\em Department of Physics, POSTECH}\\
{\em  Pohang 37673, Korea}\\
{Email: {\tt heecheol1@gmail.com}}
\\[5mm]
\noindent ${}^{3}${\em Asia Pacific Center for 
Theoretical Physics, POSTECH}\\
{\em Pohang 37673, Korea}
\\[5mm]
\noindent ${}^{4}${\em School of Physics, Korea Institute for Advanced Study}\\
{\em Seoul 02455, Korea}\\
{Email: {\tt klee@kias.re.kr}} 
\\[5mm]
\noindent ${}^{5}${\em Quantum Universe Center, Korea Institute for Advanced Study}\\
{\em Seoul 02455, Korea}\\
{Email: {\tt wxin@kias.re.kr }} 
\\[5mm]

\vspace{15mm}

\begin{abstract}
We discuss supersymmetric defects in 6d $\Ncal=(1,0)$ SCFTs with $\sorm(N_c)$ gauge group and $N_c-8$ fundamental flavors. The codimension 2 and 4 defects are engineered by coupling the 6d gauge fields to charged free fields in four and two dimensions, respectively. We find that the partition function in the presence of the codimension 2 defect on $\R^4\times \T^2$ in the Nekrasov-Shatashvili limit satisfies an elliptic difference equation which quantizes the Seiberg-Witten curve of the 6d theory. The expectation value of the codimension 4 defect appearing in the difference equation is an even (under reflection) degree $N_c$ section over the elliptic curve when $N_c$ is even, and an odd section when $N_c$ is odd. 
We also find that RG-flows of the defects and the associated difference equations  in the 6d $\sorm(2N+1)$ gauge theories triggered by Higgs VEVs of KK-momentum states provide quantum Seiberg-Witten curves for $\mathbb{Z}_2$ twisted compactifications of the 6d $\sorm(2N)$ gauge theories. 
\end{abstract}

\end{center}

\end{titlepage}
{\baselineskip=12pt
{\footnotesize
\tableofcontents
}
}
%
%
\section{Introduction}
\label{sec:intro}

Quantum vacua of 4d $\Ncal=2$ supersymmetric gauge theories are captured in terms of an algebraic curve known as the Seiberg-Witten curve \cite{Seiberg:1994rs}. Periods of this curve can be utilized to compute the prepotential of the theory, which captures non-perturbative instanton corrections, and to understand BPS states for particles carrying various charges. In this framework, a meromorphic one-form is integrated over one-cycles of the algebraic curve yielding expressions of central charges of BPS particles which solely depend on the moduli of the curve. Moreover, the Seiberg-Witten (SW) curve can be interpreted as the classical phase space of codimension 2 BPS defects in the theory. This can be visualized by lifting the picture to string theory where the curve is realized as part of a Calabi-Yau geometry and the defect becomes a Lagrangian brane that is point-like on the curve. Geometric quantization then recasts the partition function of this brane into a quantum wave-function which is annihilated by a Hamiltonian that is the quantum version of the curve \cite{Aganagic:2003qj}. Finally, when this whole setup admits a generalization to torus-compactifications of 6d superconformal field theories (SCFTs), the quantum curve becomes an \emph{elliptic quantum curve}. This then sheds new lights on properties of 6d SCFTs to which we now turn.

Recently, significant advances have been made in the understanding of six-dimensional superconformal field theories. 
Starting from the observation that 6d $\Ncal=(2,0)$ theories admit an ADE classification \cite{Witten:1995zh,Strominger:1995ac}, a classification of all 6d SCFTs has been proposed by using F-theory backgrounds compactified on local elliptic Calabi-Yau threefolds \cite{Heckman:2013pva,Heckman:2015bfa}. Supersymmetric partition functions such as the elliptic genera of self-dual strings \cite{Haghighat:2013gba,DelZotto:2016pvm,DelZotto:2018tcj} and the superconformal indices \cite{Kim:2012ava,Lockhart:2012vp,Kim:2012qf} of 6d theories have been computed. Various aspects of dualities in lower dimensional systems arising from their compactification have been explored \cite{Gaiotto:2009we,Benini:2009mz,Bah:2012dg,Gaiotto:2015usa,Razamat:2016dpl,Ohmori:2015pua,Ohmori:2015pia,Kim:2017toz,Bah:2017gph,Kim:2018bpg,Razamat:2018gro,Kim:2018lfo,Ohmori:2018ona,Razamat:2019mdt,Chen:2019njf,Pasquetti:2019hxf,Razamat:2019ukg,Razamat:2020bix,Hwang:2021xyw,Nazzal:2021tiu}. Supersymmetric defects have played another important role in investigating distinctive features of 6d SCFTs. For instance, the compactifications of 6d $\Ncal=(2,0)$ SCFTs on Riemann surfaces punctured by codimension 2 defects give rise to scenarios realizing four-dimensional class $\mathcal{S}$ theories \cite{Gaiotto:2009we,Gaiotto:2009hg,Alday:2009fs}. 
More recently \cite{Gaiotto:2012xa,Gaiotto:2015usa,Nazzal:2018brc,Nazzal:2021tiu}, novel connections between elliptic quantum integrable models and 6d SCFTs have been discovered by studying surface defects which were introduced to  4d theories in the analysis of their superconformal indices. The relevant 4d theories arise from compactifying 6d $\Ncal=(1,0)$ SCFTs on a generic Riemann surface and the surface defects amount to codimension 4 defects in the 6d theories.
In particular, the quantum elliptic Hamiltonians and their eigenvalues were shown to be related to the elliptic Seiberg-Witten curves, describing the moduli space of the 6d SCFTs compactified on a torus, in \cite{Bullimore:2014awa,Chen:2020jla,Chen:2021ivd} by quantizing the elliptic SW-curves of some simple 6d $\Ncal=(1,0)$ SCFTs in terms of partition functions with codimension 2 and 4 defects.
This motivates a natural route to formulate quantizations of elliptic Seiberg-Witten curves for 6d SCFTs using supersymmetric defects and their partition functions.

The aim of this paper is to continue to investigate supersymmetric defects and their roles in establishing the quantization of Seiberg-Witten curves for other 6d SCFTs. We focus on 6d SCFTs with $\sorm(N_c)$ gauge group that are realized on a $-4$ curve in the base $B$ of the six-dimensional F-theory geometric backgrounds. The Seiberg-Witten curves for these SCFTs were obtained in \cite{Haghighat:2018dwe} by taking the thermodynamic limit of the $\Omega$-background partition functions, with parameters $\epsilon_{1,2}$. To obtain the quantum SW-curves, we use the properties of the defect operators along the same lines as \cite{Chen:2021ivd}.
We introduce half-BPS codimension 2 and 4 defects coupled to the bulk 6d $\sorm(N_c)$ gauge symmetry and compute their partition function in the $\Omega$-background using localization. This involves new proposals of 2d $\Ncal=(0,4)$ gauged linear sigma models (GLSMs) for the worldsheet of self-dual strings in the presence of the defects as well as the associated brane realizations. With this, we show that the codimension 2 defect partition function for $\sorm(N_c)$ theory with $N_f=N_c-8$ fundamental flavors in the Nekrasov-Shatashvili (NS) limit $\epsilon_2\rightarrow 0$, denoted as $\Psi(x) $, is annihilated by a difference operator
\begin{align}
\left[
e^{-\epsilon_1 \tfrac{\partial}{\partial x}}
+ \vartheta_1(2x)\vartheta_1(\epsilon_1+2x)^2\vartheta_1(2\epsilon_1+2x)
\prod_{l=1}^{N_f}\vartheta_1\left(x+\tfrac{\epsilon_1}{2}\pm m_l\right)
\cdot e^{\epsilon_1 \tfrac{\partial}{\partial x}}
- \chi(x)\right] \Psi(x)=0 \ , 
\end{align}
which is basically the twisted chiral ring relation of the 6d/4d coupled system on a torus with $\epsilon_1$-deformation.
The difference equation we illustrate here is further detailed in \eqref{eq:quantum-curve} and \eqref{eq:Bn_curve}. Furthermore, we find that the difference equation in the classical limit $\epsilon_1\rightarrow 0$ reduces to the Seiberg-Witten curve of the 6d $\sorm(N_c)$ gauge theory obtained in \cite{Haghighat:2018dwe}. Therefore, we propose that the difference equations provide a quantization of the Seiberg-Witten geometries for a family of the 6d $\sorm(N_c)$ gauge theories realized on a $-4$ curve.
This proposal is further verified by carefully analyzing the path integral over all instantonic string backgrounds (without defects) in the Nekrasov-Shatashvili limit.

The 6d $\sorm(N_c)$ gauge theories when compactified on a circle also admit 5-brane web constructions in Type IIB string theory. In particular, the 5-brane web suggests an exotic type of Higgs branch renormalization group (RG) flows of the $\sorm(2N+1)$ gauge theories which in the UV leads to the circle compactifications of the $\sorm(2N)$ gauge theories with a $\mathbb{Z}_2$ outer-automorphism twist. These RG-flows are  a slight extension of the ordinary Higgs branch flows: they are triggered by giving vacuum expectation values (VEVs) to chiral operators carrying Kaluza-Klein (KK) momentum charge so that they only take place after a circle compactification of the 6d theory. We study these RG-flows in the presence of the BPS defects. At the end of the RG-flow, the codimension 2 and 4 defects in the UV $\sorm(2N+1)$ gauge theory are reduced to those in the $\mathbb{Z}_2$ twisted compactification of the $\sorm(2N)$ gauge theory. This implies that the defect partition functions in the $\mathbb{Z}_2$ twisted $\sorm(2N)$ gauge theory in infrared satisfy a difference equation which naturally inherits the difference equation for the defect operators in the UV theory, as is shown in \eqref{eq:DnZ2_curve}. Therefore, we argue that the RG-flows provide the quantization of the Seiberg-Witten curves for the circle compactification $\sorm(2N)$ gauge theories twisted by $\mathbb{Z}_2$ action.

The organization of this paper is as follows. In Section \ref{sec:braneweb} we review key features of the 6d $\sorm(N_c)$ gauge theories coupled to codimension 2 and 4 defects and their 5-brane web realizations as well as Higgs branch RG-flows.  In Section \ref{sec:DN} we compute the defect partition functions in the 6d $\sorm(2N)$ gauge theory using the 2d GLSM constructions of self-dual strings interacting with the BPS defects and show that the partition functions satisfy a difference equation which quantizes the elliptic Seiberg-Witten curve of $\sorm(2N)$ type. In Section \ref{sec:BN} and \ref{sec:DN-twist} we study the defects and the quantum Seiberg-Witten curves of $\sorm(2N+1)$ and $\mathbb{Z}_2$ twisted $\sorm(2N)$ type, respectively, by utilizing Higgs branch flows.
In Section \ref{sec:conc} we present our conclusions. Several appendices complement the main text. Appendix \ref{app:functions} recalls the definition of theta functions. Appendix \ref{app:Higgs} provides a field theory description of the relevant Higgs mechanism. Appendices \ref{app:NS_finitie_sector} and \ref{app:ell_genera}  summarize computational details of the elliptic genera. 

\section{Brane realizations}
\label{sec:braneweb}
The theories considered are 6d $\Ncal=(1,0)$ $\sorm(N_c)$ gauge theories with $N_f=N_c-8$ fundamental flavors and one tensor multiplet. In terms of a F-theory realization, these theories correspond to a $-4$ curve with an $\sormL(N_c)$ gauge algebra \cite{Heckman:2013pva,Heckman:2015bfa}. 
Another realization is given in terms of brane configurations in Type II superstring theory \cite{Brunner:1997gk,Hanany:1997gh,Brunner:1997gf,Hanany:1999sj}. To begin with, the 6d brane configuration involving NS5, D6, and D8-branes in the presence of orientifolds is reviewed. Since the partition function is computed for the 6d theory on $\T^2\times \R^4_{\epsilon_1,\epsilon_2}$, one may also consider the 5d theories obtained via compactification on a circle. In the second half, the corresponding 5-brane webs are discussed, following \cite{Hayashi:2015vhy,Kim:2019dqn}. One particularly convenient feature is that the 5-brane web approach allows to consider 6d $\sorm(N_c)$ theories twisted by a $\Z_2$ outer automorphism along the circle.

Besides the brane realization of the theories of interests, it is also possible to realize codimension 2 defects as well as codimension 4 defects via the insertion of additional D-branes.
\subsection{6d perspective}
\label{subsec:6d_branes}
The 6d $\Ncal=(1,0)$ $\sorm(N_c)$ theories with $N_f =N_c-8$ fundamental flavors can be realized as world-volume theories of intersecting branes in Type IIA superstring theory \cite{Brunner:1997gk,Hanany:1997gh,Brunner:1997gf,Hanany:1999sj}. The NS5-D6-D8 brane configuration with O6-orientifolds preserves 8 supercharges and is summarized in Table \ref{tab:branes_6d}. For later purposes, the 6 dimensional world-volume is not flat space, but rather an $\Omega$-deformed background of the form $\T^2\times \R^4_{\epsilon_1,\epsilon_2}$. Here $\T^2$ fills the $x^{0,1}$ directions, and $\R^4_{\epsilon_1,\epsilon_2}$ occupies the directions along $x^{2,3,4,5}$.

\begin{table}[t]
\begin{center}
 \begin{tabular}{c|cccccccccc}
 \toprule 
  IIA & $x^0$ &  $x^1$ & $x^2$ & $x^3$ & $x^4$ & $x^5$ & $x^6$ & $x^7$ & $x^8$ & $x^9$ \\ \midrule
NS5 & $\times$ & $\times$ & $\times$ & $\times$ & $\times$ & $\times$ & \\
D6/O6& $\times$ & $\times$ & $\times$ & $\times$ & $\times$ & $\times$ &  $\times$ \\
D8& $\times$ & $\times$ & $\times$ & $\times$ & $\times$ & $\times$ &   & $\times$ & $\times$ &  $\times$\\
\midrule
D4& $\times$ & $\times$ & $\times$ & $\times$ &  &  &  & $\times$ \\
\midrule
D$4^\prime$& $\times$ & $\times$ & &  &  &  &  & $\times$ & $\times$ & $\times$ \\
\bottomrule
 \end{tabular}
 \caption{The Type IIA brane configuration for the 6d $\Ncal=(1,0)$ $\sorm(N_c)$ theories with $N_f=N_c-8$ flavors and one tensor multiplet. Here $\times$ denote directions filled by the various branes and O6-planes. The codimension 2 defect is realized by the D4-brane; while the D$4^\prime$-brane introduces a codimension 4 defect.}
 \label{tab:branes_6d}
 \end{center}
\end{table}

To begin with, consider the 6d $\Ncal=(1,0)$ $\sorm(2N)$ world-volume theory realized via a stack of $N$ full D6-branes on top of an O$6^-$ orientifold. In the presence of the O6-plane, a full D6-brane can split into two half D6-branes which can only move away from the orientifold as a pair.
The stack of D6s is suspended between two half NS5-branes, which are stuck on the O$6^-$-plane. The fundamental flavors originate from $N_f =2N-8$ full D8-branes which intersect the stack of D6-branes. Due to the orientifold, the $N_f$ full D8-branes can split into $4N-16$ half D8-branes \cite{Feng:2000eq,Bertoldi:2002nn}. The configuration is depicted in Figure \ref{subfig:6d_branes_SOeven}. Generically, for a configuration of NS5, D$p$, and D$(p{+}2)$-branes with $3\leq p\leq 6$, it is important to recall that the character of an O$p$-orientifold changes as follows \cite{Evans:1997hk,Hanany:1999sj,Hanany:2000fq}: firstly, an $\mathrm{O}p^{\pm}$ becomes an $\mathrm{O}p^\mp$ when passing through a half NS5; likewise, a $\widetilde{\mathrm{O}p}^\pm$ turns into $\widetilde{\mathrm{O}p}^\mp$. Secondly, an $Op^\pm$ becomes an $\widetilde{\mathrm{O}p}^\pm$ when passing through a half D$(p{+}2)$, and vice versa.

The transition to $\sorm(2N-1)$ is achieved via a Higgs mechanism. In the brane configuration, one opens up a Higgs branch direction by moving one half D8-brane to the left and right each. 
According to the analysis of \cite{Hanany:1996ie}, whenever a NS5 passes through a D8-brane a D6-brane is either created or annihilated. The presence of the orientifold implies analogous brane creation and annihilation effects whenever a half NS5 passes through a half D8-brane, see for instance \cite{Feng:2000eq} or \cite[App.\ A]{Cabrera:2019dob}.
To illustrate the point, consider a full D8-brane on a O$6^-$-plane, which then splits into two half D8s. The orientifold in between the two half D8-branes becomes an $\widetilde{\mathrm{O}6}^-$. If there is a half NS5 to either side, one can consider moving a half NS5 through a half D8 as follows:
\begin{align}
\raisebox{-.5\height}{
 \begin{tikzpicture}
 \draw[dashed] (-1,0)--(2,0);
  \ns{0,0}
  \draw (1,1)--(1,-1) (1.5,1)--(1.5,-1);
  \node at (-0.5,-0.75) {$+$};
  \node at (0.5,-0.75) {$-$};
  \node at (1.25,-0.75) {$\widetilde{-}$};
  \node at (1.75,-0.75) {$-$};
  \node at (0,0.5) {\tiny{half NS5}};
  \node at (2.1,0.85) {\tiny{half D8s}};
  \draw[dotted,thin,->] (2,0.8)--(1.1,0.3);
  \draw[dotted,thin,->] (2.1,0.7)--(1.6,0.3);
 \end{tikzpicture}
 }
 \qquad \longleftrightarrow \qquad
 \raisebox{-.5\height}{
 \begin{tikzpicture}
 \draw[dashed] (-1,0)--(2,0);
 \draw (0,0.15)--(1,0.15) (0,-0.15)--(1,-0.15); 
  \ns{1,0}
  \draw (0,1)--(0,-1) (1.5,1)--(1.5,-1);
  \node at (-0.5,-0.75) {$+$};
  \node at (0.5,-0.75) {$\widetilde{+}$};
  \node at (1.25,-0.75) {$\widetilde{-}$};
  \node at (1.75,-0.75) {$-$};
  \node at (0.5,0.3) {\tiny{1 D6}};
 \end{tikzpicture}
 }
\end{align}
where the right-hand side configuration exhibits a new full D6-brane, due to brane creation.
Applying the same logic to the configuration in Figure \ref{subfig:6d_branes_SOeven}, one arrives at the configuration in Figure \ref{subfig:HW_transition}, where the created D6-branes are depicted in red. These D6-branes, connecting the outermost half D8-branes with a half NS5-brane, can reconnect with a gauge D6-brane, which is suspended between the two half NS5-branes. This opens up the Higgs branch directions that allows to Higgs towards $\sorm(2N-1)$ as shown in Figure \ref{subfig:open_Higgs_direction}. Moving the brane along the Higgs branch directions off to infinity, the resulting effective brane configuration is depicted in Figure \ref{subfig:6d_branes_SOodd}, which has a $\sorm(2N-1)$ world-volume theory with $N_f=2N-9$ fundamental flavors.

The field theory description of this Higgsing is detailed in Appendix \ref{app:Higgs}.
\begin{figure}[t]
\centering
 \begin{subfigure}{0.475\textwidth}
 \centering
  \begin{tikzpicture}
  \draw[dashed] (-1,0)--(5,0);
  \draw (0,0.1)--(4,0.1) (0,-0.1)--(4,-0.1);
  \draw (1+0.5,1)--(1+0.5,-1) (1.1+0.5,1)--(1.1+0.5,-1) (1.2+0.5,1)--(1.2+0.5,-1) (1.3+0.5,1)--(1.3+0.5,-1);
  \draw (2+0.5,1)--(2+0.5,-1) (1.9+0.5,1)--(1.9+0.5,-1) (1.8+0.5,1)--(1.8+0.5,-1) (1.7+0.5,1)--(1.7+0.5,-1);
  \node at (2,0.25) {$\cdots$};
  \node at (2,-0.25) {$\cdots$};
  \ns{0,0}
  \ns{4,0}
  \node at (-0.5,-1) {\footnotesize{O$6^+$}};
  \node at (0.5,-1) {\footnotesize{O$6^-$}};
  \node at (3.5,-1) {\footnotesize{O$6^-$}};
  \node at (4.5,-1) {\footnotesize{O$6^+$}};
  \node at (0.75,0.25) {\footnotesize{$N$ D6}};
  \draw[decoration={brace,mirror,raise=10pt},decorate,thick]
  (1.5-0.2,-0.75) -- node[below=15pt] {\footnotesize{$4N-16$ half D8}} (2.5+0.2,-0.75);
\draw[->] (4,0.35)--(4.5,0.35);
\draw[->] (4,0.35)--(4,0.35+0.5);
\node at (4.75,0.4) {\footnotesize{$x^6$}};
\node at (4.5,0.8) {\footnotesize{$x^{7,8,9}$}};
 \end{tikzpicture}
 \caption{}
 \label{subfig:6d_branes_SOeven}
 \end{subfigure}
\hfill
\begin{subfigure}{0.475\textwidth}
\centering
 \begin{tikzpicture}
  \draw[dashed] (-1,0)--(5,0);
  \draw (0,0.1)--(4,0.1) (0,-0.1)--(4,-0.1);
  \draw (-0.75,1)--(-0.75,-1) (1.1+0.5,1)--(1.1+0.5,-1) (1.2+0.5,1)--(1.2+0.5,-1) (1.3+0.5,1)--(1.3+0.5,-1);
  \draw (4.75,1)--(4.75,-1) (1.9+0.5,1)--(1.9+0.5,-1) (1.8+0.5,1)--(1.8+0.5,-1) (1.7+0.5,1)--(1.7+0.5,-1);
  \node at (2,0.25) {$\cdots$};
  \node at (2,-0.25) {$\cdots$};
  \draw[red] (-0.75,0.1)--(0,0.1) (-0.75,-0.1)--(0,-0.1);
  \draw[red] (-0.75,0.1)--(0,0.1) (-0.75,-0.1)--(0,-0.1);
  \draw[red] (4,0.1)--(4.75,0.1) (4,-0.1)--(4.75,-0.1);
  \draw[red] (4,0.1)--(4.75,0.1) (4,-0.1)--(4.75,-0.1);
  \ns{0,0}
  \ns{4,0}
  \node at (-1.1,-1) {\footnotesize{$\mathrm{O}6^+$}};
  \node at (-0.25,-1) {\footnotesize{$\widetilde{\mathrm{O}6}^+$}};
  \node at (0.5,-1) {\footnotesize{$\widetilde{\mathrm{O}6}^-$}};
  \node at (3.5,-1) {\footnotesize{$\widetilde{\mathrm{O}6}^-$}};
  \node at (4.25,-1) {\footnotesize{$\widetilde{\mathrm{O}6}^+$}};
  \node at (5.1,-1) {\footnotesize{$\mathrm{O}6^+$}};
  \node at (0.75,0.25) {\footnotesize{$N$ D6}};
  \draw[decoration={brace,mirror,raise=10pt},decorate,thick]
  (1.5-0.2,-0.75) -- node[below=15pt] {\footnotesize{$4N-18$ half D8}} (2.5+0.2,-0.75);
 \end{tikzpicture}
 \caption{}
 \label{subfig:HW_transition}
\end{subfigure}
\\
\begin{subfigure}{0.475\textwidth}
\centering
 \begin{tikzpicture}
  \draw[dashed] (-1,0)--(5,0);
  \draw (0,0.1)--(4,0.1) (0,-0.1)--(4,-0.1);
  \draw (-0.75,1)--(-0.75,-1) (1.1+0.5,1)--(1.1+0.5,-1) (1.2+0.5,1)--(1.2+0.5,-1) (1.3+0.5,1)--(1.3+0.5,-1);
  \draw (4.75,1)--(4.75,-1) (1.9+0.5,1)--(1.9+0.5,-1) (1.8+0.5,1)--(1.8+0.5,-1) (1.7+0.5,1)--(1.7+0.5,-1);
  \node at (2,0.25) {$\cdots$};
  \node at (2,-0.25) {$\cdots$};
  \draw[red] (-0.75,0.75)--(4.75,0.75) (-0.75,-0.75)--(4.75,-0.75);
  \ns{0,0}
  \ns{4,0}
  \node at (-1.1,-1) {\footnotesize{$\mathrm{O}6^+$}};
  \node at (-0.25,-1) {\footnotesize{$\widetilde{\mathrm{O}6}^+$}};
  \node at (0.5,-1) {\footnotesize{$\widetilde{\mathrm{O}6}^-$}};
  \node at (3.5,-1) {\footnotesize{$\widetilde{\mathrm{O}6}^-$}};
  \node at (4.25,-1) {\footnotesize{$\widetilde{\mathrm{O}6}^+$}};
  \node at (5.1,-1) {\footnotesize{$\mathrm{O}6^+$}};
  \node at (0.75,0.25) {\footnotesize{$N{-}1$ D6}};
  \draw[decoration={brace,mirror,raise=10pt},decorate,thick]
  (1.5-0.2,-0.75) -- node[below=15pt] {\footnotesize{$4N-18$ half D8}} (2.5+0.2,-0.75);
 \end{tikzpicture}
 \caption{}
 \label{subfig:open_Higgs_direction}
\end{subfigure}
\hfill
\begin{subfigure}{0.475\textwidth}
\centering
 \begin{tikzpicture}
  \draw[dashed] (-1,0)--(5,0);
  \draw (0,0.1)--(4,0.1) (0,-0.1)--(4,-0.1);
  \draw (1.1+0.5,1)--(1.1+0.5,-1) (1.2+0.5,1)--(1.2+0.5,-1) (1.3+0.5,1)--(1.3+0.5,-1);
  \draw (1.9+0.5,1)--(1.9+0.5,-1) (1.8+0.5,1)--(1.8+0.5,-1) (1.7+0.5,1)--(1.7+0.5,-1);
  \node at (2,0.25) {$\cdots$};
  \node at (2,-0.25) {$\cdots$};
  \ns{0,0}
  \ns{4,0}
  \node at (-0.5,-1) {\footnotesize{$\widetilde{\mathrm{O}6}^+$}};
  \node at (0.5,-1) {\footnotesize{$\widetilde{\mathrm{O}6}^-$}};
  \node at (3.5,-1) {\footnotesize{$\widetilde{\mathrm{O}6}^-$}};
  \node at (4.5,-1) {\footnotesize{$\widetilde{\mathrm{O}6}^+$}};
  \node at (0.75,0.25) {\footnotesize{$N{-}1$ D6}};
  \draw[decoration={brace,mirror,raise=10pt},decorate,thick]
  (1.5-0.2,-0.75) -- node[below=15pt] {\footnotesize{$4N-18$ half D8}} (2.5+0.2,-0.75);
 \end{tikzpicture}
 \caption{}
 \label{subfig:6d_branes_SOodd}
\end{subfigure}
\caption{Higgsing of $\sorm(2N)$ to $\sorm(2N-1)$. The half D6-branes are denoted by horizontal solid lines, half D8-branes are vertical solid lines, and half NS5-branes are denoted by $\bigotimes$. The NS5-branes are drawn on the O6 orientifold which is denoted by horizontal dashed lines, and are separated along the $x^6$ direction. In \subref{subfig:6d_branes_SOeven}, the brane configuration for $\sorm(2N)$ is shown. Moving a half D8-brane across each of the half NS5-branes leads to brane creation as shown in \subref{subfig:HW_transition}, where the created D6 is drawn in red. In \subref{subfig:open_Higgs_direction}, a Higgs branch directions is opened up by aligning a gauge D6 with the two created D6 segments.  Moving the brane along the Higgs branch off to infinity leads to the brane configuration for $\sorm(2N-1)$, as displayed in \subref{subfig:6d_branes_SOodd}. }
\label{fig:branes_6d}
\end{figure}

\subsubsection{Codimension 2 defect}
In order to introduce a codimension 2 defect in a $\sorm(N_c-2)$ theory, a partial Higgs mechanism of the form $\sorm(N_c) \to \sorm(N_c-2)$ is utilized. This is demonstrated for $\sorm(2N)$ in this section and works analogously for $\sorm(2N+1)$. 

\begin{figure}[t]
 \begin{subfigure}{0.475\textwidth}
\centering
\begin{tikzpicture}
  \draw[dashed] (-1,0)--(5,0);
  \draw (0,0.1)--(4,0.1) (0,-0.1)--(4,-0.1);
  \draw (-0.75,1)--(-0.75,-1) (-0.75+0.1,1)--(-0.75+0.1,-1) (1.2+0.5,1)--(1.2+0.5,-1) (1.3+0.5,1)--(1.3+0.5,-1);
  \draw (4.75,1)--(4.75,-1) (4.75-0.1,1)--(4.75-0.1,-1) (1.8+0.5,1)--(1.8+0.5,-1) (1.7+0.5,1)--(1.7+0.5,-1);
  \node at (2,0.25) {$\cdots$};
  \node at (2,-0.25) {$\cdots$};
  \draw[red] (-0.75,0.1)--(0,0.1) (-0.75,-0.1)--(0,-0.1);
  \draw[red] (-0.75,0.1)--(0,0.1) (-0.75,-0.1)--(0,-0.1);
  \draw[red] (4,0.1)--(4.75,0.1) (4,-0.1)--(4.75,-0.1);
  \draw[red] (4,0.1)--(4.75,0.1) (4,-0.1)--(4.75,-0.1);
  \ns{0,0}
  \ns{4,0}
  \node at (-1.1,-1) {\footnotesize{$\mathrm{O}6^+$}};
  \node at (-0.25,-1) {\footnotesize{$\mathrm{O}6^+$}};
  \node at (0.5,-1) {\footnotesize{$\mathrm{O}6^-$}};
  \node at (3.5,-1) {\footnotesize{$\mathrm{O}6^-$}};
  \node at (4.25,-1) {\footnotesize{$\mathrm{O}6^+$}};
  \node at (5.1,-1) {\footnotesize{$\mathrm{O}6^+$}};
  \node at (0.75,0.25) {\footnotesize{$N$ D6}};
  \draw[decoration={brace,mirror,raise=10pt},decorate,thick]
  (1.5-0.2,-0.75) -- node[below=15pt] {\footnotesize{$4N-20$ half D8}} (2.5+0.2,-0.75);
 \end{tikzpicture}
\caption{}
\label{subfig:6d_branes_aux_HW}
 \end{subfigure}
\hfill
\begin{subfigure}{0.475\textwidth}
  \centering
\begin{tikzpicture}
  \draw[dashed] (-1,0)--(5,0);
  \draw (0,0.1)--(4,0.1) (0,-0.1)--(4,-0.1);
  \draw (-0.75,1)--(-0.75,-1) (-0.75+0.1,1)--(-0.75+0.1,-1) (1.2+0.5,1)--(1.2+0.5,-1) (1.3+0.5,1)--(1.3+0.5,-1);
  \draw (4.75,1)--(4.75,-1) (4.75-0.1,1)--(4.75-0.1,-1) (1.8+0.5,1)--(1.8+0.5,-1) (1.7+0.5,1)--(1.7+0.5,-1);
  \node at (2,0.25) {$\cdots$};
  \node at (2,-0.25) {$\cdots$};
  \draw[red] (-0.75,0.75)--(4.75,0.75) (-0.75,-0.75)--(4.75,-0.75);
  \draw[blue,thick] (4,-0.75)--(4,0.75);
  \ns{0,0}
  \ns{4,0}
  \node at (-1.1,-1) {\footnotesize{$\mathrm{O}6^+$}};
  \node at (-0.25,-1) {\footnotesize{$\mathrm{O}6^+$}};
  \node at (0.5,-1) {\footnotesize{$\mathrm{O}6^-$}};
  \node at (3.5,-1) {\footnotesize{$\mathrm{O}6^-$}};
  \node at (4.25,-1) {\footnotesize{$\mathrm{O}6^+$}};
  \node at (5.1,-1) {\footnotesize{$\mathrm{O}6^+$}};
  \node at (0.75,0.25) {\footnotesize{$N{-}1$ D6}};
  \node[blue] at (4.25,0.45) {\footnotesize{D4}};
  \draw[decoration={brace,mirror,raise=10pt},decorate,thick]
  (1.5-0.2,-0.75) -- node[below=15pt] {\footnotesize{$4N-20$ half D8}} (2.5+0.2,-0.75);
 \end{tikzpicture}
\caption{}
\label{subfig:6d_branes_defect_Higgs}
 \end{subfigure}
 \caption{Inclusion of codimension 2 defect. In \subref{subfig:6d_branes_aux_HW}, half D8-branes are moved across half NS5-branes which leads to brane creation, which is indicated by the red segments. In \subref{subfig:6d_branes_defect_Higgs}, a Higgs branch direction opens up by aligning the created D6-branes with a gauge D6-brane. If an additional D4-brane is suspended between the red D6-brane, moving along the Higgs branch direction, and the residual brane configuration, then a codimension 2 defect is present for the $\sorm(2N-2)$ theory. }
 \label{fig:6d_defect_Higgs}
\end{figure}
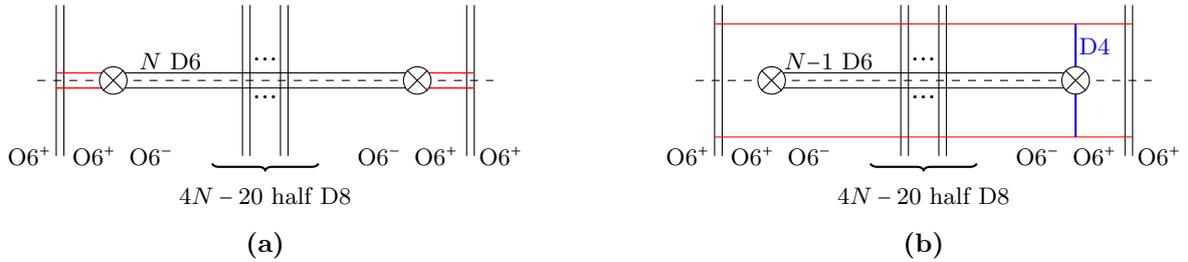

\paragraph{Brane construction.}
The starting point is the brane configuration in Figure \ref{subfig:6d_branes_SOeven}. To Higgs the theory to $\sorm(2N-2)$, one has to move two half D8-branes to the left and right each. Taking brane creation into account, one arrives at the configuration displayed in Figure \ref{subfig:6d_branes_aux_HW}. Now, a new Higgs branch direction opens up by aligning a full gauge D6-brane with the newly created D6-branes on the left and right, respectively. A codimension 2 defect is introduced via an additional D4-brane which is suspended between one of the half NS5 and the D6-brane that is moved along the Higgs branch.
Schematically, this defect Higgsing is depicted in Figure \ref{subfig:6d_branes_defect_Higgs}. 
Since the brane configuration shows the covering space of the orientifold, the schematic also contains the mirror D4.

In more detail, the defect D4 occupies space-times directions as indicated in Table \ref{tab:branes_6d}. In view of the $\Omega$-background, the codimension 2 defect wraps $\R^2_{\epsilon_1}$ inside $\R^4_{\epsilon_1,\epsilon_2}$ and is point-like in the $\R^2_{\epsilon_2}$ plane. Consequently, the D4 reduces the original eight supercharges to four and, moreover, breaks the rotation symmetry $\sorm(4)_{2345} \to \sorm(2)_{23} \times \sorm(2)_{45}$ on $\R^4_{2345}$. 

\paragraph{Field theory.}
In the field theory, the defect can be introduced by giving a position-dependent vacuum expectation value (VEV) to the moment map operator breaking $\sprm(2N-8)$ flavor symmetry to $\sprm(2N-10)$ and by taking an IR limit, which is analogous to the defect construction in \cite{Gaiotto:2012xa}.
Similar types of defect Higgsing in 6d theories have been considered in \cite{Chen:2020jla,Chen:2021ivd}.
The field theory description of the underlying standard Higgs process is detailed in Appendix \ref{app:Higgs}.

\paragraph{Remark.}
The possible Higgs transitions of an $\sorm(N_c)$ gauge theory from brane configuration faces known subtleties \cite{Cabrera:2019dob,Bourget:2020gzi}. For instance, the brane configuration in Figure \ref{subfig:HW_transition} for Higgsing $\sorm(2N)$ to $\sorm(2N-1)$ is not much different compare to the Higgsing of $\sorm(2N)$ to $\sorm(2N-2)$ in Figure \ref{subfig:6d_branes_aux_HW}. The reason is that the underlying brane transition does neither create or annihilate a physical brane. Put differently, starting from $\sorm(2N)$, the transition to $\sorm(2N-1)$ is expected to be reached along a $2N-8$ dimensional subspace of the full Higgs branch of $\sorm(2N)$. Likewise, the Higgsing $\sorm(2N-1)$ to $\sorm(2N-2)$ is expected along a $2N-9$ dimensional subspace of the Higgs branch of $\sorm(2N-1)$. However, the Higgs branch generically opened up in the configuration of Figure \ref{subfig:open_Higgs_direction} is of (quaternionic) dimension $4N-17$. Therefore, it is so far not clear how to distinguish the two partial Higgs mechanisms in the brane system.

Nevertheless, once the defect D4-brane is added into the setup, the two configurations are distinct. This is because the D4-brane is attached to a half NS5-brane such that one cannot simply move a D8-brane through as before. This becomes particularly evident in the 5-brane web discussed in Section \ref{subsec:5d_branes}. To preview the key-point, the Higgsing $\sorm(2N)\to\sorm(2N-1)$ is manifest in a 5-brane confined to an orientifold, see Figure \ref{fig:details_Higgs_5brane_web}; while the Higgsing  $\sorm(2N)\to\sorm(2N-2)$ is realized by a 5-brane away from an orientifold, see Figure \ref{fig:5d_branes_defect_Higgs}. Therefore, the latter transition is accompanied with a mass parameter, whereas the first is not.

\subsubsection{Codimension 4 defect}
Moving on, a codimension 4 defects can be introduced in various ways, which are discussed in turn.
\paragraph{Brane construction}
A codimension 4 defect for the 6d $\sorm(N_c)$ theories with $N_f = N_c-8$ can be introduced via an additional D$4^\prime$-brane which occupies $x^{0,1,7,8,9}$ as shown in   Table \ref{tab:branes_6d}. In contrast to the D$4$ defect, the D$4^\prime$ does not break the $\sorm(4)_{2345}$ rotation symmetry on $\R^4_{2345}$ as it is located at the origin of $\R^4_{2345}$.

\begin{figure}[t]
\centering
 \begin{subfigure}{0.3275\textwidth}
 \centering
  \begin{tikzpicture}
  \draw (0,0.1)--(4,0.1) (0,-0.1)--(4,-0.1);
  \draw (1+0.5,1)--(1+0.5,-1) (1.1+0.5,1)--(1.1+0.5,-1) (1.2+0.5,1)--(1.2+0.5,-1) (1.3+0.5,1)--(1.3+0.5,-1);
  \draw (2+0.5,1)--(2+0.5,-1) (1.9+0.5,1)--(1.9+0.5,-1) (1.8+0.5,1)--(1.8+0.5,-1) (1.7+0.5,1)--(1.7+0.5,-1);
  \node at (2,0.25) {$\cdots$};
  \node at (2,-0.25) {$\cdots$};
  \draw (-0.4,-1)--(-0.4,1) (4+0.4,-1)--(4+0.4,1);
    \draw (-0.6,-1)--(-0.6,1) (4+0.6,-1)--(4+0.6,1);
    \draw (-0.4,0.75)--(4.4,0.75) (-0.6,-0.75)--(4.6,-0.75);
    \draw[red,thick] (4.1,0)--(4.1,0.75) (3.9,0)--(3.9,-0.75);
  \ns{0,0}
  \ns{4,0}
  \node at (0.75,0.25) {\footnotesize{$N$ D6}};
  \node[red] at (3.8,0.375) {\footnotesize{D4}};
  \node[red] at (3.6,-0.375) {\footnotesize{D4}};
  \draw[decoration={brace,mirror,raise=10pt},decorate,thick]
  (1.5-0.2,-0.75) -- node[below=15pt] {\footnotesize{some D8}} (2.5+0.2,-0.75);
 \end{tikzpicture}
 \caption{}
 \label{subfig:6d_codim4_a}
 \end{subfigure}
\hfill
\begin{subfigure}{0.3275\textwidth}
\centering
 \begin{tikzpicture}
  \draw (0,0.1)--(4,0.1) (0,-0.1)--(4,-0.1);
  \draw (1+0.5,1)--(1+0.5,-1) (1.1+0.5,1)--(1.1+0.5,-1) (1.2+0.5,1)--(1.2+0.5,-1) (1.3+0.5,1)--(1.3+0.5,-1);
  \draw (2+0.5,1)--(2+0.5,-1) (1.9+0.5,1)--(1.9+0.5,-1) (1.8+0.5,1)--(1.8+0.5,-1) (1.7+0.5,1)--(1.7+0.5,-1);
  \node at (2,0.25) {$\cdots$};
  \node at (2,-0.25) {$\cdots$};
  \draw (-0.4,-1)--(-0.4,1) (4+0.4,-1)--(4+0.4,1);
    \draw (-0.6,-1)--(-0.6,1) (4+0.6,-1)--(4+0.6,1);
    \draw (-0.4,0.75)--(4.4,0.75) (-0.6,-0.75)--(4.6,-0.75);
    \draw[red,thick] (4,-0.75)--(4,0.75);
  \ns{0,0}
  \ns{4,0}
  \node at (0.75,0.25) {\footnotesize{$N$ D6}};
  \draw[decoration={brace,mirror,raise=10pt},decorate,thick]
  (1.5-0.2,-0.75) -- node[below=15pt] {\footnotesize{some D8}} (2.5+0.2,-0.75);
 \end{tikzpicture}
 \caption{}
 \label{subfig:6d_codim4_b}
\end{subfigure}
\begin{subfigure}{0.3275\textwidth}
\centering
 \begin{tikzpicture}
  \draw (0,0.1)--(3,0.1) (0,-0.1)--(3,-0.1);
  \draw (1+0.5,1)--(1+0.5,-1) (1.1+0.5,1)--(1.1+0.5,-1) (1.2+0.5,1)--(1.2+0.5,-1) (1.3+0.5,1)--(1.3+0.5,-1);
  \draw (2+0.5,1)--(2+0.5,-1) (1.9+0.5,1)--(1.9+0.5,-1) (1.8+0.5,1)--(1.8+0.5,-1) (1.7+0.5,1)--(1.7+0.5,-1);
  \node at (2,0.25) {$\cdots$};
  \node at (2,-0.25) {$\cdots$};
  \draw (-0.4,-1)--(-0.4,1) (4+0.4,-1)--(4+0.4,1);
    \draw (-0.6,-1)--(-0.6,1) (4+0.6,-1)--(4+0.6,1);
    \draw (-0.4,0.75)--(4.4,0.75) (-0.6,-0.75)--(4.6,-0.75);
    \draw[red,thick] (4,-0.75)--(4,0.75);
    \draw[blue,thick] (3,0)--(4,0);
  \ns{0,0}
  \ns{3,0}
  \node at (0.75,0.25) {\footnotesize{$N$ D6}};
  \node[blue] at (3.5,0.25) {\footnotesize{D2}};
  \draw[decoration={brace,mirror,raise=10pt},decorate,thick]
  (1.5-0.2,-0.75) -- node[below=15pt] {\footnotesize{some D8}} (2.5+0.2,-0.75);
 \end{tikzpicture}
 \caption{}
 \label{subfig:6d_codim4_c}
\end{subfigure}
\\
\begin{subfigure}{0.3275\textwidth}
\centering
 \begin{tikzpicture}
  \draw (0,0.1)--(3,0.1) (0,-0.1)--(3,-0.1);
  \draw (1+0.5,1)--(1+0.5,-1) (1.1+0.5,1)--(1.1+0.5,-1) (1.2+0.5,1)--(1.2+0.5,-1) (1.3+0.5,1)--(1.3+0.5,-1);
  \draw (2+0.5,1)--(2+0.5,-1) (1.9+0.5,1)--(1.9+0.5,-1) (1.8+0.5,1)--(1.8+0.5,-1) (1.7+0.5,1)--(1.7+0.5,-1);
  \node at (2,0.25) {$\cdots$};
  \node at (2,-0.25) {$\cdots$};
    \draw[blue,thick] (3,0)--(4,0);
  \ns{0,0}
  \ns{3,0}
  \node at (0.75,0.25) {\footnotesize{$N$ D6}};
  \draw[decoration={brace,mirror,raise=10pt},decorate,thick]
  (1.5-0.2,-0.75) -- node[below=15pt] {\footnotesize{some D8}} (2.5+0.2,-0.75);
 \end{tikzpicture}
 \caption{}
 \label{subfig:6d_codim4_d}
\end{subfigure}
\begin{subfigure}{0.3275\textwidth}
\centering
 \begin{tikzpicture}
  \draw (0,0.1)--(3,0.1) (0,-0.1)--(3,-0.1);
  \draw (1+0.5,1)--(1+0.5,-1) (1.1+0.5,1)--(1.1+0.5,-1) (1.2+0.5,1)--(1.2+0.5,-1) (1.3+0.5,1)--(1.3+0.5,-1);
  \draw (2+0.5,1)--(2+0.5,-1) (1.9+0.5,1)--(1.9+0.5,-1) (1.8+0.5,1)--(1.8+0.5,-1) (1.7+0.5,1)--(1.7+0.5,-1);
  \node at (2,0.25) {$\cdots$};
  \node at (2,-0.25) {$\cdots$};
    \draw[blue,thick] (3,0)--(4,0);
    \ThreeB{4,0}
  \ns{0,0}
  \ns{3,0}
  \node at (0.75,0.25) {\footnotesize{$N$ D6}};
  \node at (4,0.25) {\footnotesize{D$4^\prime$}};
  \draw[decoration={brace,mirror,raise=10pt},decorate,thick]
  (1.5-0.2,-0.75) -- node[below=15pt] {\footnotesize{some D8}} (2.5+0.2,-0.75);
 \end{tikzpicture}
 \caption{}
 \label{subfig:6d_codim4_e}
\end{subfigure}
\hfill
\begin{subfigure}{0.3275\textwidth}
\centering
 \begin{tikzpicture}
  \draw (0,0.1)--(4,0.1) (0,-0.1)--(4,-0.1);
  \draw (1+0.5,1)--(1+0.5,-1) (1.1+0.5,1)--(1.1+0.5,-1) (1.2+0.5,1)--(1.2+0.5,-1) (1.3+0.5,1)--(1.3+0.5,-1);
  \draw (2+0.5,1)--(2+0.5,-1) (1.9+0.5,1)--(1.9+0.5,-1) (1.8+0.5,1)--(1.8+0.5,-1) (1.7+0.5,1)--(1.7+0.5,-1);
  \node at (2,0.25) {$\cdots$};
  \node at (2,-0.25) {$\cdots$};
\ThreeB{3,0}
  \ns{0,0}
  \ns{4,0}
  \node at (0.75,0.25) {\footnotesize{$N$ D6}};
  \draw[decoration={brace,mirror,raise=10pt},decorate,thick]
  (1.5-0.2,-0.75) -- node[below=15pt] {\footnotesize{some D8}} (2.5+0.2,-0.75);
 \end{tikzpicture}
 \caption{}
 \label{subfig:6d_codim4_f}
\end{subfigure}
\caption{Cartoon for the realization of a codimension 4 defect via tuning of two codimension 2 defects. For simplicity, the O6-plane is omitted in this discussion. In \subref{subfig:6d_codim4_a}, two codimension 2 defects are introduced via defect Higgsing that leave behind D4-branes. In \subref{subfig:6d_codim4_b}, the defect D4-branes are tuned such that they can reconnect and form a single brane. The tuning is such that the resulting D4-brane can move along the D6-branes in $x^6$ direction, but leaves behind a D2-brane extended in $x^{0,1,6}$ as in \subref{subfig:6d_codim4_c}. In \subref{subfig:6d_codim4_d}, the D4 and the D6-branes have been moved off to infinite. In \subref{subfig:6d_codim4_e}, the semi-infinite D2 is terminated on a D$4^\prime$-brane, which is extended along $x^{0,1,7,8,9}$. In \subref{subfig:6d_codim4_f}, the D$4^\prime$ is moved through the NS5 upon which the D2 is annihilated.}
\label{fig:6d_codim4_double_Higgs}
\end{figure}
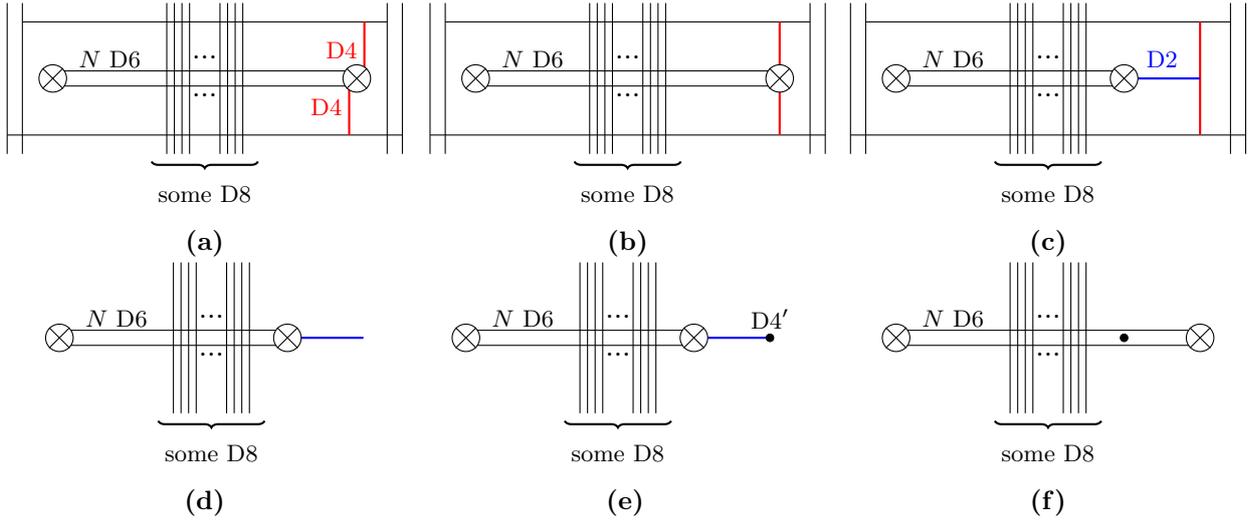

Alternatively, the codimension 4 defect can be introduced by two consecutive defect Higgsings. In order to demonstrate the idea, the O6-plane is omitted for the moment. Starting from a brane configuration as in Figure \ref{subfig:6d_codim4_a}, two D6-branes, which have been moved onto the Higgs branch, are connected to the remaining brane configuration via D4-branes. Each of which is associated with a defect parameter. These two defects can be tuned suitably such that the defect D4-branes can reconnect, as in Figure \ref{subfig:6d_codim4_b}. However, reconnecting is not enough; one needs to tune the parameters such that the D4 can be moved along the D6-branes, but leaves behind a D2-brane connected to the residual brane configuration, see Figure \ref{subfig:6d_codim4_c}. This tuning is analogous to the modification in the defect Higgsing which leaves behind a D4-brane when a D6-brane moves away from the brane configuration. Finally, moving the defect D4-brane, together with the D6-branes which are on the Higgs branch, off to infinity leaves the brane configuration in Figure \ref{subfig:6d_codim4_d}. Crucially, the semi-infinite D2 is extended in $x^{0,1,6}$ directions such that it is of codimension 4 for the 6d world-volume theory. Moreover, the Poincar\'{e} symmetry $\sorm(4)_{2345}$, which was broken by the codimension 2 defect D4-branes, is restored. Hence, the D2-brane realizes a codimension 4 defect.

One may compare this codimension 4 defect to the one introduced by a D$4^\prime$-brane as in Table \ref{tab:branes_6d}, see for instance \cite{Agarwal:2018tso}. For this, one notices that the semi-infinite D2 can be terminated one a D$4^\prime$ without breaking supersymmetry, see Figure \ref{subfig:6d_codim4_e}. This is because the NS5-D2-D$4^\prime$ system forms a Hanany-Witten setup \cite{Hanany:1996ie}. Therefore, the D$4^\prime$-brane can be moved through the NS5-brane, upon which the D2-brane is annihilated and one ends up in the configuration of Figure \ref{subfig:6d_codim4_f}. 
\paragraph{Field theory.}
One can construct this codimension 4 defect in the field theory as follows. First, a codimension 2 defect supporting $x^{2,3}$ directions is introduced by giving a position-dependent VEV to a moment map operator breaking $\sprm(2N-4)$ flavor symmetry to $\sprm(2N-6)$ and another codimension 2 defect of the same type along the same plane is introduced so that the flavor symmetry is further broken to $\sprm(2N-8)$. These two codimension 2 defects amount to the two independent D4-branes in Figure \ref{subfig:6d_codim4_a}. The theory then has the emergent $\urm(2)$ defect flavor symmetry rotating two codimension 2 defects and the associated 4d scalar fields carry the defect flavor charges. Now a position-dependent VEV along the $x^{4,5}$ plane is turned on for the 4d scalar fields such that the $\surm(2)\subset \urm(2)$ defect symmetry is broken. Eventually, this introduces a codimension 4 defect described by the D2-brane in the brane configuration.
\subsection{5d perspective}
\label{subsec:5d_branes}
T-dualizing the 6d setup of Table \ref{tab:branes_6d} along the $x^1$ direction allows to derive 5-brane webs for the relevant 5d KK theories \cite{Hayashi:2015vhy,Kim:2019dqn} with space-time occupation as in Table \ref{tab:branes_5d}. For later purposes, the five-dimensional world-volume theory is placed on an $\Omega$-background of the form $S^1\times \R^4_{\epsilon_1,\epsilon_2}$. The circle extends along $x^0$ and the $\R^4_{\epsilon_1,\epsilon_2}$ occupies $x^{2,3,4,5}$, as in the 6d origin.

A crucial advantage is that there exists even a 5-brane construction for the $\mathbb{Z}_2$-twisted theories. To begin with, the construction is briefly reviewed.

\begin{table}[t]
\begin{center}
 \begin{tabular}{c|cccccccccc}
 \toprule 
  IIB & $x^0$ &  $x^1$ & $x^2$ & $x^3$ & $x^4$ & $x^5$ & $x^6$ & $x^7$ & $x^8$ & $x^9$ \\ \midrule
NS5 & $\times$ & $\times$ & $\times$ & $\times$ & $\times$ & $\times$ & \\
D5/O5& $\times$ & & $\times$ & $\times$ & $\times$ & $\times$ &  $\times$ \\
D7& $\times$ &  & $\times$ & $\times$ & $\times$ & $\times$ &   & $\times$ & $\times$ &  $\times$\\
\midrule
D3& $\times$ &  & $\times$ & $\times$ &  &  &  & $\times$ \\
\midrule
D$3^\prime$& $\times$ &  &  &  &  &  &  & $\times$ & $\times$ & $\times$ \\
\bottomrule
 \end{tabular}
 \caption{The brane configuration for the circle compactification of 6d $\Ncal=(1,0)$ $\sorm(N_c)$ theories with $N_f=N_c-8$ flavors. Note that a generic $(p,q)$ 5-brane is depicted as line in the $(x^1,x^6)$ plane with slope determined by $p,q$. Besides D7-branes, there may also be $[p,q]$ 7-branes, on which $(p,q)$ 5-branes can end. The codimension 2 defect is realized by the D3-brane; while the D$3^\prime$-brane introduces a codimension 4 defect.}
 \label{tab:branes_5d}
 \end{center}
\end{table}

\begin{figure}[t]
 \centering 
 \begin{subfigure}{0.275\textwidth}
 \centering 
 \scalebox{0.5}{
 \begin{tikzpicture}
  \draw[dashed] (0,0)--(8,0);
  \draw (1,0)--(2,0.5)--(2.5,1)--(2.5,1.5)--(3,2)--(3,2.5)--(3.5,3)--(3.5,3.5);
  \draw (7,0)--(6,0.5)--(5.5,1)--(5.5,1.5)--(5,2)--(5,2.5)--(4.5,3)--(4.5,3.5);
  \draw (2,0.5)--(6,0.5) (2.5,1)--(5.5,1) (3,2)--(5,2) (3.5,3)--(4.5,3);
    \draw (2.5,1.5)--(2,1.5) (3,2.5)--(2,2.5);
    \draw (5.5,1.5)--(6,1.5) (5,2.5)--(6,2.5);
    \SevenB{6,2.5}
    \SevenB{6,1.5}
    \SevenB{2,2.5}
    \SevenB{2,1.5}
%
    \node at (4,3.75) {$\vdots$};
\begin{scope}[shift={(8,7.5)},rotate=180]
 \draw[dashed] (0,0)--(8,0);
  \draw (1,0)--(2,0.5)--(2.5,1)--(2.5,1.5)--(3,2)--(3,2.5)--(3.5,3)--(3.5,3.5);
  \draw (7,0)--(6,0.5)--(5.5,1)--(5.5,1.5)--(5,2)--(5,2.5)--(4.5,3)--(4.5,3.5);
  \draw (2,0.5)--(6,0.5) (2.5,1)--(5.5,1) (3,2)--(5,2) (3.5,3)--(4.5,3);
    \draw (2.5,1.5)--(2,1.5) (3,2.5)--(2,2.5);
    \draw (5.5,1.5)--(6,1.5) (5,2.5)--(6,2.5);
    \SevenB{6,2.5}
    \SevenB{6,1.5}
    \SevenB{2,2.5}
    \SevenB{2,1.5}
\end{scope}
\begin{scope}[shift={(0,7.5)}]
  \draw (1,0)--(2,0.5)--(2.5,1)--(2.5,1.5)--(3,2)--(3,2.5)--(3.5,3)--(3.5,3.5);
  \draw (7,0)--(6,0.5)--(5.5,1)--(5.5,1.5)--(5,2)--(5,2.5)--(4.5,3)--(4.5,3.5);
  \draw (2,0.5)--(6,0.5) (2.5,1)--(5.5,1) (3,2)--(5,2) (3.5,3)--(4.5,3);
    \draw (2.5,1.5)--(2,1.5) (3,2.5)--(2,2.5);
    \draw (5.5,1.5)--(6,1.5) (5,2.5)--(6,2.5);
    \SevenB{6,2.5}
    \SevenB{6,1.5}
    \SevenB{2,2.5}
    \SevenB{2,1.5}
%
    \node at (4,3.75) {$\vdots$};
\begin{scope}[shift={(8,7.5)},rotate=180]
 \draw[dashed] (0,0)--(8,0);
  \draw (1,0)--(2,0.5)--(2.5,1)--(2.5,1.5)--(3,2)--(3,2.5)--(3.5,3)--(3.5,3.5);
  \draw (7,0)--(6,0.5)--(5.5,1)--(5.5,1.5)--(5,2)--(5,2.5)--(4.5,3)--(4.5,3.5);
  \draw (2,0.5)--(6,0.5) (2.5,1)--(5.5,1) (3,2)--(5,2) (3.5,3)--(4.5,3);
    \draw (2.5,1.5)--(2,1.5) (3,2.5)--(2,2.5);
    \draw (5.5,1.5)--(6,1.5) (5,2.5)--(6,2.5);
    \SevenB{6,2.5}
    \SevenB{6,1.5}
    \SevenB{2,2.5}
    \SevenB{2,1.5}
\end{scope}
  \draw[decoration={brace,mirror,raise=10pt},decorate,thick]
  (1.75,6.25) -- node[left=10pt] {\LARGE{$N{-}4$}} (1.75,1.25);
    \draw[decoration={brace,mirror,raise=10pt},decorate,thick]
  (6.25,1.25) -- node[right=10pt] {\LARGE{$N{-}4$}} (6.25,6.25);
\end{scope}
%
    \draw[decoration={brace,mirror,raise=10pt},decorate,thick]
  (1.75,6.25) -- node[left=10pt] {\LARGE{$N{-}4$}} (1.75,1.25);
    \draw[decoration={brace,mirror,raise=10pt},decorate,thick]
  (6.25,1.25) -- node[right=10pt] {\LARGE{$N{-}4$}} (6.25,6.25);
%
    \node at (0.5,-0.35) {\LARGE{$\mathrm{O}5^+$}};
    \node at (4,-0.35) {\LARGE{$\mathrm{O}5^-$}};
    \node at (7.5,-0.35) {\LARGE{$\mathrm{O}5^+$}};
    \node at (0.5,7.75) {\LARGE{$\mathrm{O}5^+$}};
    \node at (4,7.75) {\LARGE{$\mathrm{O}5^-$}};
    \node at (7.5,7.75) {\LARGE{$\mathrm{O}5^+$}};
    \node at (0.5,15.35) {\LARGE{$\mathrm{O}5^+$}};
    \node at (4,15.35) {\LARGE{$\mathrm{O}5^-$}};
    \node at (7.5,15.35) {\LARGE{$\mathrm{O}5^+$}};
%
 \node at (2,0) {$\slash\slash$};
 \node at (6,0) {$\slash\slash$};
 \node at (2,15) {$\slash\slash$};
 \node at (6,15) {$\slash\slash$};
 \end{tikzpicture}
 }
 \caption{}
 \label{subfig:5d_branes_SOeven_double}
 \end{subfigure}
\hfill
 \begin{subfigure}{0.675\textwidth}
 \centering
 \begin{tikzpicture}
  \draw[dashed] (0,0)--(8,0);
  \draw (1,0)--(2,0.5)--(2.5,1)--(2.5,1.5)--(3,2)--(3,2.5)--(3.5,3)--(3.5,3.5);
  \draw (7,0)--(6,0.5)--(5.5,1)--(5.5,1.5)--(5,2)--(5,2.5)--(4.5,3)--(4.5,3.5);
  \draw (2,0.5)--(6,0.5) (2.5,1)--(5.5,1) (3,2)--(5,2) (3.5,3)--(4.5,3);
    \draw (2.5,1.5)--(2,1.5) (3,2.5)--(2,2.5);
    \draw (5.5,1.5)--(6,1.5) (5,2.5)--(6,2.5);
    \SevenB{6,2.5}
    \SevenB{6,1.5}
    \SevenB{2,2.5}
    \SevenB{2,1.5}
%
    \node at (4,3.75) {$\vdots$};
\begin{scope}[shift={(8,7.5)},rotate=180]
 \draw[dashed] (0,0)--(8,0);
  \draw (1,0)--(2,0.5)--(2.5,1)--(2.5,1.5)--(3,2)--(3,2.5)--(3.5,3)--(3.5,3.5);
  \draw (7,0)--(6,0.5)--(5.5,1)--(5.5,1.5)--(5,2)--(5,2.5)--(4.5,3)--(4.5,3.5);
  \draw (2,0.5)--(6,0.5) (2.5,1)--(5.5,1) (3,2)--(5,2) (3.5,3)--(4.5,3);
    \draw (2.5,1.5)--(2,1.5) (3,2.5)--(2,2.5);
    \draw (5.5,1.5)--(6,1.5) (5,2.5)--(6,2.5);
    \SevenB{6,2.5}
    \SevenB{6,1.5}
    \SevenB{2,2.5}
    \SevenB{2,1.5}
\end{scope}
%
    \draw[decoration={brace,mirror,raise=10pt},decorate,thick]
  (1.75,6.25) -- node[left=10pt] {\footnotesize{$N{-}4$}} (1.75,1.25);
    \node at (0.5,-0.25) {\footnotesize{$\mathrm{O}5^+$}};
    \node at (4,-0.25) {\footnotesize{$\mathrm{O}5^-$}};
    \node at (7.5,-0.25) {\footnotesize{$\mathrm{O}5^+$}};
    \node at (0.5,7.75) {\footnotesize{$\mathrm{O}5^+$}};
    \node at (4,7.75) {\footnotesize{$\mathrm{O}5^-$}};
    \node at (7.5,7.75) {\footnotesize{$\mathrm{O}5^+$}};
 \draw [line join=round,decorate, decoration={zigzag, segment length=4,amplitude=.9,post=lineto,post length=2pt},orange] (5,0.5)--(5,1) (4,1)--(4,2) (4.25,2)--(4.25,3) (5,6.5)--(5,7) (4,5.5)--(4,6.5) (4.25,4.5)--(4.25,5.5);
 \draw [line join=round,decorate, decoration={zigzag, segment length=4,amplitude=.9,post=lineto,post length=2pt},orange] (3,1) .. controls (3.24,-0.25) .. (3.4,0.5);
 \draw [line join=round,decorate, decoration={zigzag, segment length=4,amplitude=.9,post=lineto,post length=2pt},orange] (3,6.5) .. controls (3.24,7.75) .. (3.4,7);
\node[orange] at (2.75,0.25) {\tiny{$W_1$}};
\node[orange] at (4.75,0.75) {\tiny{$W_0$}};
\node[orange] at (3.75,1.5) {\tiny{$W_2$}};
\node[orange] at (4,2.5) {\tiny{$W_3$}};
\node[orange] at (2.75,0.25+7) {\tiny{$W_N$}};
\node[orange] at (4.75-0.15,0.75+6) {\tiny{$W_{N-1}$}};
\node[orange] at (3.75-0.25,1.5+4.5) {\tiny{$W_{N-2}$}};
\node[orange] at (4-0.25,2.5+2.5) {\tiny{$W_{N-3}$}};
%
\draw[dotted] (6,0.5)--(8,0.5) (5.5,1)--(8,1) (5,2)--(8,2) (4,3)--(8,3) 
(6,0.5+6.5)--(8,0.5+6.5) (5.5,1+5.5)--(8,1+5.5) (5,2+3.5)--(8,2+3.5) (4,3+1.5)--(8,3+1.5);
\draw[dotted,<->] (7,0)--(7,0.5);
\draw[dotted,<->] (6.25,0.5)--(6.25,1);
\draw[dotted,<->] (6.5,1)--(6.5,2);
\draw[dotted,<->] (6.4,2)--(6.4,3);
\draw[dotted,<->] (7,0+7)--(7,0.5+7);
\draw[dotted,<->] (6.25,0.5+6)--(6.25,1+6);
\draw[dotted,<->] (6.5,1+4.5)--(6.5,2+4.5);
\draw[dotted,<->] (6.4,2+2.5)--(6.4,3+2.5);
\node at (7.5,0.25) {\tiny{$\phi_1-\phi_0$}};
\node at (7,0.75) {\tiny{$2\phi_0-\phi_2$}};
\node at (7.75,1.5) {\tiny{$2\phi_2-\phi_0-\phi_1 -\phi_3$}};
\node at (7.5,2.5) {\tiny{$2\phi_3-\phi_2-\phi_4$}};
\node at (7.75,0.25+7) {\tiny{$\phi_N-\phi_{N{-}1}$}};
\node at (7.25,0.75+6) {\tiny{$2\phi_{N{-}1}-\phi_{N{-}2}$}};
\node at (8.25,1.5+4.5) {\tiny{$2\phi_{N{-}2}-\phi_{N{-}3}-\phi_{N{-}1} -\phi_{N}$}};
\node at (7.75,2.5+2.5) {\tiny{$2\phi_{N{-}3}-\phi_{N{-}4}-\phi_{N{-}2}$}};
\draw[->] (0.5-1,6.5)--(1-1,6.5);
\draw[->] (0.5-1,6.5)--(0.5-1,7);
\node at (1.25-1,6.5) {\footnotesize{$x^6$}};
\node at (0.8-1,7.05) {\footnotesize{$x^1$}};
 \end{tikzpicture}
 \caption{}
 \label{subfig:5d_branes_SOeven_half}
 \end{subfigure}
\caption{5-brane web for the 6d $\sorm(2N)$ gauge theory compactified on a circle. In \subref{subfig:5d_branes_SOeven_double}, the covering space of the 5-brane web is shown; that is the T-dual of Figure \ref{subfig:6d_branes_SOeven}, which contains two O5-planes and the $x^1$ direction (here the vertical direction) is a circle. For the remainder of the paper, only half of 5-brane web is drawn as in \subref{subfig:5d_branes_SOeven_half}, because it contains all relevant information. The flavor D5-branes have been terminated on (half) 7-branes, denoted by $\bigcirc$. These (half) $[1,0]$ 7-branes are T-dual to the half D8-branes from the Type IIA configuration. The monodromy cuts of the 7-branes have been omitted.}
\label{fig:5d_branes_SOeven}
\end{figure}
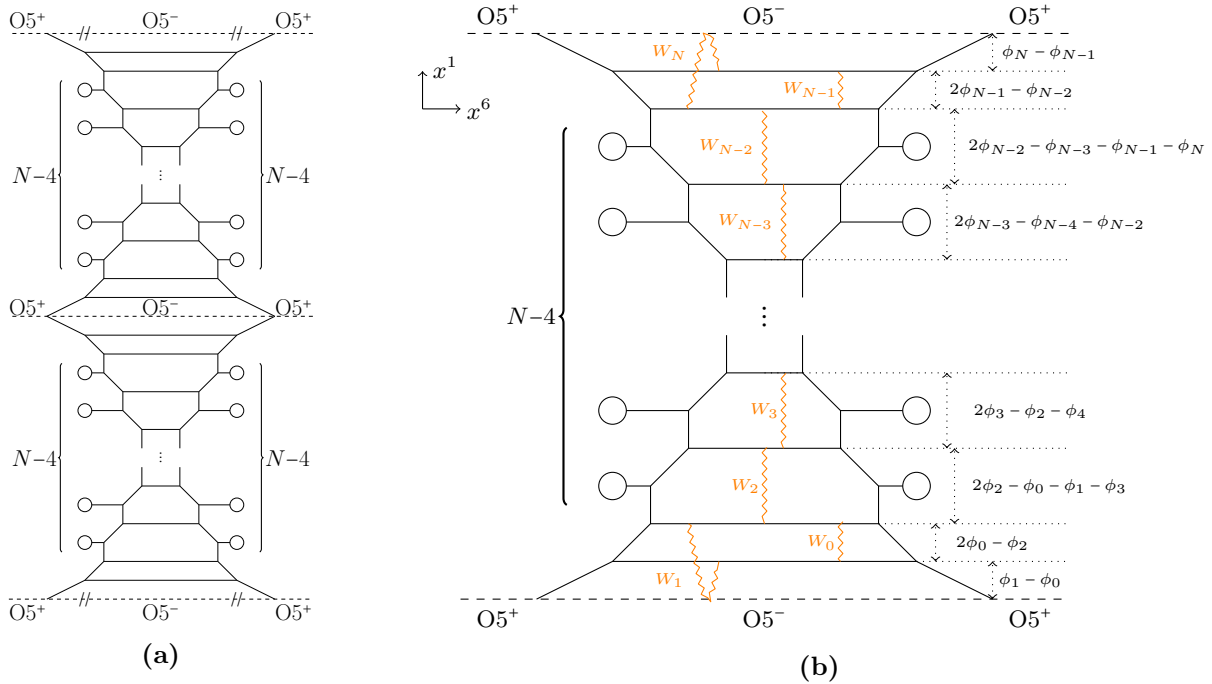

\paragraph{5-brane web for SO(even).}
Returning to Figure \ref{subfig:6d_branes_SOeven}, T-dualizing the O6-plane leads to two O5-planes in the 5-brane web, which is depicted in Figure \ref{subfig:5d_branes_SOeven_double}. The figure shows the covering space, where the top and bottom O5-plane are identified such that the configuration has 2 orientifolds. Consequently, the top half of the 5-brane web is the mirror image of the bottom half; thus, it is sufficient to display only one half. The 5-brane web in Figure \ref{subfig:5d_branes_SOeven_half} displays $N$ internal half D5-branes, which give rise to a $\sorm(2N)$ gauge group due to the O5 orientifold action. The external half D5-branes are terminated on half $[1,0]$ 7-branes, i.e.\ half D7-branes. The two sets of $N-4$ flavor branes result in $N_f= 2N-8$ fundamental flavors of the $\sorm(2N)$ gauge group.
From the 5-brane web in Figure \ref{subfig:5d_branes_SOeven_half}, the masses of the $W_i$-bosons, which are stretched between adjacent D5-branes and the orientifold, are given by
\begin{align}
\begin{pmatrix}
 m_{W_0} \\ m_{W_1} \\ \vdots \\ m_{W_N}
\end{pmatrix}
= C_{\widehat{D}_N^{(1)}} 
\begin{pmatrix}
 \phi_0 \\ \phi_1 \\ \vdots \\ \phi_N
\end{pmatrix}
\end{align}
where $ C_{\widehat{D}_N^{(1)}} $ is the (untwisted) affine Cartan matrix of $D_N$ gauge algebra. The $\phi_i$ denotes the dynamical Coulomb branch parameters, which are chosen such that the simple roots of $D_N$ are expressed in the Dynkin basis. The two D-type bifurcations of the $\widehat{D}_N^{(1)}$ Dynkin diagram are due to the boundary conditions, see for example \cite{Hanany:2001iy}: firstly, F1 cannot end on O5${}^-$; secondly, F1 cannot end on the mirror image of the same D5 it started from.

\begin{figure}[t]
 \centering
 \begin{subfigure}{0.475\textwidth}
\centering  
\scalebox{0.8}{
\begin{tikzpicture}
  \draw[dashed] (0,0)--(10,0);
  \draw (1,0)--(2,0.5)--(3.5,1)--(4.5,1.5)--(5,2);
  \draw (6,2)--(6,1.5)--(6.5,1)--(8.5,0);
  \draw (3.5,1)--(6.5,1) (4.5,1.5)--(6,1.5);
  \draw (0.5,0.5)--(2,0.5);
  \draw[red,dotted,thick] (0,0.5)--(0.5,0.5);
  \SevenB{0.5,0.5} 
 \node at (0.5,-0.35) {\footnotesize{$\mathrm{O}5^+$}};
 \node at (3,-0.35) {\footnotesize{$\mathrm{O}5^-$}};
 \node at (9.5,-0.35) {\footnotesize{$\mathrm{O}5^+$}};
\node at (5.5,2) {$\vdots$};
 \end{tikzpicture}
 }
 \caption{}
 \label{subfig:5d_Higgs_a}
 \end{subfigure}
\hfill
\begin{subfigure}{0.475\textwidth}
\centering  
\scalebox{0.8}{
\begin{tikzpicture}
  \draw[dashed] (0,0)--(10,0);
  \draw (1,0)--(2,0.5)--(3.5,1)--(4.5,1.5)--(5,2);
  \draw (6,2)--(6,1.5)--(6.5,1)--(8.5,0);
  \draw (3.5,1)--(6.5,1) (4.5,1.5)--(6,1.5);
  \draw[red,dotted,thick] (0,0.5)--(4.5,0.5);
  \SevenB{4.5,0.5} 
 \node at (0.5,-0.35) {\footnotesize{$\mathrm{O}5^+$}};
 \node at (3,-0.35) {\footnotesize{$\mathrm{O}5^-$}};
 \node at (9.5,-0.35) {\footnotesize{$\mathrm{O}5^+$}};
\node at (5.5,2) {$\vdots$};
 \end{tikzpicture}
 }
 \caption{}
 \label{subfig:5d_Higgs_b}
 \end{subfigure}
 \\
\begin{subfigure}{0.475\textwidth}
\centering  
\scalebox{0.8}{
\begin{tikzpicture}
  \draw[dashed] (0,0)--(10,0);
  \draw (2,0)--(3.5,0.5)--(4.5,1)--(5,1.5);
  \draw (6,1.5)--(6,1)--(6.5,0.5)--(7.5,0);
  \draw (3.5,0.5)--(6.5,0.5) (4.5,1)--(6,1);
  \draw[red,dotted,thick] (0,0.1)--(4,0.1) (0,-0.1)--(5,-0.1);
  \SevenB{4,0}
  \SevenB{5,0} 
 \node at (0.25,-0.35) {\footnotesize{$\mathrm{O}5^+$}};
 \node at (3,-0.35) {\footnotesize{$\mathrm{O}5^-$}};
 \node at (4.5,-0.35) {\footnotesize{$\widetilde{\mathrm{O}5}^-$}};
 \node at (7,-0.35) {\footnotesize{$\mathrm{O}5^-$}};
 \node at (9.5,-0.35) {\footnotesize{$\mathrm{O}5^+$}};
\node at (5.5,1.5) {$\vdots$};
 \end{tikzpicture}
 }
 \caption{}
 \label{subfig:5d_Higgs_c}
 \end{subfigure}
\hfill
\begin{subfigure}{0.475\textwidth}
\centering  
\scalebox{0.8}{
\begin{tikzpicture}
  \draw[dashed] (0,0)--(10,0);
  \draw (2,0)--(4,1)--(4.5,1.5);
  \draw (6,1.5)--(6,1)--(7,0);
  \draw (2.275,0.15)--(6.85,0.15) (4,1)--(6,1);
  \draw[red,dotted,thick] (0,0.075)--(4,0.075) (0,-0.075)--(5,-0.075);
  \SevenB{4,0}
  \SevenB{5,0} 
 \node at (0.25,-0.35) {\footnotesize{$\mathrm{O}5^+$}};
 \node at (3,-0.35) {\footnotesize{$\mathrm{O}5^-$}};
 \node at (4.5,-0.35) {\footnotesize{$\widetilde{\mathrm{O}5}^-$}};
 \node at (7,-0.35) {\footnotesize{$\mathrm{O}5^-$}};
 \node at (9.5,-0.35) {\footnotesize{$\mathrm{O}5^+$}};
\node at (5.25,1.5) {$\vdots$};
 \end{tikzpicture}
 }
 \caption{}
 \label{subfig:5d_Higgs_d}
 \end{subfigure}
 \\
\begin{subfigure}{0.475\textwidth}
\centering  
\scalebox{0.8}{
\begin{tikzpicture}
  \draw[dashed] (0,0)--(10,0);
  \draw (2,0)--(4,1)--(4.5,1.5);
  \draw (6,1.5)--(6,1)--(7,0);
  \draw (2.275,0.15)--(4,0.15)  (5,0.15)--(6.85,0.15) (4,1)--(6,1);
  \draw[red,dotted,thick] (0,0.075)--(4,0.075) (0,-0.075)--(5,-0.075);
  \SevenB{4,0}
  \SevenB{5,0} 
 \node at (0.25,-0.35) {\footnotesize{$\mathrm{O}5^+$}};
 \node at (3,-0.35) {\footnotesize{$\mathrm{O}5^-$}};
 \node at (4.5,-0.35) {\footnotesize{$\widetilde{\mathrm{O}5}^-$}};
 \node at (7,-0.35) {\footnotesize{$\mathrm{O}5^-$}};
 \node at (9.5,-0.35) {\footnotesize{$\mathrm{O}5^+$}};
\node at (5.25,1.5) {$\vdots$};
 \end{tikzpicture}
 }
 \caption{}
 \label{subfig:5d_Higgs_e}
 \end{subfigure}
\hfill
\begin{subfigure}{0.475\textwidth}
\centering  
\scalebox{0.8}{
\begin{tikzpicture}
  \draw[dashed] (0,0)--(10,0);
  \draw (2,0)--(4,1)--(4.5,1.5);
  \draw (6,1.5)--(6,1)--(7,0);
  \draw  (4,1)--(6,1);
  \draw[red,dotted,thick] (0,0.075)--(1,0.075) (0,-0.075)--(8,-0.075);
  \SevenB{1,0}
  \SevenB{8,0}
 \node at (0.25,-0.35) {\footnotesize{$\mathrm{O}5^+$}};
 \node at (1.5,-0.35) {\footnotesize{$\widetilde{\mathrm{O}5}^+$}};
 \node at (4.5,-0.35) {\footnotesize{$\widetilde{\mathrm{O}5}^-$}};
 \node at (7.5,-0.35) {\footnotesize{$\widetilde{\mathrm{O}5}^+$}};
 \node at (9.5,-0.35) {\footnotesize{$\mathrm{O}5^+$}};
\node at (5.25,1.5) {$\vdots$};
 \end{tikzpicture}
 }
 \caption{}
 \label{subfig:5d_Higgs_f}
 \end{subfigure}
 \caption{Higgsing in the 5-brane web. In \subref{subfig:5d_Higgs_a}, a single half D7 has been moved towards one of the O5-planes as compared to the configuration in Figure \ref{subfig:5d_branes_SOeven_half}. The red dotted line denotes the half monodromy cut of the 7-brane. In \subref{subfig:5d_Higgs_b}, the half D7 has been moved inside the 5-brane web. During this process, the half 5-brane that connected the D7 and the web has been annihilated. In \subref{subfig:5d_Higgs_c}, the half D7 has been moved onto the orientifold plane, where it meets its mirror image. Subsequently, the two half D7-branes can be separated along the O5. Following charge conservation, the orientifold between the two half D7 is a $\widetilde{\mathrm{O}5}^-$. In \subref{subfig:5d_Higgs_d}, a half D5 gauge brane (together with its mirror half brane) has been moved onto the orientifold. The D5 can be split along the half D7-branes, such that a Higgs branch direction opens up. In \subref{subfig:5d_Higgs_e}, the full D5 segment between the half D7-branes has been moved along the Higgs branch direction off to infinity. In \subref{subfig:5d_Higgs_f}, the half D7-branes have been moved out of the 5-brane web, during which the attached 5-branes have been annihilated. As a result, the 5-brane web is now on top of a $\widetilde{\mathrm{O}5}$-plane.}
 \label{fig:details_Higgs_5brane_web}
\end{figure}
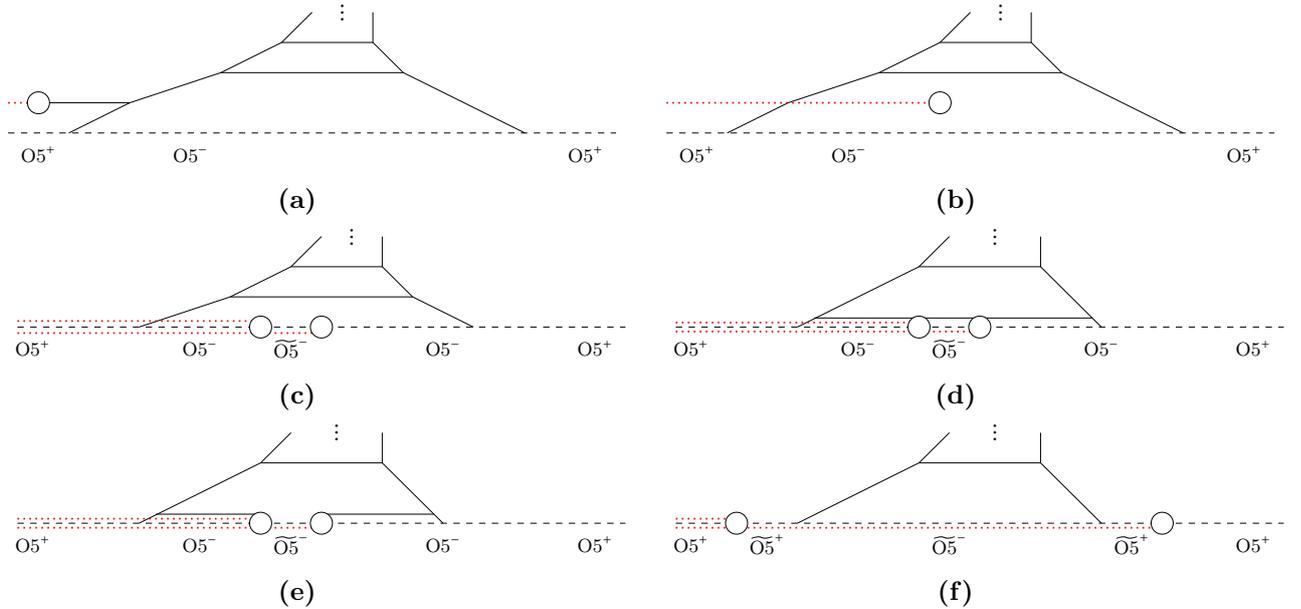

\paragraph{Higgsing to SO(odd).}
First, one can repeat the Higgsing from $\sorm(2N)$ to $\sorm(2N-1)$ in the 5-brane web. As before, one may align a flavor half D5 with an internal gauge half D5 as displayed in Figure \ref{subfig:5-brane_Higgs_SOodd}, and the 
resulting half brane (depicted in red) can be moved onto the top O5-plane. Note that there is the corresponding mirror brane approaching the O5-plane from above. The details of this Higgs mechanism in the 5-brane web are summarized in Figure \ref{fig:details_Higgs_5brane_web}.
This tuning of parameters in the 5-brane web corresponds to a Higgsing of a fundamental flavor, which is charged under the gauge symmetry. Once the associated half 7-brane and its mirror image are on the orientifold, one can move one half 7-brane to the left-hand side and the other half brane to the right-hand side. Pushing them off to $\pm \infty$ then realizes the Higgs flow to $\sorm(2N-1)$ theory, and in the process the character of the O5-plane changes accordingly.
In comparison to the discussion above, adjusting the position of two half 7-branes is in one-to-one correspondence with the tuning of two half D8-branes in the 6d setup of Figure \ref{subfig:HW_transition}. 

To make the discussion more concrete, one can assign mass and Coulomb parameter to the 5-brane webs in Figure \ref{fig:details_Higgs_5brane_web} as follows:
\begin{align}
      \raisebox{-.5\height}{
 \begin{tikzpicture}
 \begin{scope}[rotate=180]
  \draw[dashed] (0,0)--(8,0);
  \draw (1,0)--(2,0.5)--(2.5,1)--(3.5,1.5)--(4,2);
  \draw (5,2)--(5,1.5)--(6,0.5)--(7,0);
  \draw (2,0.5)--(6,0.5) (3.5,1.5)--(5,1.5);
  \draw (0.5,1)--(2.5,1);
 \draw[dotted] (6,0.5)--(8,0.5) (5,1.5)--(8,1.5);
 \draw[dotted,<->] (7,0)--(7,0.5);
 \draw[dotted,<->] (6.25,0.5)--(6.25,1.5);
 \node at (8,0.25) {\tiny{$\phi_N-\phi_{N-1}$}};
 \node at (7.5,1) {\tiny{$2\phi_{N-1}-\phi_{N-2}$}};
 \draw[dotted] (0,0.5)--(2,0.5) (0,1.5)--(3.5,1.5) (0,1)--(0.5,1);
 \draw[dotted,<->] (2,0.5)--(2,1);
 \draw[dotted,<->] (2.5,1)--(2.5,1.5);
 \node at (0.5,0.75) {\tiny{$\phi_{N-1} -\phi_N-m$}};
 \node at (1,1.25) {\tiny{$\phi_{N-1} +\phi_{N}+\phi_{N-2} +m$}};
 \end{scope}
 \end{tikzpicture}
 }
 \label{eq:5-brane_normal_Higgs}
\end{align}
and the Higgs mechanism implies that the mass of the fundamental hypermultiplet and the mass of the W-boson are tuned to zero, i.e.
\begin{align}
    \phi_N-\phi_{N-1}=0 \qquad \text{and} \qquad m=0\,.
    \label{eq:Higgs_D_to_B}
\end{align}
The analogous field theory derivation is given in \eqref{eq:Higgs_cond_D_to_B}.
In addition, it is straightforward to verify that the masses of the W-bosons of the 5-brane web in Figure \ref{subfig:5-brane_Higgs_SOodd}, after performing the transition in Figure \ref{fig:details_Higgs_5brane_web}, are given by the (untwisted) affine Cartan matrix of $B_{N-1}$ algebra,
\begin{align}
 \begin{pmatrix}
 m_{W_0} \\ m_{W_1} \\ \vdots \\ m_{W_{N-1}}
\end{pmatrix}
= C_{\widehat{B}_{N-1}^{(1)}} 
\begin{pmatrix}
 \phi_0 \\ \phi_1 \\ \vdots \\ \phi_{N-1}
\end{pmatrix} \,.
\end{align}
To see this, one can explicitly draw the fundamental strings as in Figure \ref{subfig:5-brane_SOodd_algebra}; the affine $B_{N-1}$ algebra arises by recalling that F1 can end on $\widetilde{\mathrm{O5}}^-$ due to the stuck half D5-brane, cf.\ \cite{Hanany:2001iy}. Hence, the 5-brane web depicted in \ref{subfig:5-brane_SOodd_algebra} gives rise to an $\sorm(2N-1)$ theory with $2N-7$ fundamental flavors.
\paragraph{Higgsing to twisted SO(even).}
As shown in \cite{Kim:2019dqn}, the 5-brane web allows another type of Higgsing. The Higgs mechanism considered so far assigns a large VEV to the scalar fields in the fundamental flavor hypermultiplets, which are charged under ordinary gauge symmetries. However, one may also turn on a VEV for scalar modes which carry KK-momentum.

In terms of the 5-brane web in Figure \ref{subfig:5-brane_Higgs_SOodd}, one has two options for the Higgsing of another fundamental flavor. On the one hand, one may move another flavor brane towards the $\widetilde{\mathrm{O}5}$-plane at the top. By analogy to the transition discussed in Figure \ref{fig:details_Higgs_5brane_web}, the top orientifold would change back to O5-plane. Hence, realizing the Higgsing to $\sorm(2N-1) \to \sorm(2N-2)$. One the other hand, the flavor brane can be moved towards the orientifold at the bottom, which is detailed now.

Since the 5-brane web in Figure \ref{fig:5d_branes_SOeven} originates from the 6d theory compactified on a circle of radius $R$, the distance between the two O5-planes is set by $R$. One may choose to restore the $R$ dependence in the 5-brane web, see also \cite{Kim:2019dqn,Kim:2021cua}. For instance, the vertical distance to the bottom O5-plane can be parametrised to contain explicit $\tfrac{1}{2R}$ dependence. More explicitly, reparametrising the two bottom-most distances in Figure \ref{fig:5d_branes_SOeven} via
\begin{align}
    \phi_1 - \phi_0 \to  \phi_1 - \phi_0  - \tfrac{1}{2R} 
    \; ,\qquad
    2\phi_0 - \phi_2 \to 2\phi_0 - \phi_2 + \tfrac{1}{R} \,,
\end{align}
while keeping all other distances unchanged, implies that the vertical distance between the two O5-planes is simply $\tfrac{1}{2R}$.
Next, moving, for example, a flavor brane on the right-hand side towards the bottom O5-plane, one can align it with the internal D5-brane closed to the O5-plane, as shown in Figure \ref{subfig:5-brane_Higgs_twisted}. 
Locally, the 5-brane web looks like 
\begin{align}
      \raisebox{-.5\height}{
 \begin{tikzpicture}
  \draw[dashed] (0,0)--(8,0);
  \draw (1,0)--(2,0.5)--(2.5,1)--(3.5,1.5)--(4,2);
  \draw (5,2)--(5,1.5)--(6,0.5)--(7,0);
  \draw (2,0.5)--(6,0.5) (3.5,1.5)--(5,1.5);
  \draw (0.5,1)--(2.5,1);
 \draw[dotted] (6,0.5)--(8,0.5) (5,1.5)--(8,1.5);
 \draw[dotted,<->] (7,0)--(7,0.5);
 \draw[dotted,<->] (6.25,0.5)--(6.25,1.5);
 \node at (8,0.25) {\tiny{$\phi_1-\phi_0-\tfrac{1}{2R}$}};
 \node at (7.5,1) {\tiny{$2\phi_0-\phi_2+\tfrac{1}{R}$}};
 \draw[dotted] (0,0.5)--(2,0.5) (0,1.5)--(3.5,1.5) (0,1)--(0.5,1);
 \draw[dotted,<->] (2,0.5)--(2,1);
 \draw[dotted,<->] (2.5,1)--(2.5,1.5);
 \node at (0.5,0.75) {\tiny{$\phi_0 -\phi_1+\frac{1}{R} -m$}};
 \node at (1,1.25) {\tiny{$\phi_0 +\phi_1+\phi_2 +m$}};
 \end{tikzpicture}
 }
 \label{eq:5-brane_twisted_Higgs}
\end{align}
and aligning the left flavor brane with the bottom-most internal gauge brane implies
\begin{align}
 m=\phi_0 -\phi_1+\tfrac{1}{R} \,.
\end{align}
Then moving these recombined 5-brane onto the orientifold means that some combination of Coulomb parameter $\phi$ is set to zero. However, the position of flavor brane depends on gauge and flavor holonomies as well as the KK-scale $R^{-1}$.
More concretely, pushing the resulting brane in \eqref{eq:5-brane_twisted_Higgs} onto the bottom O5-planes is realized by tuning the parameters as 
\begin{align}
 \phi_1 -\phi_0  = \tfrac{1}{2R}\quad  \Rightarrow \quad  m = \tfrac{1}{2R} \,.
 \label{eq:Higgs_twisted_D}
\end{align}
 Hence, tuning the brane onto the orientifold plane corresponds to tuning the gauge and flavor holonomies such that they cancel the KK mass, which is proportional to $R^{-1}$. The resulting brane web in Figure \ref{subfig:5-brane_twisted_algebra} has been argued \cite{Kim:2019dqn} to give rise to the $\mathbb{Z}_2$-twisted compactification of the $\sorm(2N-2)$ theory with $N_f=2N-10$ fundamental flavors. 
 For instance, it is straightforward to verify that the masses of the W-bosons in the 5-brane web of Figure \ref{subfig:5-brane_twisted_algebra}, are given by the Cartan matrix of the twisted affine $D_{N-1}$ algebra
 \begin{align}
 \begin{pmatrix}
 m_{W_1} \\ m_{W_2} \\ \vdots \\ m_{W_{N-1}}
\end{pmatrix}
= C_{\widehat{D}_{N-1}^{(2)}} 
\begin{pmatrix}
 \phi_1 \\ \phi_2 \\ \vdots \\ \phi_{N-1}
\end{pmatrix} \,.
\end{align}
 As above, the two non-simply laced edges in $\widehat{D}_{N-1}^{(2)}$ originate from the fact that F1 can end on $\widetilde{\mathrm{O5}}^-$.
 
\begin{figure}[t]
\centering
 \begin{subfigure}{0.475\textwidth}
  \centering
  \begin{tikzpicture}
  \draw[dashed] (0,0)--(8,0);
  \draw (1,0)--(2,0.5)--(2.5,1)--(2.5,1.5)--(3,2)--(3,2.5)--(3.5,3)--(3.5,3.5);
  \draw (7,0)--(6,0.5)--(5.5,1)--(5.5,1.5)--(5,2)--(5,2.5)--(4.5,3)--(4.5,3.5);
  \draw (2,0.5)--(6,0.5) (2.5,1)--(5.5,1) (3,2)--(5,2) (3.5,3)--(4.5,3);
    \draw (2.5,1.5)--(2,1.5) (3,2.5)--(2,2.5);
    \draw (5.5,1.5)--(6,1.5) (5,2.5)--(6,2.5);
    \SevenB{6,2.5}
    \SevenB{6,1.5}
    \SevenB{2,2.5}
    \SevenB{2,1.5}
%
    \node at (4,3.75) {$\vdots$};
\begin{scope}[shift={(8,7)},rotate=180]
 \draw[dashed] (0,-0.5)--(8,-0.5);
  \draw (0,-0.5)--(2,0.5)--(3,1.5)--(3,2)--(3.5,2.5)--(3.5,3);
  \draw (7,-0.5)--(6,0)--(5.5,0.5)--(5.5,1)--(5,1.5)--(5,2)--(4.5,2.5)--(4.5,3);
  \draw (2,0.5)--(5.5,0.5) (3,1.5)--(5,1.5) (3.5,2.5)--(4.5,2.5);
    \draw (3,2)--(2,2);
    \draw (5,2)--(6,2) (5.5,1)--(6,1);
    \SevenB{6,2}
    \SevenB{6,1}
    \SevenB{2,2}
    \draw[red] (0,0)--(6,0);
    \SevenB{0,0}
\end{scope}
%
    \draw[decoration={brace,mirror,raise=10pt},decorate,thick]
  (6.25,1.25) -- node[right=10pt] {\footnotesize{$N{-}5$}} (6.25,5.25);
    \draw[decoration={brace,mirror,raise=10pt},decorate,thick]
  (1.75,6.25) -- node[left=10pt] {\footnotesize{$N{-}4$}} (1.75,1.25);
    \node at (0.5,-0.25) {\footnotesize{$\mathrm{O}5^+$}};
    \node at (4,-0.25) {\footnotesize{$\mathrm{O}5^-$}};
    \node at (7.5,-0.25) {\footnotesize{$\mathrm{O}5^+$}};
    \node at (0.5,7.75) {\footnotesize{$\mathrm{O}5^+$}};
    \node at (4,7.75) {\footnotesize{$\mathrm{O}5^-$}};
 \end{tikzpicture}
  \caption{}
  \label{subfig:5-brane_Higgs_SOodd}
 \end{subfigure}
\hfill
\begin{subfigure}{0.475\textwidth}
  \centering
  \begin{tikzpicture}
  \draw[dashed] (0,0)--(8,0);
  \draw (1,0)--(2,0.5)--(2.5,1)--(2.5,1.5)--(3,2)--(3,2.5)--(3.5,3)--(3.5,3.5);
  \draw (7,0)--(6,0.5)--(5.5,1)--(5.5,1.5)--(5,2)--(5,2.5)--(4.5,3)--(4.5,3.5);
  \draw (2,0.5)--(6,0.5) (2.5,1)--(5.5,1) (3,2)--(5,2) (3.5,3)--(4.5,3);
    \draw (2.5,1.5)--(2,1.5) (3,2.5)--(2,2.5);
    \draw (5.5,1.5)--(6,1.5) (5,2.5)--(6,2.5);
    \SevenB{6,2.5}
    \SevenB{6,1.5}
    \SevenB{2,2.5}
    \SevenB{2,1.5}
%
    \node at (4,3.75) {$\vdots$};
\begin{scope}[shift={(8,7)},rotate=180]
 \draw[dashed] (0,-0.5)--(8,-0.5);
  \draw (0,-0.5)--(2,0.5)--(3,1.5)--(3,2)--(3.5,2.5)--(3.5,3);
  \draw (6.5,-0.5)--(5.5,0.5)--(5.5,1)--(5,1.5)--(5,2)--(4.5,2.5)--(4.5,3);
  \draw (2,0.5)--(5.5,0.5) (3,1.5)--(5,1.5) (3.5,2.5)--(4.5,2.5);
    \draw (3,2)--(2,2);
    \draw (5,2)--(6,2) (5.5,1)--(6,1);
    \SevenB{6,2}
    \SevenB{6,1}
    \SevenB{2,2}
\end{scope}
%
  \node at (0.5,-0.25) {\footnotesize{$\mathrm{O}5^+$}};
    \node at (4,-0.25) {\footnotesize{$\mathrm{O}5^-$}};
    \node at (7.5,-0.25) {\footnotesize{$\mathrm{O}5^+$}};
    \node at (0.5,7.75) {\footnotesize{$\widetilde{\mathrm{O}5}^+$}};
    \node at (4,7.75) {\footnotesize{$\widetilde{\mathrm{O}5}^-$}};
 \draw [line join=round,decorate, decoration={zigzag, segment length=4,amplitude=.9,post=lineto,post length=2pt},orange] (5,0.5)--(5,1) (4,1)--(4,2) (4.25,2)--(4.25,3) (3,6.5)--(3,7.5) (4,5.5)--(4,6.5) (4.25,4.5)--(4.25,5.5);
 \draw [line join=round,decorate, decoration={zigzag, segment length=4,amplitude=.9,post=lineto,post length=2pt},orange] (3,1) .. controls (3.24,-0.25) .. (3.4,0.5);
\node[orange] at (2.75,0.25) {\tiny{$W_1$}};
\node[orange] at (4.75,0.75) {\tiny{$W_0$}};
\node[orange] at (3.75,1.5) {\tiny{$W_2$}};
\node[orange] at (4,2.5) {\tiny{$W_3$}};
\node[orange] at (3.75-0.15,7) {\tiny{$W_{N-1}$}};
\node[orange] at (3.75-0.25,1.5+4.5) {\tiny{$W_{N-2}$}};
\node[orange] at (4-0.25,2.5+2.5) {\tiny{$W_{N-3}$}};
%
\draw[dotted] (6,0.5)--(8,0.5) (5.5,1)--(8,1) (5,2)--(8,2) (4,3)--(8,3) 
 (5.5,1+5.5)--(8,1+5.5) (5,2+3.5)--(8,2+3.5) (4,3+1.5)--(8,3+1.5);
\draw[dotted,<->] (7,0)--(7,0.5);
\draw[dotted,<->] (6.25,0.5)--(6.25,1);
\draw[dotted,<->] (6.5,1)--(6.5,2);
\draw[dotted,<->] (6.4,2)--(6.4,3);
\draw[dotted,<->] (4.75,0.5+6)--(4.75,1.5+6);
\draw[dotted,<->] (6.5,1+4.5)--(6.5,2+4.5);
\draw[dotted,<->] (6.4,2+2.5)--(6.4,3+2.5);
\node at (7.5,0.25) {\tiny{$\phi_1-\phi_0$}};
\node at (7,0.75) {\tiny{$2\phi_0-\phi_2$}};
\node at (7.25,1.5) {\tiny{$\substack{2\phi_2-\phi_0 \\ -\phi_1 -\phi_3}$}};
\node at (7.5,2.5) {\tiny{$2\phi_3-\phi_2-\phi_4$}};
\node at (5.75,1+6) {\tiny{$2\phi_{N{-}1}-\phi_{N{-}2}$}};
\node at (7.5,1.5+4.5) {\tiny{$\substack{2\phi_{N{-}2}-\phi_{N{-}3} \\ -2\phi_{N{-}1}}$}};
\node at (7.25,2.5+2.5) {\tiny{$\substack{2\phi_{N{-}3}-\phi_{N{-}4} \\ -\phi_{N{-}2}}$}};
\end{tikzpicture}
  \caption{}
  \label{subfig:5-brane_SOodd_algebra}
 \end{subfigure}
 \caption{Partial Higgs mechanisms for the brane web displayed in Figure \ref{fig:5d_branes_SOeven}. In \subref{subfig:5-brane_Higgs_SOodd}, $\sorm(2N)$ is Higgsed to $\sorm(2N-1)$ via moving the red brane segment onto the top O5-plane. The associated fundamental flavor is only charged under the gauge symmetry. In \subref{subfig:5-brane_SOodd_algebra}, the brane web for the $\sorm(2N-1)$ theory with $2N-9$ fundamentals is shown after the transition of Figure \ref{fig:details_Higgs_5brane_web}. The W-boson masses are functions of the dynamical K\"ahler parameter and the dependence is mediated via the Cartan matrix of $\widehat{B}_{N-1}^{(1)}$.}
\end{figure}
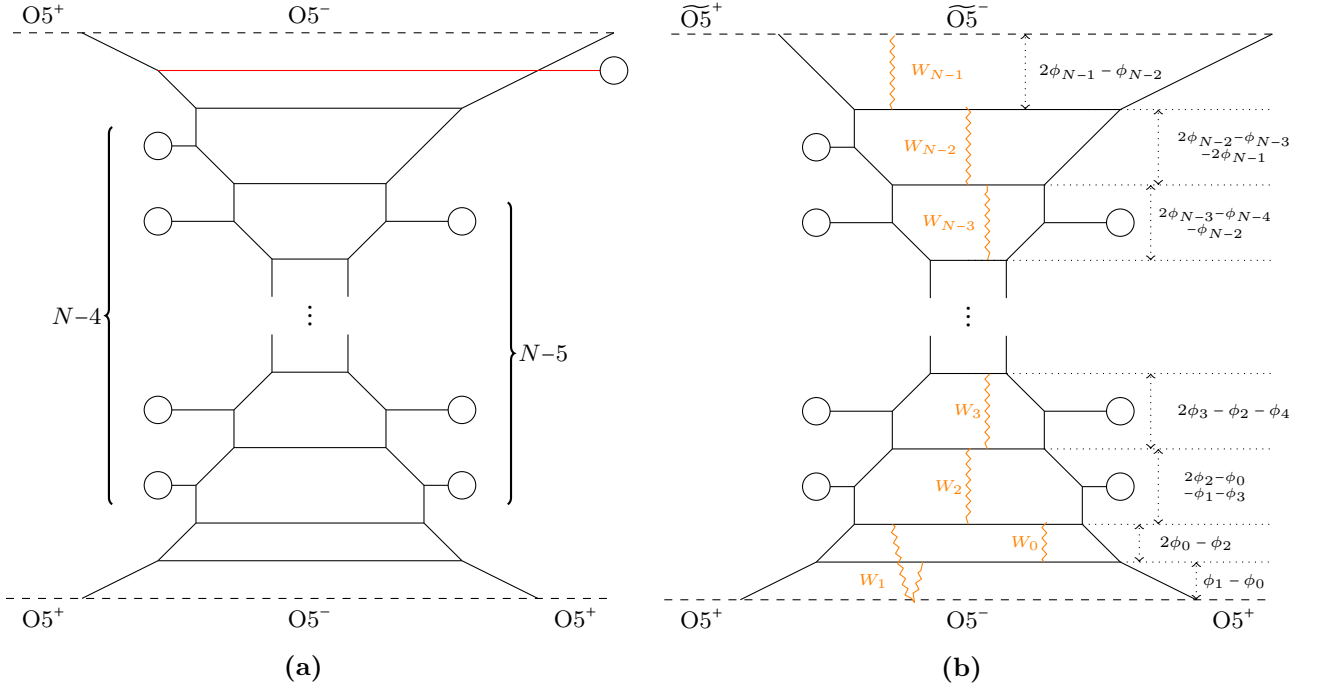

\begin{figure}
 \begin{subfigure}{0.475\textwidth}
  \centering
  \begin{tikzpicture}
 \draw[dashed] (0,-0.5)--(8,-0.5);
  \draw (0,-0.5)--(2,0.5)--(3,1.5)--(3,2)--(3.5,2.5)--(3.5,3);
  \draw (7,-0.5)--(6,0)--(5.5,0.5)--(5.5,1)--(5,1.5)--(5,2)--(4.5,2.5)--(4.5,3);
  \draw (2,0.5)--(5.5,0.5) (3,1.5)--(5,1.5) (3.5,2.5)--(4.5,2.5);
    \draw (3,2)--(2,2);
    \draw (5,2)--(6,2) (5.5,1)--(6,1);
    \SevenB{6,2}
    \SevenB{6,1}
    \SevenB{2,2}
    \draw[red] (0,0)--(6,0);
    \SevenB{0,0}
    \node at (4,3.25) {$\vdots$};
\begin{scope}[shift={(8,6.5)},rotate=180]
 \draw[dashed] (0,-0.5)--(8,-0.5);
  \draw (0,-0.5)--(2,0.5)--(3,1.5)--(3,2)--(3.5,2.5)--(3.5,3);
  \draw (6.5,-0.5)--(5.5,0.5)--(5.5,1)--(5,1.5)--(5,2)--(4.5,2.5)--(4.5,3);
  \draw (2,0.5)--(5.5,0.5) (3,1.5)--(5,1.5) (3.5,2.5)--(4.5,2.5);
    \draw (3,2)--(2,2);
    \draw (5,2)--(6,2) (5.5,1)--(6,1);
    \SevenB{6,2}
    \SevenB{6,1}
    \SevenB{2,2}
\end{scope}
%
    \draw[decoration={brace,mirror,raise=10pt},decorate,thick]
  (6.25,0.75) -- node[right=10pt] {\footnotesize{$N{-}5$}} (6.25,4.75);
    \draw[decoration={brace,mirror,raise=10pt},decorate,thick]
  (1.75,5.75) -- node[left=10pt] {\footnotesize{$N{-}5$}} (1.75,1.75);
    \node at (4,-0.75) {\footnotesize{$\mathrm{O}5^-$}};
    \node at (7.5,-0.75) {\footnotesize{$\mathrm{O}5^+$}};
    \node at (0.5,7.25) {\footnotesize{$\widetilde{\mathrm{O}5}^+$}};
    \node at (4,7.25) {\footnotesize{$\widetilde{\mathrm{O}5}^-$}};
 \end{tikzpicture}
  \caption{}
  \label{subfig:5-brane_Higgs_twisted}
 \end{subfigure}
 \hfill
 \begin{subfigure}{0.475\textwidth}
  \centering
  \begin{tikzpicture}
  \draw[dashed] (0,-0.5)--(8,-0.5);
  \draw (0,-0.5)--(2,0.5)--(3,1.5)--(3,2)--(3.5,2.5)--(3.5,3);
  \draw (6.5,-0.5)--(5.5,0.5)--(5.5,1)--(5,1.5)--(5,2)--(4.5,2.5)--(4.5,3);
  \draw (2,0.5)--(5.5,0.5) (3,1.5)--(5,1.5) (3.5,2.5)--(4.5,2.5);
    \draw (3,2)--(2,2);
    \draw (5,2)--(6,2) (5.5,1)--(6,1);
    \SevenB{6,2}
    \SevenB{6,1}
    \SevenB{2,2}
    \node at (4,3.25) {$\vdots$};
\begin{scope}[shift={(8,6.5)},rotate=180]
 \draw[dashed] (0,-0.5)--(8,-0.5);
  \draw (0,-0.5)--(2,0.5)--(3,1.5)--(3,2)--(3.5,2.5)--(3.5,3);
  \draw (6.5,-0.5)--(5.5,0.5)--(5.5,1)--(5,1.5)--(5,2)--(4.5,2.5)--(4.5,3);
  \draw (2,0.5)--(5.5,0.5) (3,1.5)--(5,1.5) (3.5,2.5)--(4.5,2.5);
    \draw (3,2)--(2,2);
    \draw (5,2)--(6,2) (5.5,1)--(6,1);
    \SevenB{6,2}
    \SevenB{6,1}
    \SevenB{2,2}
\end{scope}
%
    \node at (4,-0.75) {\footnotesize{$\widetilde{\mathrm{O}5}^-$}};
    \node at (7.5,-0.75) {\footnotesize{$\widetilde{\mathrm{O}5}^+$}};
    \node at (0.5,7.25) {\footnotesize{$\widetilde{\mathrm{O}5}^+$}};
    \node at (4,7.25) {\footnotesize{$\widetilde{\mathrm{O}5}^-$}};
 \draw [line join=round,decorate, decoration={zigzag, segment length=4,amplitude=.9,post=lineto,post length=2pt},orange] (3,-0.5)--(3,0.5) (4,0.5)--(4,1.5) (4.25,1.5)--(4.25,2.5) (3,6)--(3,7) (4,5)--(4,6) (4.25,4)--(4.25,5);
\node[orange] at (2.5,0.25) {\tiny{$W_1$}};
\node[orange] at (3.75,1) {\tiny{$W_2$}};
\node[orange] at (4,2) {\tiny{$W_3$}};
\node[orange] at (3.75-0.15,6.5) {\tiny{$W_{N-1}$}};
\node[orange] at (3.75-0.25,1+4.5) {\tiny{$W_{N-2}$}};
\node[orange] at (4-0.25,2+2.5) {\tiny{$W_{N-3}$}};
%
\draw[dotted] (5.5,0.5)--(8,0.5) (5,1.5)--(8,1.5) (4,2.5)--(8,2.5) 
 (5.5,1+5)--(8,1+5) (5,2+3)--(8,2+3) (4,3+1)--(8,3+1);
\draw[dotted,<->] (4.25,-0.5)--(4.25,0.5);
\draw[dotted,<->] (6.5,0.5)--(6.5,1.5);
\draw[dotted,<->] (6.4,1.5)--(6.4,2.5);
\draw[dotted,<->] (4.75,0+6)--(4.75,1+6);
\draw[dotted,<->] (6.5,0.5+4.5)--(6.5,1.5+4.5);
\draw[dotted,<->] (6.4,1.5+2.5)--(6.4,2.5+2.5);
\node at (5,0) {\tiny{$2\phi_1-\phi_2$}};
\node at (7.25,1) {\tiny{$\substack{2\phi_2-\phi_3 \\ -2\phi_1}$}};
\node at (7.5,2) {\tiny{$2\phi_3-\phi_2-\phi_4$}};
\node at (5.75,0.5+6) {\tiny{$2\phi_{N{-}1}-\phi_{N{-}2}$}};
\node at (7.5,1+4.5) {\tiny{$\substack{2\phi_{N{-}2}-\phi_{N{-}3} \\ -2\phi_{N{-}1}}$}};
\node at (7.25,2+2.5) {\tiny{$\substack{2\phi_{N{-}3}-\phi_{N{-}4} \\ -\phi_{N{-}2}}$}};
 \end{tikzpicture}
  \caption{}
  \label{subfig:5-brane_twisted_algebra}
 \end{subfigure}
 \caption{Partial Higgs mechanisms for the brane web of Figure \ref{subfig:5-brane_SOodd_algebra}. In \subref{subfig:5-brane_Higgs_twisted}, the $\sorm(2N{-}1)$ gauge symmetry is Higgsed to $\sorm(2N-2)$ via moving the red brane segment onto the bottom O5-plane. The associated fundamental flavor is charged under the KK scale. In \subref{subfig:5-brane_twisted_algebra}, the brane web for the $\Z_2$-twisted compactification of an $\sorm(2N-2)$ theory with $2N-10$ fundamentals is shown. The W-boson masses are functions of the dynamical K\"ahler parameter and the dependence is mediated via the Cartan matrix of $\widehat{D}_{N-1}^{(2)}$.}
\end{figure}
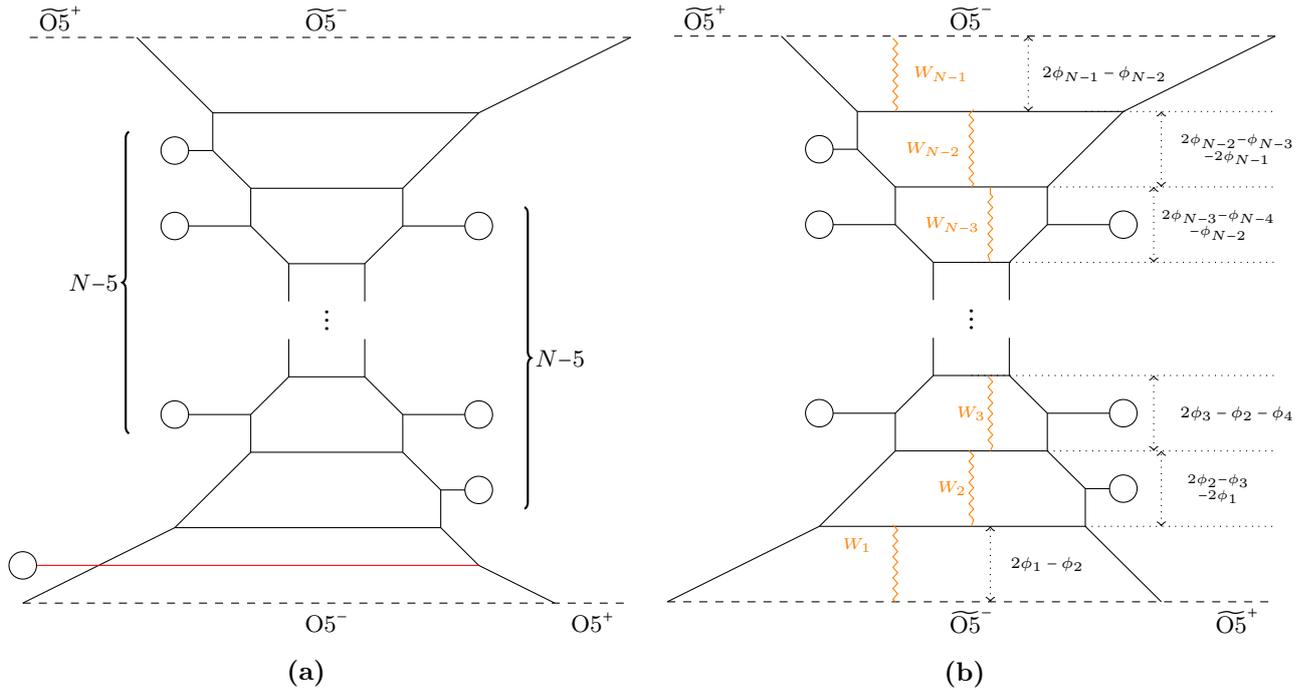

\begin{figure}[t]
\centering
 \begin{tikzpicture}
  \draw[dashed] (0,0)--(8,0);
  \draw (1,0)--(2,0.5)--(2.5,1)--(2.5,1.5)--(3,2)--(3,2.5)--(3.5,3)--(3.5,3.5);
  \draw (7,0)--(6,0.5)--(5.5,1)--(5.5,1.5)--(5,2)--(5,2.5)--(4.5,3)--(4.5,3.5);
  \draw (2,0.5)--(6,0.5) (2.5,1)--(5.5,1) (3,2)--(5,2) (3.5,3)--(4.5,3);
    \draw (2.5,1.5)--(1,1.5) (3,2.5)--(1,2.5);
    \draw (5.5,1.5)--(7,1.5) (5,2.5)--(7,2.5);
    \SevenB{1,1.5}
    \SevenB{1,2.5}
    \SevenB{7,1.5}
    \SevenB{7,2.5}    
%
    \node at (4,3.75) {$\vdots$};
\begin{scope}[shift={(8,7)},rotate=180]
 \draw[dashed] (0,0)--(8,0);
  \draw (0.5,0)--(2.5,1)--(3,1.5)--(3,2)--(3.5,2.5)--(3.5,3);
  \draw (7.5,0)--(5.5,1)--(5,1.5)--(5,2)--(4.5,2.5)--(4.5,3);
  \draw (2.5,1)--(5.5,1) (3,1.5)--(5,1.5) (3.5,2.5)--(4.5,2.5);
    \draw (3,2)--(1,2);
    \draw (5,2)--(7,2);
    \SevenB{1,2}
    \SevenB{7,2}
    \draw[red] (0,0.75)--(8,0.75);
    \SevenB{0,0.75}
    \SevenB{8,0.75}
    \draw[thick,blue] (1,0.25)--(4,0.75);
\end{scope}
%
    \draw[decoration={brace,mirror,raise=10pt},decorate,thick]
  (7.25,1.25) -- node[right=10pt] {\footnotesize{$N-5$}} (7.25,5.25);
      \draw[decoration={brace,mirror,raise=10pt},decorate,thick]
  (0.75,5.25) -- node[left=10pt] {\footnotesize{$N-5$}} (0.75,1.25);
    \node at (0.5,-0.25) {\footnotesize{$\mathrm{O}5^+$}};
    \node at (4,-0.25) {\footnotesize{$\mathrm{O}5^-$}};
    \node at (7.5,-0.25) {\footnotesize{$\mathrm{O}5^+$}};
    \node at (0.5,7.75) {\footnotesize{$\mathrm{O}5^+$}};
    \node at (4,7.75) {\footnotesize{$\mathrm{O}5^-$}};
    \node at (7.5,7.75) {\footnotesize{$\mathrm{O}5^+$}};
    \node[blue] at (5,6.65) {\footnotesize{D3}};
 \end{tikzpicture}
 \caption{5-brane web for $\sorm(2N-2)$ gauge theory with $N_f=2N-10$ fundamental flavors and a codimension 2 defect. The defect is realized via a D3-brane which is suspended between the 5-brane web for $\sorm(2N-2)$ and a D5-brane that moves away along the Higgs branch.}
 \label{fig:5d_branes_defect_Higgs}
\end{figure}
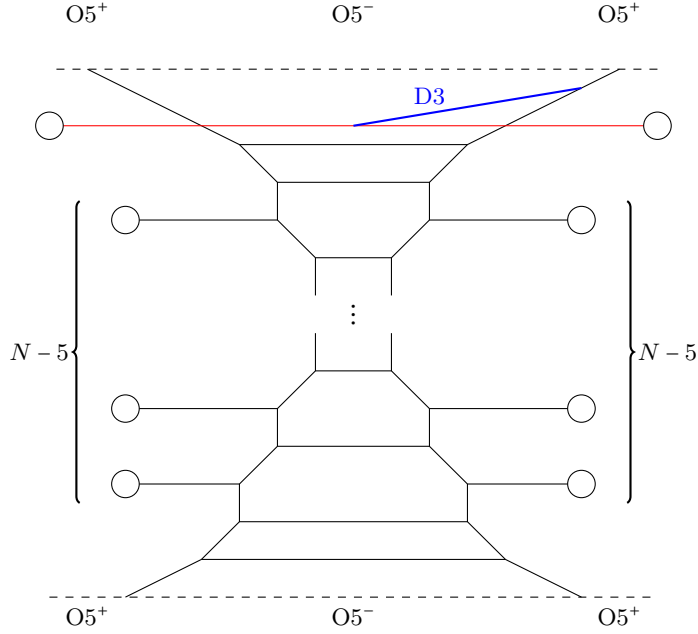

\paragraph{Higgsing to codimension 2 defect.}
Defect Higgsing proceeds as in the 6d setup, i.e.\ two (half) flavor D5-branes align with a internal (half) D5-brane. The resulting 5-brane is fully suspended between 7-branes and, hence, can be moved along the Higgs branch direction. Suspending a D3-brane between this brane and the residual 5-brane web induces a codimension 2 defect, as schematically shown in Figure \ref{fig:5d_branes_defect_Higgs}. Of course, the defect D3-brane is simply the T-dual of the defect D4-brane in the 6d setup. As such, the D3 wraps $\R^2_{\epsilon_1}$ inside $\R^4_{\epsilon_1,\epsilon_2}$ and is point-like in the $\R^2_{\epsilon_2}$ plane.
Similar types of defect Higgsings in 5-brane webs have been considered in \cite{Gaiotto:2014ina,Kimura:2017auj,Kim:2020npz}.

The defect Higgsing for $\sorm(2N-1)$ or the $\mathbb{Z}_2$-twisted $\sorm(2N)$ theory works analogously to the discussion that led to Figure \ref{fig:5d_branes_defect_Higgs}.

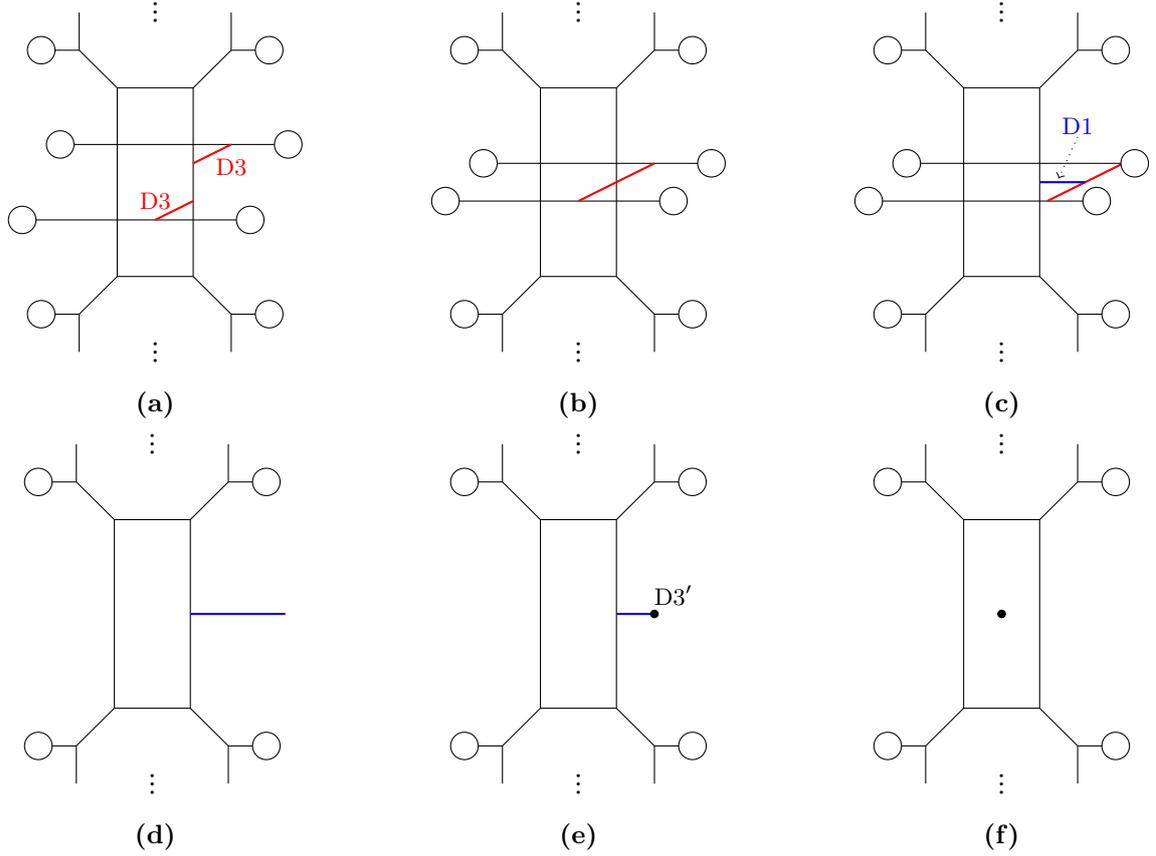
\begin{figure}[t]
 \centering
\begin{subfigure}{0.325\textwidth}
  \centering
  \begin{tikzpicture}
 \draw (1,0)--(1,0.5)--(1.5,1)--(1.5,3.5)--(1,4)--(1,4.5);
 \draw (3,0)--(3,0.5)--(2.5,1)--(2.5,3.5)--(3,4)--(3,4.5);
\node at (2,4.5) {$\vdots$};
\node at (2,0) {$\vdots$};
 \draw (1.5,1)--(2.5,1) (1.5,3.5)--(2.5,3.5);
 \draw (0.5,0.5)--(1,0.5) (3,0.5)--(3.5,0.5)  (0.5,4)--(1,4) (3,4)--(3.5,4); 
 \draw (0.25,1.75)--(3.25,1.75) (0.75,2.75)--(3.75,2.75);
 \draw[red,thick] (2,1.75)--(2.5,2) (2.5,2.5)--(3,2.75);
 \SevenB{0.5,0.5}
 \SevenB{3.5,0.5}
 \SevenB{0.5,4}
 \SevenB{3.5,4}
 \SevenB{0.25,1.75}
 \SevenB{3.25,1.75}
 \SevenB{0.75,2.75}
 \SevenB{3.75,2.75}
 \node[red] at (3,2.45) {\footnotesize{D3}};
 \node[red] at (2,2) {\footnotesize{D3}};
\end{tikzpicture}
  \caption{}
  \label{subfig:codim4_5-brane_a}
 \end{subfigure}
\hfill
\begin{subfigure}{0.325\textwidth}
  \centering
  \begin{tikzpicture}
 \draw (1,0)--(1,0.5)--(1.5,1)--(1.5,3.5)--(1,4)--(1,4.5);
 \draw (3,0)--(3,0.5)--(2.5,1)--(2.5,3.5)--(3,4)--(3,4.5);
\node at (2,4.5) {$\vdots$};
\node at (2,0) {$\vdots$};
 \draw (1.5,1)--(2.5,1) (1.5,3.5)--(2.5,3.5);
 \draw (0.5,0.5)--(1,0.5) (3,0.5)--(3.5,0.5)  (0.5,4)--(1,4) (3,4)--(3.5,4); 
 \draw (0.25,2)--(3.25,2) (0.75,2.5)--(3.75,2.5);
 \draw[red,thick] (2,2)--(2.5,2.25) (2.5,2.25)--(3,2.5);
 \SevenB{0.5,0.5}
 \SevenB{3.5,0.5}
 \SevenB{0.5,4}
 \SevenB{3.5,4}
 \SevenB{0.25,2}
 \SevenB{3.25,2}
 \SevenB{0.75,2.5}
 \SevenB{3.75,2.5}
\end{tikzpicture}
  \caption{}
  \label{subfig:codim4_5-brane_b}
 \end{subfigure}
\hfill
\begin{subfigure}{0.325\textwidth}
  \centering
  \begin{tikzpicture}
 \draw (1,0)--(1,0.5)--(1.5,1)--(1.5,3.5)--(1,4)--(1,4.5);
 \draw (3,0)--(3,0.5)--(2.5,1)--(2.5,3.5)--(3,4)--(3,4.5);
\node at (2,4.5) {$\vdots$};
\node at (2,0) {$\vdots$};
 \draw (1.5,1)--(2.5,1) (1.5,3.5)--(2.5,3.5);
 \draw (0.5,0.5)--(1,0.5) (3,0.5)--(3.5,0.5)  (0.5,4)--(1,4) (3,4)--(3.5,4); 
 \draw (0.25,2)--(3.25,2) (0.75,2.5)--(3.75,2.5);
 \draw[red,thick] (2.6,2)--(3.6,2.5);
 \draw[blue,thick] (2.5,2.25)--(3.1,2.25);
 \SevenB{0.5,0.5}
 \SevenB{3.5,0.5}
 \SevenB{0.5,4}
 \SevenB{3.5,4}
 \SevenB{0.25,2}
 \SevenB{3.25,2}
 \SevenB{0.75,2.5}
 \SevenB{3.75,2.5}
 \node[blue] at (3,3) {\footnotesize{D1}};
 \draw[dotted,->] (3,2.85)--(2.75,2.3);
\end{tikzpicture}
  \caption{}
  \label{subfig:codim4_5-brane_c}
 \end{subfigure}
 \\
\begin{subfigure}{0.325\textwidth}
  \centering
  \begin{tikzpicture}
 \draw (1,0)--(1,0.5)--(1.5,1)--(1.5,3.5)--(1,4)--(1,4.5);
 \draw (3,0)--(3,0.5)--(2.5,1)--(2.5,3.5)--(3,4)--(3,4.5);
\node at (2,4.5) {$\vdots$};
\node at (2,0) {$\vdots$};
 \draw (1.5,1)--(2.5,1) (1.5,3.5)--(2.5,3.5);
 \draw (0.5,0.5)--(1,0.5) (3,0.5)--(3.5,0.5)  (0.5,4)--(1,4) (3,4)--(3.5,4); 
 \draw[blue,thick] (2.5,2.25)--(3.75,2.25);
 \SevenB{0.5,0.5}
 \SevenB{3.5,0.5}
 \SevenB{0.5,4}
 \SevenB{3.5,4}
\end{tikzpicture}
  \caption{}
  \label{subfig:codim4_5-brane_d}
 \end{subfigure}
\hfill
\begin{subfigure}{0.325\textwidth}
  \centering
  \begin{tikzpicture}
 \draw (1,0)--(1,0.5)--(1.5,1)--(1.5,3.5)--(1,4)--(1,4.5);
 \draw (3,0)--(3,0.5)--(2.5,1)--(2.5,3.5)--(3,4)--(3,4.5);
\node at (2,4.5) {$\vdots$};
\node at (2,0) {$\vdots$};
 \draw (1.5,1)--(2.5,1) (1.5,3.5)--(2.5,3.5);
 \draw (0.5,0.5)--(1,0.5) (3,0.5)--(3.5,0.5)  (0.5,4)--(1,4) (3,4)--(3.5,4); 
 \draw[blue,thick] (2.5,2.25)--(3,2.25);
 \SevenB{0.5,0.5}
 \SevenB{3.5,0.5}
 \SevenB{0.5,4}
 \SevenB{3.5,4}
 \ThreeB{3,2.25}
 \node at (3.25,2.5) {\footnotesize{D$3^\prime$}};
\end{tikzpicture}
  \caption{}
  \label{subfig:codim4_5-brane_e}
 \end{subfigure}
 \hfill
\begin{subfigure}{ 0.325\textwidth}
  \centering
  \begin{tikzpicture}
 \draw (1,0)--(1,0.5)--(1.5,1)--(1.5,3.5)--(1,4)--(1,4.5);
 \draw (3,0)--(3,0.5)--(2.5,1)--(2.5,3.5)--(3,4)--(3,4.5);
\node at (2,4.5) {$\vdots$};
\node at (2,0) {$\vdots$};
 \draw (1.5,1)--(2.5,1) (1.5,3.5)--(2.5,3.5);
 \draw (0.5,0.5)--(1,0.5) (3,0.5)--(3.5,0.5)  (0.5,4)--(1,4) (3,4)--(3.5,4); 
 \SevenB{0.5,0.5}
 \SevenB{3.5,0.5}
 \SevenB{0.5,4}
 \SevenB{3.5,4}
 \ThreeB{2,2.25}
\end{tikzpicture}
  \caption{}
  \label{subfig:codim4_5-brane_f}
 \end{subfigure}
 \caption{The realization of a codimension 4 defect via Higgsing. The starting point is the 5-brane web for $\sorm(2N+2)$, see Figure \ref{fig:5d_branes_SOeven}. However, only the central part is relevant for the discussion such that the orientifolds are omitted here. In \subref{subfig:codim4_5-brane_a}, two codimension 2 defects are realized via defect Higgsings. In \subref{subfig:codim4_5-brane_b}, suitably adjusting the defect parameters allows to reconnect the two defect D3-branes as shown. In \subref{subfig:codim4_5-brane_c}, the resulting D3 can be move along 5-branes, which is on the defect Higgs branch, but the D3 leaves behind an attached D1-brane. In \subref{subfig:codim4_5-brane_d}, the defect D3 as well as the 5-branes on the Higgs branch have been moved off to infinity, leaving a semi-infinite D1-brane. This realizes a codimension 4 defect. In \subref{subfig:codim4_5-brane_e}, the semi-infinite D1 is terminated on a D$3^\prime$-brane, occupying $x^{0,7,8,9}$. In \subref{subfig:codim4_5-brane_f}, the D$3^\prime$ has been moved inside the brane web, upon which the D1-brane has been annihilated. This connects to the codimension 4 defect of Table \ref{tab:branes_5d}.}
\end{figure}

\paragraph{Codimension 4 defect.}
A codimension 4 defect can be introduced via a Higgsing process as well. The starting point is a the brane web for $\sorm(2N+2)$ with two different defect Higgsings, as shown in Figure \ref{subfig:codim4_5-brane_a}. 
The two defect Higgsing processes leave behind two different D3-branes connecting the residual 5-brane web and a D5-brane that has been moved away along the Higgs branch direction. 
Next, the two defects can be tuned as in Figure \ref{subfig:codim4_5-brane_b}, implying that the two D3-branes can reconnect and form a single D3-brane suspended between two D5-branes, which opens a new Higgs branch direction corresponding to motion of the D3-brane along the horizontal direction. The associated Higgs flow can leave behind a D1-brane, see Figure \ref{subfig:codim4_5-brane_c}. The D1-brane, occupying directions $x^{0,6}$, connects the residual 5-brane web and the D3-brane. Notably, the D1 is a codimension 4 object from the 5d world-volume perspective. Moving the D3 off to infinity yields the 5-brane web displayed in Figure \ref{subfig:codim4_5-brane_d}. As the D3 is removed, one observes that the $\sorm(4)_{2,3,4,5}$ space-time rotation symmetry is restored again, which hints towards the interpretation of the semi-infinite D1 as a codimension 4 defect. To consolidate this idea, one can terminate the D1-brane on a D$3^\prime$-brane, which occupies directions $x^{0,7,8,9}$, see Figure \ref{subfig:codim4_5-brane_e}. This is possible as the system of D1-D$3^\prime$-NS5 forms a Hanany-Witten configuration \cite{Hanany:1996ie}, meaning that the inclusion of the D$3^\prime$ does not break any additional supersymmetries. Moreover, the Hanany-Witten setup implies that one can move the D$3^\prime$ inside the 5-brane web, upon which the D1-brane is annihilated. Consequently, the 5-brane web in Figure \ref{subfig:codim4_5-brane_f} is the configuration of a 5d theory with a codimension 4 defect via a D$3^\prime$-brane, see for instance \cite{Assel:2018rcw}.

Comparing to the 6d setup, a codimension 4 defect is indeed realized via an additional D$3^\prime$-brane which fills $x^{0,7,8,9}$ directions. Alternatively, the semi-infinite D1 in Figure \ref{subfig:codim4_5-brane_d} originate from a semi-infinite D2 filling $x^{0,1,6}$. 

\section{6d \texorpdfstring{$\sorm(2N)$}{SO(2N)} theories}
\label{sec:DN}
In this section, we compute the partition functions of the 6d $\sorm(2N)$ gauge theories defined via a $-4$ curve in the presence of codimension 2 and 4 defects. For this  ADHM-like constructions for the self-dual strings interacting with the defects are used. We present the difference equations satisfied by the defect partition functions. We then show that the difference equation realizes a quantization of the elliptic Seiberg-Witten curve for the 6d $\sorm(2N)$ gauge theory.

\subsection{Self-dual strings and ADHM construction}
Let us consider the partition function of the 6d $\sorm(2N)$ gauge theory on a torus times the $\Omega$-deformed $\mathbb{R}^4$. This partition function can be computed by using localization and the result can be written as
\begin{align}
Z^{\rm{6d}}=Z^{\rm{6d}}_{\text{pert}}\cdot Z^{\rm{6d}}_{\text{str}} \qquad \text{with} \qquad 
Z_{\text{str}}=\sum_{k=0}^{\infty}q_{\phi}^k \oint[\rd u]Z_{k,1-\text{loop}}^{\rm{6d}}(u_I) \, ,
\label{eq:part_fct_explained}
\end{align}
where $[\rd u]=\frac{1}{2^k k!}\prod_{I=1}^k\frac{\rd u_I}{2\pi \ri}$, $q_{\phi}=\re^{\phi_0}$ is the tensor fugacity with the tensor branch parameter $\phi_0$ for counting the number of self-dual strings. Here, $Z^{\rm{6d}}_{\text{pert}}$ is the perturbative contribution which is discussed below in Section \ref{sec:Dn_pert}. $Z^{\rm{6d}}_{\rm str}$ is the contribution from the self-dual strings which is a sum over the elliptic genera of worldsheet theories on the strings. The elliptic genus at each string number is given by a contour integral over a 1-loop determinant $Z_{k,1-\text{loop}}^{\rm{6d}}$, with $Z_{0,1-\text{loop}}^{\rm{6d}}=1$.
The integration can be performed exactly by localization techniques on the 2d worldsheet gauge theory \cite{Benini:2013nda,Benini:2013xpa}.

The 2d $\Ncal=(0,4)$ quiver gauge theories for the self-dual strings in the 6d $\sorm(2N)$ gauge theory have been introduced in \cite{Haghighat:2014vxa}. For $k$-strings, the resulting theory is a $\sprm(k)$ gauge theory coupled to the bulk $\sorm(2N)$ gauge group and the $\sprm(N_f)$ flavor group as follows:
\begin{align}
    \raisebox{-.5\height}{
    \begin{tikzpicture}
  \tikzstyle{gauge} = [circle, draw,inner sep=3pt];
  \tikzstyle{flavour} = [regular polygon,regular polygon sides=4,inner sep=3pt, draw];
  \node (g1) [gauge,label=below:{$\sprm(k)$}] {};
  \node (f1) [flavour,above of=g1, label=above:{$\sorm(2N)$}] {};
  \node (f2) [flavour,right of=g1, label=right:{$\sprm(N_f)$}] {};
  \draw (g1)--(f1);
  \draw[dashed] (g1)--(f2);
  \draw (g1) to [out=140,in=220,looseness=10] (g1);
  \node at (-1,0) {\small{asym}};
  \end{tikzpicture}
    } 
    \quad \text{with } N_f=2N-8.
    \label{eq:2d_quiver}
\end{align}
Here solid/dashed lines denote hypermultiplets/Fermi multiplets, respectively.
Using this quiver description, we can compute the elliptic genera for the self-dual strings.
At $k$-string sector, the 1-loop determinant of the elliptic genus is  given by \cite{Haghighat:2014vxa}
\begin{align}\label{Z1loop_DN}
Z_{k,1-\text{loop}}^{\rm{6d}}(u_I)=\left(\frac{\vartheta _1\left(2 \epsilon _+\right)}{\vartheta _1\left(\epsilon_{1,2}\right)}\right)^k 
&\cdot \prod _{I=1}^k \vartheta
   _1\left(\pm 2 u_I\right) \vartheta _1\left(2 \epsilon _+\pm 2 u
   _I\right)\cdot \prod _{I<J}^k \frac{\vartheta _1\left(\pm
   u_I\pm u_J\right) \vartheta _1\left(2 \epsilon _+\pm u_I\pm
   u_J\right)}{\vartheta _1\left(\epsilon _1\pm u_I\pm u_J\right) \vartheta _1\left(\epsilon _2\pm u_I\pm u_J\right)}\nonumber \\
&\cdot \prod _{I=1}^k \frac{\prod _{l=1}^{2 N-8} \vartheta
   _1\left(\pm u_I+m_l\right)}{\prod _{i=1}^N \vartheta _1\left(\epsilon _+\pm
   u_I\pm \alpha_i\right)}\ ,
\end{align}
where $\epsilon_{\pm}=\frac{1}{2}(\epsilon_1\pm \epsilon_2)$, and $u_I$, $\alpha_i$, and $m_l$ denote the holonomies for the $\sprm(k)$ gauge group, the bulk $\sorm(2N)$ gauge group, and the $\sprm(N_f)$ flavor group, respectively. Here, we introduced $\vartheta_i(z)\coloneqq\frac{\ri\theta_i(z;\tau)}{\eta(\tau)}$, with the theta functions $\theta_i$ and the Dedekind eta function $\eta$ being defined in Appendix \ref{app:functions}. Also, the short-hand notations $\vartheta_i(a\pm b)\equiv\vartheta_i(a+b)\vartheta_i(a-b)$, $\vartheta_1(\epsilon_{1,2})\equiv\vartheta_1(\epsilon_{1})\vartheta_1(\epsilon_{2})$ are convenient and are used throughout the remainder of this paper.

The contour integral over the gauge holonomies $u_I$ can be evaluated using the Jeffrey-Kirwan (JK) residue prescription \cite{MR1318878}. At one-string level, the JK-poles giving non-trivial contributions are
\begin{align}\label{eq:DN_Z1_poles}
\epsilon_+\pm\alpha_i+u_1=0,\quad\quad i=1,\ldots,N \, .
\end{align}
By summing over all the corresponding residues, the one-string elliptic genus is given by
\begin{align}\label{eq:DN_Z1}
Z_1^{\rm{6d}}
&= -\frac{1}{2} \sum_{i=1}^N \Bigg(\frac{\vartheta_1(2\epsilon_+ + 2\alpha_i) \vartheta_1(4\epsilon_+ + 2\alpha_i) \prod_{l=1}^{2N-8}\vartheta_1(\epsilon_+ + \alpha_i \pm m_l)} {\vartheta_1(\epsilon_{1,2}) \prod_{j\neq i}^N  \vartheta_1(\alpha_i \pm \alpha_j) \vartheta_1(2\epsilon_+ + \alpha_i \pm \alpha_j)} + (\alpha_i \to -\alpha_i) \Bigg) \,.
\end{align}
We further computed the two- and three-string elliptic genera, but the expressions are cumbersome and we refer the reader to Appendix \ref{app:ell_genera} for details.

\subsection{Codimension 2 defect}
We can introduce a codimension 2 defect by applying the Higgsing procedure described in Section \ref{sec:braneweb}. The usual Higgsing from $\sorm(2N+2)$ to $\sorm(2N)$ gauge theory without defects can be performed by turning on VEVs of scalar fields in the moment map, see also Appendix \ref{app:Higgs}. In the elliptic genus, this Higgsing corresponds  to tuning the chemical potentials as follows:
\begin{align}
\label{eq:Higgs_Codim2_SO2N}
    m_{2N-7}+m_{2N-6}+2\epsilon_+=0\,,\quad \alpha_{N+1}+m_{2N-6}+\epsilon_+=0\,.
\end{align}
Note that the first equation is equivalent to \eqref{eq:Higgs_cond_D_to_D}.
It is straightforward to verify that after this parameter tuning, the elliptic genus of the self-dual strings in the $\sorm(2N+2)$ gauge theory reduces to the elliptic genus for the self-dual strings in the $\sorm(2N)$ theory.

Now we consider giving position dependent VEVs to introduce a codimension 2 defect. At the level of the partition function, the minimal codimension 2 defect can be introduced by tuning the parameters as
\begin{align}
\begin{aligned}
    &m_{2N-7}+m_{2N-6}+2\epsilon_++\epsilon_2=0\,,\quad \alpha_{N+1}+m_{2N-6}+\epsilon_+=0\,, \\
    &\qquad \text{with} \qquad 
    x\equiv -\alpha_{N+1}+\epsilon_2 \,,
    \end{aligned}
\label{eq:codim_two_defect}
\end{align}
which introduces a defect carrying non-trivial angular momentum along the second $\mathbb{R}^2$ plane in the spatial $\mathbb{R}^4=\mathbb{R}^2\times \mathbb{R}^2$. After this tuning, one obtains the partition function
\begin{align}
\label{eq:pt_fct_explained_defect}
Z^{\rm{6d/4d}}&=Z_{\text{pert}}^{\rm{6d}}
\cdot Z_{\text{pert}}^{\rm{4d}}
\cdot Z_{\text{str}}^{\text{6d/4d}},\quad Z_{\text{str}}^{\text{6d/4d}}=\sum_{k=0}^{\infty}q_{\phi}^k \oint[\rd{u}]Z_{k,1-\text{loop}}^{\rm{6d}}( u_I)Z_{k,1-\text{loop}}^{\rm{4d}}( u_I,x),
\end{align}
where $Z_{\text{pert}}^{\rm{4d}}$ is the perturbative contribution from the 4-dimensional defect which we consider in detail in Section \ref{sec:Dn_pert}. The elliptic genus for $k$-strings receives additional contributions from the defect given by
\begin{align}\label{eq:co-dim2-so2N}
    Z_{k,1-\text{loop}}^{\rm{4d}}( u_I,x)=\prod_{I=1}^k\frac{\vartheta_1(-\epsilon_+-x\pm  u_I)}{\vartheta_1(-\epsilon_--x\pm  u_I)},
\end{align}
and $x$ denotes the chemical potential for the $\urm(1)$ defect symmetry, as introduced in \eqref{eq:codim_two_defect}.

The defect contribution \eqref{eq:co-dim2-so2N} corresponds to the 1-loop determinant of an additional 2d $\Ncal=(2,2)$ chiral multiplet which originates from the string modes connecting the D4-brane and D2-branes suspended between the NS5-branes in Figure \ref{subfig:6d_branes_defect_Higgs}.
In fact, the partition function after the Higgsing implies that the codimension 2 defect can simply be introduced by coupling the 6d $\sorm(2N)$ gauge theory to a 4d free chiral multiplet that transforms in the fundamental representation of the $\sorm(2N)$ gauge group and has mass $x$. This is further discussed in Section \ref{sec:Dn_pert}. The 4d chiral multiplet induces a 2d $\Ncal=(2,2)$ chiral multiplet that transforms in the fundamental representation of the worldsheet $\sprm(k)$ gauge group. The $\Ncal=(2,2)$ chiral multiplet decomposes into a $\Ncal=(0,2)$ chiral multiplet and a $\Ncal=(0,2)$ Fermi multiplet. These $(0,2)$ multiplets give rise to the denominator and the numerator in \eqref{eq:co-dim2-so2N}, respectively.

One can compute the elliptic genera either via the JK-residue prescription or by tuning the parameters \eqref{eq:codim_two_defect} in the elliptic genera of the $\sorm(2N+2)$ theory. We note that the defect contribution leads to the following additional poles in the JK prescription:
\begin{align}
\label{eq:add_poles_codim2}
  -\epsilon_- - x + u_I = 0 \, .
\end{align}
At one-string, the elliptic genus is computed to read
\begin{align}
    Z_1^{\rm{6d/4d}}=&-\frac{1}{2}\frac{\vartheta_1(2x+\epsilon_1)\vartheta_1(2x+2\epsilon_1)\vartheta_1(2x-2\epsilon_2)\vartheta_1(2x+2\epsilon_-)\vartheta_1(2\epsilon_+)\prod_l^{2N-8} \vartheta_1(x\pm m_l +\epsilon_-)}{\vartheta_1(\epsilon_1)\prod_{i=1}^N\vartheta_1(x\pm \alpha_i+\epsilon_1)\vartheta_1(x\pm \alpha_i-\epsilon_2)}\nonumber\\
    & -\frac{1}{2} \sum_{i=1}^N \Bigg(\frac{\vartheta_1(2\epsilon_+ + 2\alpha_i) \vartheta_1(4\epsilon_+ + 2\alpha_i) \prod_{l=1}^{2N-8}\vartheta_1(\epsilon_+ + \alpha_i \pm m_l)} {\vartheta_1(\epsilon_{1,2}) \prod_{j\neq i}^N  \vartheta_1(\alpha_i \pm \alpha_j) \vartheta_1(2\epsilon_+ + \alpha_i \pm \alpha_j)}\cdot \frac{\vartheta_1(x+\epsilon_+\pm(\alpha_i+\epsilon_+))}{\vartheta_1(x+\epsilon_-\pm(\alpha_i+\epsilon_+))}\nonumber\\
    &\quad\quad\quad\quad + (\alpha_i \to -\alpha_i) \Bigg) \,.
\end{align}
We further computed the two- and three-string elliptic genera with codimension 2 defect by using the 2-string results of Appendix \ref{app:genera_details1}  
and the 3-string results of Appendix \ref{app:genera_details2} together with the Higgsing prescription \eqref{eq:codim_two_defect}. Since the expressions are lengthy, we refrain from presenting them here.

\subsection{Codimension 4 defect}
\label{sec:codim4_SO2N}
As explained in Section \ref{sec:braneweb}, the codimension 4 defect of the $\sorm(2N)$ theory is realized in the Type IIA brane system by inserting a D$4^\prime$-brane which occupies $x^{0,1,7,8,9}$ directions. Additional 2d degrees of freedom resulting from the defect can be conveniently read off from the brane configuration. The D$4^\prime$-D6 bound states introduce a Fermi multiplet in the fundamental representation of the $\sorm(2N)$ bulk gauge group. The  D2-D$4^\prime$ brane bound states induce a (twisted-)hypermultiplet and a Fermi-multiplet which transform in the fundamental representation of the $\sprm(k)$ worldsheet gauge group. The bound states of D2-D$4^\prime$ brane capture the self-dual string modes bound to the defect, see \cite{Tong:2014yna,Tong:2014cha,Kim:2016qqs} for more details.
From the field theory point of view, the defect corresponds to a 2d fundamental fermion coupled to the bulk $\sorm(2N)$ gauge group. The additional 2d fields contribute to the elliptic genus for $k$-strings as follows:
\begin{align}
\label{eq:Ycal_SO2N}
    \mathcal{Y}_k( u_I,x)=\prod_{i=1}^{N}\vartheta_1(x\pm \alpha_i)\cdot \prod_{I=1}^k\frac{\vartheta_1(-\epsilon_-\pm x\pm u_I)}{\vartheta_1(-\epsilon_+\pm x\pm u_I)} \ ,
\end{align}
where $x$ is the chemical potential for the $\urm(1)$ symmetry on the defect.

The same codimension 4 defect can also be obtained from the Higgsing the $\sorm(2N+4)$ theory by giving position dependent VEVs to two moment map components as illustrated in Section \ref{sec:braneweb}. This corresponds to setting the parameters in the UV $\sorm(2N+4)$ gauge theory as follows:
\begin{align}\label{Dn_cod4_Higgs}
\begin{aligned}
m_{2N-6}&=\alpha_{N+1}-\epsilon_+-\epsilon_2,\,\,\,m_{2N-7}=-\alpha_{N+1}-\epsilon_+,\,\,\,\alpha_{N+1}=x+2\epsilon_+, \\
m_{2N-4}&=\alpha_{N+2}-\epsilon_+-\epsilon_2,\,\,\,m_{2N-5}=-\alpha_{N+2}-\epsilon_+,\,\,\,\alpha_{N+2}=-x+2\epsilon_+ \, .
\end{aligned}
\end{align}
This leads to the introduction of the codimension 4 defect in the $\sorm(2N)$ gauge theory in infrared.

After the Higgsing, the partition function of the 6d/2d coupled system is given by
\begin{align}
\label{eq:pt_fct_explained_codim4}
Z^{\rm{6d/2d}}&=Z^{\rm{6d}}_{\rm pert} 
\cdot Z^{\rm{2d}}_{\rm pert} 
\cdot Z^{\rm{6d/2d}}_{\rm str}= Z^{\rm{6d}}_{\rm pert} \times q_{\phi}^{-\frac{1}{2}}\sum_{k=0}^{\infty}q_{\phi}^k \oint[\rd{u}]Z_{k,1-\text{loop}}^{\rm{6d}}( u_I)\cdot \mathcal{Y}_k( u_I,x).
\end{align}
Here we multiplied the factor $q_{\phi}^{-\frac{1}{2}}$ which comes from the Green-Schwarz contribution of the original 6d theory with the tuning of parameters \eqref{Dn_cod4_Higgs}. 
The non-trivial tensor charge of the 2d degrees of freedom can be understood in the brane system.  The codimension 4 defect is a  D2-brane connecting NS5-brane and D$4^\prime$-brane as shown in Table \ref{tab:branes_6d} and manifest as a self-dual string coupled to two-form tensor field on NS5-brane.  As NS5-brane and D$4^\prime$-brane are mutually codimension eight, if we regard the circle along $x^1$ direction to be M-circle, this D2-brane wrapping this M-circle appears fermionic. The quantization of the 1d complex spinor field with the Lagrangian  $i \psi^\dagger(\partial_0-iA_{0})\psi$ on the D4 brane, which is the wrapped NS5 brane, leads to the $A_0=\int dx^1 B_{10}$ charge to be $\pm \frac12$,  leading to $q_{\phi}^{\pm \frac12}$ factor. The D2-branes between NS5-branes generate the integer power in $q_{\phi}$, explaining the factor in \eqref{eq:pt_fct_explained_codim4}.

One can again use the JK-residue prescription to compute the elliptic genera of the self-dual strings coupled to the codimension 4 defect. The defect introduces the following new poles:
\begin{align}
\label{eq:poles_codim4_SO2N}
  -\epsilon_+ \pm x + u_I = 0 \ .
\end{align}
In the $k=0$ sector, we compute
\begin{align}\label{Dn_2d_pert}
  Z^{\rm{2d}}_{\rm pert} =  q_{\phi}^{-\frac{1}{2}}\prod_{i=1}^{N}\vartheta_1(x\pm \alpha_i) \ ,
\end{align}
which comes from the 2d fermion in the $\sorm(2N)$ fundamental representation, see also \eqref{eq:Ycal_SO2N}. In the $k=1$ string sector, one finds
\begin{align}\label{eq:Dn_Z16d2d}
    Z_1^{\rm{6d/2d}}=&\frac{1}{2}\Bigg(\frac{\vartheta_1(2x-\epsilon_{1})\vartheta_1(2x-\epsilon_{2})\vartheta_1(2x-3\epsilon_+\pm \epsilon_+)\vartheta_1(2\epsilon_+)\prod_l \vartheta_1(-x\pm m_l +\epsilon_+)}{\prod_{i=1}^N\vartheta_1(x\pm \alpha_i)\vartheta_1(x\pm \alpha_i-2\epsilon_+)}+(x\rightarrow -x)\Bigg)\nonumber\\
    & -\frac{1}{2} \sum_{i=1}^N \Bigg(\frac{\vartheta_1(2\epsilon_+ + 2\alpha_i) \vartheta_1(4\epsilon_+ + 2\alpha_i) \prod_{l=1}^{2N-8}\vartheta_1(\epsilon_+ + \alpha_i \pm m_l)} {\vartheta_1(\epsilon_{1,2}) \prod_{j\neq i}^N  \vartheta_1(\alpha_i \pm \alpha_j) \vartheta_1(2\epsilon_+ + \alpha_i \pm \alpha_j)}\cdot \frac{\vartheta_1(\pm x+\epsilon_+\pm\epsilon_-+\alpha_i)}{\vartheta_1(\pm x+\epsilon_+\pm\epsilon_++\alpha_i)}\nonumber\\
    &\quad\quad\quad\quad + (\alpha_i \to -\alpha_i) \Bigg) \,.
\end{align}
One can also compute the same result from the Higgsing procedure by tuning the parameter as described in \eqref{Dn_cod4_Higgs} in the elliptic genera of $\sorm(2N+4)$ directly.

For later purposes, we discuss the structure of the partition function of the codimension 4 defect. First of all, the 2d perturbative part of the 6d/2d partition function in \eqref{Dn_2d_pert} can be recast as
\begin{align}\label{Dn_2d_pert2}
    Z^{\rm{2d}}_{\mathrm{pert}}&=q_{\phi}^{-\frac{1}{2}}\sum_{n=0}^{N}\mathcal{W}_{0}^{(n)}\theta_2^{N-n}(2x;2\tau)\theta_3^{n}(2x;2\tau)\,, \nonumber \\
    \mathcal{W}_{0}^{(n)}&=\frac{(-1)^{N-n}}{n!(N-n)!\eta(\tau)^{2N}}\left(\prod_{j=1}^{n}\theta_2(2\alpha_j;2\tau)\prod_{j=n +1}^{N}\theta_3(2\alpha_j;2\tau)+{\mathrm{Permutations\,\, of\,\,}\alpha_j}\right)\,,
\end{align}
by using the identity
\begin{align}\label{eq:theta_identity_1}
    \theta_1(z+w;\tau)\theta_1(z-w;\tau)= \theta_3(2z;2\tau)\theta_2(2w;2\tau)-\theta_2(2z;2\tau)\theta_3(2w;2\tau)\,.
\end{align}
This means that the partition function with the codimension 4 defect, at least at the perturbative level, can be linearly decomposed to a finite series of elliptic functions which  depend only on the defect parameter $x$, while the coefficients of the decomposition are functions independent of $x$. In fact, as we checked numerically up to three-strings, the decomposition structure holds for elliptic genera at higher string numbers. Therefore, we conjecture that the full partition function of the 6d/2d coupled system takes the form of 
\begin{align}
\label{eq:cod4splitting}
    Z^{\rm{6d/2d}}=\sum_{n=0}^{N}\mathcal{W}^{(n)}\theta_2^{N-n}(2x;2\tau)\theta_3^{n}(2x;2\tau) \,, 
\qquad \text{where} \qquad 
    \mathcal{W}^{(n)}=q^{-\frac12}_\phi \sum_{k=0}^{\infty}q_{\phi}^k\mathcal{W}_{k}^{(n)} 
\end{align}
is the $x$-independent coefficient in the $(N+1)$-th order series of the elliptic functions.

It turns out that the coefficients $\mathcal{W}^{(n)}$'s are the $\frac{1}{2}$-BPS Wilson loop partition functions taking different representations of the $\sorm(2N)$ gauge group. It was recently proposed in \cite{Kim:2021gyj} that these Wilson loop partition functions satisfy the so-called blowup equation. In the case at hand, the blowup equations for the Wilson loop $\mathcal{W}^{(n)}$ of the 6d $\sorm(2N)$ gauge theory are
\begin{align}
    \theta_3^{[a]}\bigg(&-({2N-4})\epsilon_+-2\epsilon_2+\sum_{i=1}^{2N-8}m_i;4\tau\bigg) \
    \mathcal{W}_{k}^{(n)}(\alpha,m,\epsilon_1,\epsilon_2)= \\
    &\sum_{\lambda_{G}\in Q^{\vee}(G)}^{\frac{1}{2}||\lambda_G||^2+k_1+k_2=k} (-1)^{|\lambda_G|} \
    \theta_3^{[a]}\bigg(4\lambda_{G}\cdot\alpha+\sum_{i=1}^{2N-8}m_i+4k_1\epsilon_1+4k_2\epsilon_2-(2N-4-2||\lambda_{G}||^2)\epsilon_+-2\epsilon_2;4\tau\bigg)\notag\\
    &\cdot A_0(\alpha,m,\lambda_{G},\lambda_F)
    \cdot 
    \mathcal{W}_{k_1}^{(n)}\left(\alpha+\epsilon_2\lambda_{G},m-\tfrac{1}{2}\epsilon_2,\epsilon_1-\epsilon_2,\epsilon_2\right)
   \cdot
   Z^{\rm{6d}}_{k_2}\left(\alpha+\epsilon_1\lambda_{G},m-\tfrac{1}{2}\epsilon_1,\epsilon_1,\epsilon_2-\epsilon_1\right)\,, \notag
\end{align}
where $Q^{\vee}(G)$ is the co-root lattice of the gauge group $G$, $|\lambda_G|$ is the sum over components in the orthogonal base, $||\lambda_G||^2$ is the scalar product of $\lambda_\alpha$ with itself, and
\begin{align}
    \theta_3^{[a]}(z;4\tau)=\sum_{n=-\infty}^{\infty}\re^{2(n-a)^2\tau+(n-a)z},\quad\quad a=0,\frac{1}{4},\frac{1}{2},\frac{3}{4},
\end{align}
where $4a$ is the magnetic flux for the tensor parameter $\phi_0$. Here we choose the magnetic flux $\lambda_F=-\frac{1}{2}(1,\ldots,1)$ for the flavor mass. 
The function 
\begin{align}
\label{eq:D-fct_blowup}
  A_0(\alpha,m,\lambda_{G},\lambda_F)\equiv  A_V(\alpha,\lambda_{G},\lambda_F)A_H(\alpha,m,\lambda_{G},\lambda_F)  
\end{align}
comes from the perturbative contribution of the 6d partition function $Z^{\rm 6d}_{\rm pert}$ without the Green-Schwarz term.
The explicit expressions of $A_{V,H}$ in terms of theta functions are rather elaborate and we refer the reader to \cite[eq.\ (3.7)--(3.8)]{Gu:2020fem} for the details.

One can compute the exact expression of the Wilson loop partition function $\mathcal{W}_{k}^{(n)}$ at each string order by using three equations with any three over $a\in\{0,\frac{1}{4},\frac{1}{2},\frac{3}{4}\}$. At one string level, we have the solutions
\begin{align}\label{eq:Dn_W1}
    \mathcal{W}_{1}^{(n)}=-\sum_{i<j,i,j=1}^{N}&\sum_{r=\pm 1,s=\pm 1}\frac{\prod_{l=1}^{2N-8}\vartheta_1(r\alpha_i-m_l+\epsilon_+)\vartheta_1(s\alpha_j-m_l+\epsilon_+)}{\vartheta_1(r\alpha_i+s \alpha_j)\vartheta_1(r\alpha_i+s \alpha_j+\epsilon_1)\vartheta_1(r\alpha_i+s \alpha_j+\epsilon_1)\vartheta_1(r\alpha_i+s \alpha_j+2\epsilon_+)}\nonumber\\
    &\cdot \prod_{k\neq i,j}\prod_{p=\pm 1}{\vartheta_1(r\alpha_i+p \alpha_k)^{-1}\vartheta_1(s\alpha_j+p \alpha_k)}^{-1}
    \cdot 
    \mathcal{W}_{0}^{(n)}\left(\alpha_l+(r\delta_{l,i}+s\delta_{l,j})\epsilon_2\right)\nonumber\\
    &\cdot \frac{
    D_{a_1,a_2,a_3}\left(4(r\alpha_i+s\alpha_j+\epsilon_1+\epsilon_2),4\epsilon_1,4\epsilon_2;-(2N-4)\epsilon_+-2\epsilon_2+\sum_{l=1}^{2N-8}m_l \right) }{ D_{a_1,a_2,a_3}\left(0,4\epsilon_1,4\epsilon_2;-(2N-4)\epsilon_+-2\epsilon_2+\sum_{l=1}^{2N-8}m_l\right)},
\end{align}
where $\delta_{i,j}$ is the Kronecker delta and
\begin{align}\label{eq:Da1a2a3}
    D_{a_1,a_2,a_3}(z_1,z_2,z_3;z)=
    \det\left(
\begin{array}{ccc}
 \theta_3^{[a_1]}(z+z_1;4\tau) & \theta_3^{[a_1]}(z+z_2;4\tau) & \theta_3^{[a_1]}(z+z_3;4\tau) \\
 \theta_3^{[a_2]}(z+z_1;4\tau) & \theta_3^{[a_2]}(z+z_2;4\tau) & \theta_3^{[a_2]}(z+z_3;4\tau) \\
 \theta_3^{[a_3]}(z+z_1;4\tau) & \theta_3^{[a_3]}(z+z_2;4\tau) & \theta_3^{[a_3]}(z+z_3;4\tau) \\
\end{array}
\right),
\end{align}
with $a_{1,2,3}$ arbitrary three different values chosen from $\{0,\frac{1}{4},\frac{1}{2},\frac{3}{4}\}$. It is interesting to remark that even though the function $D_{a_1,a_2,a_3}(z_1,z_2,z_3;z)$ depends on the choice of three $a_i$'s, the final result of $\mathcal{W}_1^{(n)}$ does not depend on the choice of the $a_i$'s. We have verified that the results \eqref{eq:Dn_W1} from the blowup equations, together with \eqref{eq:cod4splitting}, indeed agree with the one-string elliptic genus \eqref{eq:Dn_Z16d2d} of the codimension 4 partition functions that we obtained using localization.

\subsection{Quantum curve and difference equation}\label{sec:DN_curve}
In this section, we study the quantum Seiberg-Witten curve of the 6d $\sorm(2N)$ theory via the defect partition function computed in the previous subsection. We propose a difference equation satisfied by the defect partition functions and relate it to the quantum Seiberg-Witten curve.

\subsubsection{Perturbative contributions}\label{sec:Dn_pert}
We first consider the perturbative part of the codimension 2 defect partition function of the $\sorm(2N)$ theory with $2N-8$ flavors and derive the difference equation it satisfies. The perturbative contributions to the 6d partition function on $\mathbb{R}^4\times \T^2$ without defect are given as follows:
\begin{align}
\label{eq:perturbative_partition_function}
Z^{\rm 6d}_{\rm pert}=Z^{\rm 6d}_{\rm eff}\cdot\PE\bigg[-\frac{1+p q}{(1-p)(1-q)(1-Q)} &\sum_{i<j}^N\left(A_i A_j+A_i A_j^{-1}+(A_i A_j)^{-1}Q+(A_i A_j^{-1})^{-1}Q\right)\\
&+\frac{\sqrt{pq}}{(1-p)(1-q)(1-Q)}\sum_{i=1}^N(A_i+A_i^{-1}Q)\sum_{l=1}^{2N-8}\left(M_l+M_l^{-1}\right)\bigg]\,, \notag
\end{align}
where $p=\re^{\epsilon_{1}}$, $q=\re^{\epsilon_{2}}$, $Q=\re^{\tau}$, $A_i=\re^{\alpha_i}$ and $M_l=\re^{m_l}$, and all terms independent on $A_i$ have been dropped. Here, $Z^{\rm 6d}_{\rm eff}= \exp( \frac{1}{\epsilon_1\epsilon_2}F_{\rm eff} )$ and $F_{\rm eff}$ is the effective prepotential for the 6d theory evaluated on $\Omega$-background which is given by
\begin{align}
F_{\rm eff}= \frac{1}{6}\sum_{i<j}(\alpha_i\pm\alpha_j)^3-\frac{1}{12}\sum_{i=1}^{N}\sum_{j=1}^{2N-8}(\alpha_i\pm m_j)^3+\frac{1}{2}\phi_0\sum_{i=1}^{N}\alpha_i^2+\ldots,
\end{align}
with the short-hand notation $(a\pm b)^3\equiv (a+ b)^3+(a- b)^3$. The ``$\ldots$" part is irrelevant to our discussion and is therefore omitted. The first two terms in the effective prepotential come from the fermion 1-loop calculations and the third term is the classical contribution from the Green-Schwarz term. See \cite{Kim:2020hhh} for details.

At the level of the partition function, the $\sorm(2N)$ theory can be Higgsed from $\sorm(2N+2)$ theory by the following assignment to the parameters:
\begin{align}
    A_{N+1}=\frac{q}{X}\,,\quad M_{2N-7}=\frac{\sqrt{q}}{X\sqrt{p }}\,,\quad M_{2N-6}=\frac{X}{\sqrt{pq}q}\,.
    \label{eq:normal_higgs}
\end{align}
which follows from \eqref{eq:Higgs_Codim2_SO2N}.
In addition, one needs to remove the contributions from Goldstone bosons
\begin{align}
    Z_{\rm G.B.}=\PE\bigg[ &\frac{\sqrt{pq}}{(1-p)(1-q)(1-Q)}(qX^{-1}+q^{-1}X Q)\sum_{l=1}^{2N-8}(M_l+M_l^{-1})\notag\\
    &+\frac{1+pq}{(1-p)(1-q)(1-Q)}(q^2X^{-2}+q^{-2}X^{2}Q)\bigg]\,.
\end{align}
We note that the fugacity $X$ in the above equations are the fugacity for the $U(1)$ symmetry acting only on the Goldstone modes.
Comparing the appearing $4(2N-8)+4 = 2(4N-15)$ massless chiral fields with the expectation from the Higgs mechanism discussed in Section \ref{subsec:6d_branes}, one observes that the sub-space of the Higgs branch where $\sorm(2N+2)$ is broken to $\sorm(2N)$ has quaternionic dimension $4N-13$. The missing degrees of freedom are simply a consequence of neglecting $A_i$ independent terms in \eqref{eq:perturbative_partition_function}.

To introduce a codimension 2 defect into the Higgsed theory one instead tunes the parameters as in \eqref{eq:codim_two_defect}. In terms of $A_i$ and $M_l$, we set the parameters as
\begin{align}
    A_{N+1}=\frac{q}{X}\,,\quad M_{2N-7}=\frac{1}{X\sqrt{pq}}\,,\quad M_{2N-6}=\frac{X}{\sqrt{pq}q}\,.
    \label{eq:codim_two_defect_higgs}
\end{align}
After this Higgsing and subtracting $Z_{\rm G.B.}$ from the result, we end up with the codimension 2 defect partition function of the $\sorm(2N)$ theory. The perturbative piece with the defect (in the limit $q\rightarrow 1$) can be written as
\begin{align}
    Z^{\rm 6d/4d}_{\rm pert}(x) &= Z^{\rm 6d}_{\rm pert} \cdot Z^{\rm 4d}_{\rm pert} \ , \nonumber \\
    Z^{\rm{4d}}_{\rm pert}(x)&=Z_{\rm eff}^{\rm4d}\cdot\PE\bigg[\sum_{i=1}^N\frac{X^{-1} A_i-pXA_i+QX^{-1}A_i^{-1}-pQXA_i^{-1}}{(1-p)(1-Q)}\bigg] =Z_{\rm eff}^{\rm4d}\cdot \prod_{i=1}^N\frac{\Gamma_e(X^{-1}A_i)}{\Gamma_e(pXA_i)} \, ,\notag\\
    Z^{\rm{4d}}_{\rm eff}(x)&=\exp\left[-\frac{(x+\tfrac{\epsilon_1}{2})}{\epsilon_1}\left(\sum_{i=1}^N\alpha_i+\frac{\phi_0}{2}-\frac{N\tau}{6}\right) + \ldots\right]\,,
\end{align}
where $\Gamma_{\mathrm{e}}(Z)$ is the elliptic gamma function
\begin{align}
   \Gamma_{\mathrm{e}}(Z)\equiv\PE\bigg[\frac{Z-pQZ^{-1}}{(1-p)(1-Q)}\bigg] \ .
\end{align}
Here, we have omitted the terms independent of the dynamical K\"ahler parameters, and
$Z_{\rm pert}^{6d}$ is the perturbative contribution of the 6d $\sorm(2N)$ theory.
The 4d part is precisely the perturbative contributions from a 4d $\Ncal=1$ fundamental chiral multiplet coupled to the bulk $\sorm(2N)$ gauge field. Unlike the Higgsing in (\ref{eq:normal_higgs}), the resulting partition function now depend on $X$ which becomes the fugacity for the $U(1)$ symmetry rotating the 4d chiral multiplet. 

By using the identity
\begin{align}
    \Gamma_{\mathrm{e}}(p X)=-Q^{-\frac{1}{12}}X^{\frac{1}{2}}\vartheta_1(x)\Gamma_{\mathrm{e}}( X),
\end{align}
one can verify that the perturbative part of the defect partition function satisfies the difference equation \footnote{We ignore an additional sign factor $(-1)^N$, which is irrelevant to our discussion.}
\begin{align}\label{eq:difference-pert}
    Y\cdot \lim_{\epsilon_2\rightarrow0} Z^{\rm{6d/4d}}_{\rm pert}(x)&= q_{\phi}^{-\frac{1}{2}}\prod_{i=1}^N\vartheta_1 (x\pm \alpha_i)\cdot \lim_{\epsilon_2\rightarrow0}Z^{\rm{6d/4d}}_{\rm pert}(x) = Z^{\rm 2d}_{\rm pert}(x) \cdot \lim_{\epsilon_2\rightarrow0}Z^{\rm{6d/4d}}_{\rm pert}(x)\ ,
\end{align}
where $Z^{\rm 2d}_{\rm pert}(x)$ is the perturbative part of the codimension 4 defect partition function defined in \eqref{Dn_2d_pert}.
Here, we have defined a shift operator as 
\begin{align}
    Y={\rm exp}\left(-\epsilon_1\frac{\partial}{\partial x}\right)
    \label{eq:def_shift_op}\,,
\end{align}
such that $Y\cdot f(x)=f(x-\epsilon_1)$. Note that the shift operator $Y$ perturbatively acts as an operator inserting a codimension 4 defect with parameter $x$ into the 6d/4d coupled system.

\subsubsection{Quantum curve}
\label{subsec:Quantum_curve}
Now we generalize the difference equation \eqref{eq:difference-pert} satisfied by the perturbative part to the full partition function of the 6d/4d coupled system including all self-dual string contributions. This eventually establishes the quantization of the Seiberg-Witten curve of the 6d $\sorm(2N)$ gauge theory.

As noticed, the action of the difference operator $Y$ on the defect partition function in the NS-limit $\epsilon_2\rightarrow0$ is associated with the appearance of a codimension 4 defect. Hence, it is natural to expect that the left-hand side of \eqref{eq:difference-pert} generalizes to include self-dual string corrections such that after the resulting difference operator acts on the 6d/4d partition function, the right-hand side of \eqref{eq:difference-pert} generalizes to include the self-dual string contributions of the codimension 4 defect partition function. Therefore, a legitimate ansatz for the full difference equation takes the form
\begin{align}
  \left[ Y  + V(Y,x) \right] \Psi(x) = \chi(x) \Psi(x) \ , 
  \label{eq:ansatz}
\end{align}
where $V(Y,x)$ denotes the self-dual string corrections to the shift operator, which is subject to the constraint $V(Y,x)\rightarrow 0$ in the perturbative limit. Moreover, 
\begin{align}
    \Psi(x)\equiv \lim_{\epsilon_2\rightarrow 0} \frac{Z^{\rm{6d/4d}}(x)}{Z^{\rm{6d}}}\ , \quad \chi(x) \equiv  \lim_{\epsilon_2 \rightarrow 0} \frac{Z^{\rm{6d/2d}}(x)}{Z^{\rm{6d}}} \ ,
    \label{eq:norm_def_pf}
\end{align}
are the NS-limits of the expectation values of the codimension 2 defect and codimension 4 defect, respectively.
In fact, based on the ansatz \eqref{eq:ansatz}, we find that that the normalized defect partition function $\Psi(x)$ satisfies the following difference equation:
\begin{align}\label{eq:quantum-curve}
  \left[ Y  + \vartheta_1(2x)\vartheta_1(\epsilon_1+2x)^2\vartheta_1(2\epsilon_1+2x)\prod_{l=1}^{2N-8}\vartheta_1(x+\tfrac{\epsilon_1}{2}\pm m_l)Y^{-1} \right]  \Psi(x) = \chi(x) \Psi(x) \ ,
\end{align}
which we have verified up to 3-string order by using the explicit elliptic genus results of Appendix \ref{app:genera_details2}. 

Also, based on the conjecture \eqref{eq:cod4splitting}, the function $\chi(x)$ for the $\sorm(2N)$ theory can be alternatively written as
\begin{align}
    \chi(x)=\sum_{n=0}^{N}H_n\theta_2^{N-n}(2x;2\tau)\theta_3^{n}(2x;2\tau), \label{eq:chieven}
\end{align}
where $H_n$ is the normalized $\frac{1}{2}$-BPS Wilson loop in the NS-limit 
\begin{align}\label{eq:hamiltonian}
    H_n(\phi,\alpha_i,m_l;\epsilon_{1})= \lim_{\epsilon_2\rightarrow 0}\frac{\mathcal{W}^{(n)}(\phi,\alpha_i,m_l;\epsilon_{1,2})}{Z^{\rm{6d}}_{\rm {str}}} \ .
\end{align}
The function \eqref{eq:chieven} is an even section of a degree $2N$ line bundle over the elliptic curve as $\theta_2$ and $\theta_3$ are both even under the reflection $x \mapsto -x$. In fact, the number of even and odd sections of a degree $d$ line bundle $L$ is given by \cite{AbelianVarieties}
\begin{equation} \label{eq:Levenodd}
    h^0(L)_{\pm} = \frac{1}{2} h^0(L) \pm 2^{-s},
\end{equation}
where $s=1$ if $d$ is odd and $s=0$ if $d$ is even. In our present case where $d=2N$, the formula gives 
\begin{equation}
    h^0(L)_{+} = \frac{1}{2}(2N) + 2^0 = N + 1,
\end{equation}
a basis of which is given by $\theta_2^{N-n}(2x;2\tau)\theta_3^n(2x;2\tau)$ for $n=0,\ldots, N$. Thus, we see that all $H_n$ appearing in \eqref{eq:chieven} are independent and do not mix, and we can associate them to the $N+1$ nodes of the affine $\widehat{D}_N$ Dynkin diagram.

In the classical limit $\epsilon_1\rightarrow0$, the difference equation \eqref{eq:quantum-curve} reduces to the elliptic Seiberg-Witten curve for the 6d $\sorm(2N)$ theory which was obtained in \cite{Haghighat:2018dwe} by analyzing the elliptic genera of self-dual strings in the thermodynamic limit.
Therefore, we claim that the difference equation \eqref{eq:quantum-curve} realizes a quantization of the elliptic Seiberg-Witten curve of the 6d $\sorm(2N)$ gauge theory. Under this quantization, two coordinates $X=\re^{x}$ and $Y$ on the elliptic curve are promoted to the non-commuting quantum operators with the relation $YX=pXY$.

We can also reformulate the difference equation \eqref{eq:quantum-curve}, by factoring out the perturbative part from the codimension 2 defect partition function, as an equation only acting on the elliptic genera of the self-dual strings:
\begin{align}\label{eq:Dn_curve}
  &\left[Y+q_{\phi}\frac{\vartheta_1(2x)\vartheta_1(\epsilon_1+2x)^2\vartheta_1(2\epsilon_1+2x)\prod_{l=1}^{2N-8}\vartheta_1(x+\frac{\epsilon_1}{2}\pm m_l)}{\prod_{i=1}^{N}\vartheta_1(x\pm \alpha_i)\vartheta_1(x+\epsilon_1\pm \alpha_i)}Y^{-1}-\widetilde\chi(x)\right]\widetilde{\Psi}(x) = 0 \ , \\
  &\qquad \text{with} \qquad 
  \widetilde\Psi(x) = \lim_{\epsilon_2\rightarrow 0}\frac{Z_{\rm str}^{\mathrm{6d/4d}}(x;\,\epsilon_1,\,\epsilon_2)}{Z_{\rm str}^{\mathrm{6d}}(x;\,\epsilon_1,\,\epsilon_2)}
 \qquad \text{and} \qquad 
  \widetilde\chi(x) = \lim_{\epsilon_2\rightarrow 0}\frac{Z_{\rm str}^{\mathrm{6d/2d}}(x;\,\epsilon_1,\,\epsilon_2)}{Z_{\rm str}^{\mathrm{6d}}(x;\,\epsilon_1,\,\epsilon_2)}
\label{eq:normalised_pf_str}
\end{align}
where $\widetilde\Psi(x) $ and $\widetilde\chi(x) $ are the NS-limits of the normalized elliptic genera of self-dual strings in the presence of the codimension 2 and codimension 4 defect, respectively.
In the following subsections, we provide other arguments or partial proofs of \eqref{eq:Dn_curve} by, firstly, studying the action of the difference equation on the finite sector of the self-dual string contributions and, secondly, by performing a saddle point analysis of the path integral representation in the NS-limit.

\subsubsection{Finite sector under NS-limit}
In this subsection, we focus on a subsector of the normalized codimension 2 or 4 defect partition functions and verify the proposed difference equation. Recall that the defect partition functions $Z_{\rm str}^{\mathrm{6d/4d}}(\epsilon_1,\,\epsilon_2)$ and $Z_{\rm str}^{\mathrm{6d/2d}}(\epsilon_1,\,\epsilon_2)$ are divergent for $\epsilon_2$ going to zero, and thus need to be normalized as in \eqref{eq:normalised_pf_str} to render the NS-limit finite.
The normalized defect partition function of the $k$-th string order in general depends on the derivatives of theta function $\theta_1^{(1)}$, $ \theta_1^{(2)}$, $\ldots$, $\theta_1^{(k)}$, which appear due to the normalization. On the other hand, there are also contributions in $Z_{\rm str}^{\mathrm{6d/4d}}(\epsilon_1,\,\epsilon_2)$ and  $Z_{\rm str}^{\mathrm{6d/2d}}(\epsilon_1,\,\epsilon_2)$, which are \emph{finite} in the NS-limit. We observe that all these contributions are from the poles originating from the defect part, and only in terms of theta function $\theta_1$, but not its derivatives. Therefore, we can decompose the normalized partition function into two sectors,
\begin{align}
\widetilde\Psi\equiv \widetilde\Psi_{\mathrm{finite}}[ \theta_1]+ \widetilde\Psi_{\mathrm{etc}}[\theta_1^{(1)},\, \theta_1^{(2)},\, \dots]\,,
\end{align} 
where $\widetilde\Psi_{\mathrm{finite}}$ denotes terms that originate from terms in $Z_{\rm str}^{\mathrm{6d/4d}}(\epsilon_1,\,\epsilon_2)$ that are intrinsically finite in the NS-limit; while $\widetilde\Psi_{\mathrm{etc}}$ contains the remaining terms that are only finite because of the normalization. One notes in particular that the two sectors are distinct and do not mix. Consequently, both $\widetilde\Psi_{\mathrm{finite}}$ and $\widetilde\Psi_{\mathrm{etc}}$ need to satisfy the difference equation separately. It is straightforward to verify analytically that the difference equation is satisfy for the finite sector $\widetilde\Psi_{\mathrm{finite}}[ \theta_1]$. Here, we present the results up to 3-instanton order for the partition functions with codimension 2 or 4 defects. The computational details are provided in Appendix \ref{app:NS_finitie_sector}.

\paragraph{Codimension 2 defect.} The finite sector of the codimension 2 defect partition function up to 3-instanton is given by
\begin{align}
\label{eq:codim_2_finite}
\widetilde\Psi_{\mathrm{finite}}=1+\frac{q_\phi}{2}\widetilde\Psi_1+\frac{q_\phi^2}{8}\widetilde\Psi_2+\frac{q_\phi^3}{48}\widetilde\Psi_3+\mathcal{O}(q_{\phi}^4),
\end{align}
where
\begin{subequations}
\begin{align}
\widetilde\Psi_1&=Q(x)\,,\\
\widetilde\Psi_2&=-Q(x)^2+Q(x)Q(x-\epsilon_1)+3Q(x)Q(x+\epsilon_1)\,,\\
\widetilde\Psi_3&=3Q(x-2\epsilon_1)Q(x-\epsilon_1)Q(x)-Q(x-\epsilon_1)^2Q(x)-4Q(x-\epsilon_1)Q(x)^2-2Q(x)^2Q(x+\epsilon)\notag\\
&+3Q(x)^3-Q(x)Q(x+\epsilon_1)^2+2Q(x-\epsilon_1)Q(x)Q(x+\epsilon_1)+3Q(x)Q(x+\epsilon_1)Q(x+2\epsilon_1)\notag\\
&+12Q(x)Q(x+\epsilon_1)Q(x+2\epsilon_1)+4Q(x-\epsilon_1)Q(x)Q(x+\epsilon_1)-4Q(x)Q(x+\epsilon_1)^2\,,
\end{align}
\end{subequations}
and we have introduced the following function:
\begin{align}
Q(x)=\vartheta_1(2x)\vartheta_1(2x+\epsilon_1)^2\vartheta_1(2x+2\epsilon_1)\frac{\prod_{l=1}^{2N-8}\vartheta_1(x\pm m_l+\frac{\epsilon_1}{2})}{\prod_{i=1}^N\vartheta_1(x\pm\alpha_i)\vartheta_1(x\pm\alpha_i+\epsilon_1)}\,.
\label{eq:Q-fct}
\end{align}
As it turns out, this appearing function $Q$ is precisely the correction to the difference equation \eqref{eq:Dn_curve} due to the self-dual strings.

\paragraph{Codimension 4 defect.} The finite sector of the codimension 4 defect partition function up to 3-instanton is given by
\begin{align}
\label{eq:codim_4_finite}
\tilde\chi_{\mathrm{finite}}=1+\frac{q_\phi}{2}\tilde\chi_1+\frac{q_\phi^2}{8}\tilde\chi_2+\frac{q_\phi^3}{48}\tilde\chi_3+\mathcal{O}(q_{\phi}^4),
\end{align}
where
\begin{subequations}
\begin{align}
\tilde\chi_1&=Q(x)+Q(x-\epsilon_1)\,,\\
\tilde\chi_2&=-Q(x)^2+Q(x-2\epsilon_1)Q(x-\epsilon_1)-Q(x-\epsilon_1)^2+Q(x)Q(x+\epsilon_1) \,,\\
\tilde\chi_3&=3Q(x-3\epsilon_1)Q(x-2\epsilon_1)Q(x-\epsilon_1)+3Q(x)Q(x+\epsilon_1)Q(x+2\epsilon_1)-Q(x-2\epsilon_1)^2Q(x-\epsilon_1)\notag\\
&-Q(x)Q(x+\epsilon_1)^2-Q(x)^2Q(x-\epsilon_1)-Q(x-\epsilon_1)^2Q(x)-4Q(x-2\epsilon_1)Q(x-\epsilon_1)^2\notag\\
&-4Q(x+\epsilon_1)Q(x)^2+3Q(x)^3+3Q(x-\epsilon_1)^3\,,
\end{align}
\end{subequations}
with $Q$ as defined in \eqref{eq:Q-fct}.
\paragraph{Difference equation in finite sector.}
Collecting all the results, one can verify that the finite sectors of the codimension 2 or 4 defect partition functions in \eqref{eq:codim_2_finite} and \eqref{eq:codim_4_finite}, respectively, satisfy the following difference equation,
\begin{align}
\left(Y+q_\phi Q(x)\cdot Y^{-1}\right)\widetilde\Psi_{\mathrm{finite}}(x)=\tilde\chi_{\mathrm{finite}}(x)\cdot \widetilde\Psi_{\mathrm{finite}}(x)\,,
\end{align}
which is nothing else than \eqref{eq:Dn_curve} acting on the finite sector
\subsection{Path integral representation}
\label{sec:path_integral_representation}
In this subsection, we derive the quantum curve from saddle point approach in a path integral formalism developed in \cite{Chen:2020jla,Chen:2021ivd}. We start from the $k$-string elliptic genus,
\begin{align}
Z_{k}^{\sprm(k)}
&=\frac{1}{2^k k!}
 \int \prod_{I=1}^k \diff {u_I} \left( \frac{ 2\pi \eta^2 \vartheta_1(2 
\epsilon_+)}{\vartheta_1( \epsilon_1) \vartheta_1( \epsilon_2)} \right)^k
\cdot \prod_{I=1}^k \left(\vartheta_1(\pm 2{u_I})\vartheta_1(\pm 2{u_I}+2\epsilon_+)\frac{\prod_{l=1}^{2N-8}\vartheta_1(\pm {u_I}+m_l)}{\prod_{i=1}^N\vartheta_1(\epsilon_+\pm{u_I}\pm\alpha_i)}\right)\notag\\
& \qquad \qquad \qquad \qquad\qquad \qquad \qquad  \quad 
\cdot\prod_{1\leq I< J\leq k}
\frac{\vartheta_1 (\pm {u_I}\pm {u_J})  \vartheta_1 
(\pm {u_I}\pm {u_J}+2\epsilon_+)}{\vartheta_1(\pm {u_I}\pm {u_J}+\epsilon_1) 
\vartheta_1(\pm {u_I}\pm {u_J}+\epsilon_2)}\,,\notag\\
&=\int \prod_{I=1}^k \diff {u_I} \left( \frac{ 2\pi \eta^2 \vartheta_1(2 
\epsilon_+)}{\vartheta_1( \epsilon_1) \vartheta_1( \epsilon_2)} \right)^k
\cdot \prod_{I=1}^k \left(\vartheta_1(\pm 2{u_I})\vartheta_1(\pm 2{u_I}+2\epsilon_+)\frac{\prod_{l=1}^{2N-8}\vartheta_1(\pm {u_I}+m_l)}{\prod_{i=1}^N\vartheta_1(\epsilon_+\pm{u_I}\pm\alpha_i)}\right)\notag\\
&\qquad \qquad\qquad\qquad\qquad \qquad \quad  \cdot\prod_{I\neq J}^k\left(
\frac{\vartheta_1 (\pm {u_I}\pm {u_J})  \vartheta_1 
(\pm {u_I}\pm {u_J}+2\epsilon_+)}{\vartheta_1(\pm {u_I}\pm {u_J}+\epsilon_1) 
\vartheta_1(\pm {u_I}\pm {u_J}+\epsilon_2)}\right)^{\frac{1}{2}} \,,\notag\\
&\equiv\frac{1}{2^k k!}\int \prod_{I=1}^k \diff {u_I} \,\Delta^k
\cdot \prod_{I=1}^k \mathcal Q({u_I})\cdot \prod_{I\neq J}D({u_I}, {u_J})^{\frac{1}{2}}\,,
\end{align}
where $2^k k!$ is the order of the Weyl group of $\sprm(k)$ and $\vartheta_1(z)\equiv \frac{\ri\theta_1(z)}{\eta}$ is defined for convenience, as above. Moreover, the following notation has been introduced: 
\begin{subequations}
\begin{align}
D(u, u^\prime) &=\frac{\vartheta_1 (\pm u\pm u^\prime)  \vartheta_1 
(\pm u\pm u^\prime+2\epsilon_+)}{\vartheta_1(\pm u\pm u^\prime+\epsilon_1) 
\vartheta_1(\pm u\pm u^\prime+\epsilon_2)}\,,\\
\mathcal Q(u) &=\vartheta_1(\pm 2u)\vartheta_1(\pm 2u+2\epsilon_+)\frac{\prod_{l=1}^{2N-8}\vartheta_1(\pm u+m_l)}{\prod_{i=1}^N\vartheta_1(\epsilon_+\pm u\pm\alpha_i)}\,,\label{eq:mathcalQ}\\
\Delta &=\frac{2\pi \eta^2 \vartheta_1(2 
\epsilon_+)}{\vartheta_1( \epsilon_1) \vartheta_1( \epsilon_2)} \,.
\end{align}
\label{eq:def_D_Q_Delta}
\end{subequations}
Next, we introduce a density function
\begin{align}
\bar\rho(u)\equiv \frac{1}{\Delta}\sum_{i=1}^k\delta(u-u_I)\,,
\label{eq:density_function}
\end{align}
such that the integrand in the partition function can be recast as
\begin{align}
\prod_{I=1}^k &\mathcal Q({u_I})\cdot \prod_{I\neq J}D({u_I}, {u_J})^{\frac{1}{2}} 
=\exp\left(\sum_{I=1}^k\log \mathcal Q({u_I})+\frac{1}{2}\sum_{I\neq J}\log D({u_I}, {u_J})\right)\notag\\
&=\int\!\mathcal D\rho\;\delta\left(\rho(u)-\bar\rho\right)\cdot\exp\left(\frac{\Delta^2}{2}\int\! \diff u \, \diff u^\prime\,\rho(u^\prime)\log D(u, u^{\prime})\rho(u)+\Delta\int\!\diff u\, \rho(u)\log \mathcal Q(u)\right)\notag\\
&=\int\!\mathcal D\rho\,\mathcal D\lambda\,\exp\bigg(\frac{\Delta^2}{2}\int\! \diff u \, \diff u^\prime\,\rho(u^\prime)\log D(u, u^{\prime})\rho(u)+\Delta\int\!\diff u\,\rho(u)\log \mathcal Q(u)\notag\\
&\qquad \qquad\qquad\qquad +\ri\int\!\diff u\,\lambda(u)\cdot\left(\rho(u)-\bar\rho\right)\bigg)\,,
\end{align}
where the Fourier transform of the delta distribution
\begin{align}
\delta(\rho(u)-\bar\rho)=\int\!\mathcal D\lambda\,\exp\left(\ri\int\! \diff u\, \lambda(u)\left(\rho(u)-\bar\rho\right)\right)
\end{align}
has been used. Therefore, the partition function can be rewritten as
\begin{align}
Z_{\rm str}^{\rm{6d}}&=\sum_{k=0}^{\infty}q_\phi^{k}Z_{k}^{\sprm(k)}\notag\\
&=\int\!\mathcal D\rho\,\mathcal D\lambda\,\sum_{k=0}^{\infty}\frac{q_\phi^{k}}{2^k k!}\int\!\left(\prod_{I=1}^k \frac{\diff {u_I}}{2\pi \ri}\right)\cdot\Delta^k \re^{-\ri\int\! \diff u\,\lambda(u)\cdot\bar\rho(u)}\\
&\cdot \exp\left(\frac{\Delta^2}{2}\int\! \diff u \, \diff u^\prime\,\rho(u^\prime)\log D(u, u^{\prime})\rho(u)+\Delta\int\!\diff u\, \rho(u)\log \mathcal Q(u)+\ri\!\int\!\diff u\,\lambda(u)\cdot\bar\rho(u)\right)\,.\notag
\end{align}
The sum over $k$ factor can be evaluated
\begin{align}
\begin{aligned}
\sum_{k=0}^{\infty}\frac{q_\phi^{k}}{2^k k!}\int\!\left(\prod_{I=1}^k \frac{\diff {u_I}}{2\pi \ri}\right)\cdot\Delta^k \re^{-\ri\int\!\diff u\,\lambda(u)\cdot\bar\rho(u)}
&=\sum_{k=0}^\infty\frac{1}{k!}\left(\frac{q_\phi\Delta}{2}\int\!\frac{\diff u}{2\pi \ri}\re^{-\frac{\ri}{\Delta}\lambda(u)}\right)^k \\
&=\exp\left(\frac{q_\phi\Delta}{2}\int\!\frac{\diff u}{2\pi \ri}\re^{-\frac{\ri}{\Delta}\lambda(u)}\right)\, ,
\end{aligned}
\end{align}
such that the expression becomes
\begin{align}
\begin{aligned}
Z_{\rm str}^{\rm{6d}}=\int\!\mathcal D\rho\,\mathcal D\lambda\,\exp&\bigg(\frac{\Delta^2}{2}\int\! \diff u \, \diff u^\prime\,\rho(u^\prime)\log D(u, u^{\prime})\rho(u) \\
&+\Delta\int\!\diff u\, \rho(u)\log \mathcal Q(u)+\ri\!\int\!\diff u\,\lambda(u)\cdot\rho(u)+\frac{q_\phi\Delta}{4\pi \ri}\int\! \diff u\,\re^{-\frac{\ri}{\Delta}\lambda(u)}\bigg)\,.
\end{aligned}
\end{align}
One can further employ a shift in the auxiliary variable $\lambda(u)$ as
\begin{align}
\lambda(u)=\lambda^\prime(u)-\ri\Delta\log(-q_\phi)\,,
\end{align}
to remove the explicit $q_\phi$ dependence of the last term. Finally, one ends up with
\begin{align}
\begin{aligned}
Z_{\rm str}^{\rm{6d}}=&\int\!\mathcal D\rho\,\mathcal D\lambda^\prime\,\exp\bigg(\frac{\Delta^2}{2}\int\! \diff u \,\diff u^\prime\,\rho(u^\prime)\log D(u, u^{\prime})\rho(u)\\
&+\Delta\int\!\diff u\, \rho(u)\log\left(-q_\phi \mathcal Q(u)\right)+\ri\!\int\!\diff u\,\lambda^\prime(u)\cdot\rho(u)-\frac{\Delta}{4\pi \ri}\int\! \diff u\,\re^{-\frac{\ri}{\Delta}\lambda^\prime(u)}\bigg)\,.
\end{aligned}
\label{eq:path_integral_reprensentation_of_instanton_partition_function}
\end{align}
Now, consider the leading order approximation of \eqref{eq:path_integral_reprensentation_of_instanton_partition_function} in the NS-limit $\epsilon_2\rightarrow 0$. It is useful to evaluate the small $\epsilon_2$ expansions of the following terms: 
\begin{subequations}
\label{eq:leading_order_approximations}
\begin{align}
\Delta &=\epsilon_2^{-1}+\mathcal O(1)\,,\\
\log D(u, u^\prime) &=\left(K(u, u^\prime;\, \epsilon_1)-K(u, u^\prime;\, 0)\right)\cdot\epsilon_2+\mathcal O(\epsilon_2^2)\,,\\
\mathcal Q(u) &=\mathcal Q_0(u)+\mathcal O(\epsilon_2)\,,
\end{align}
\end{subequations}
with the following short-hand notations:
\begin{subequations}
\begin{align}
K(u, u^\prime;\, \beta) &\equiv \frac{\partial}{\partial\beta}\log\vartheta_1(\pm u\pm u^\prime+\beta)\notag\\
&=\frac{\vartheta_1^{\prime}(u+u^\prime+\beta)}{\vartheta_1(u+u^\prime+\beta)}-\frac{\vartheta_1^{\prime}(u+u^\prime-\beta)}{\vartheta_1(u+u^\prime-\beta)}+\frac{\vartheta_1^{\prime}(u-u^\prime+\beta)}{\vartheta_1(u-u^\prime+\beta)}-\frac{\vartheta_1^{\prime}(u-u^\prime-\beta)}{\vartheta_1(u-u^\prime-\beta)} \,,\\\notag
\mathcal Q_0(u)&\equiv \mathcal Q(u)|_{\epsilon_2=0} =\vartheta_1(\pm 2u)\vartheta_1(\pm 2u+\epsilon_1)\frac{\prod_{l=1}^{2N-8}\vartheta_1(\pm u+m_l)}{\prod_{i=1}^N\vartheta_1(\frac{\epsilon_1}{2}+\pm u\pm\alpha_i)}\notag\\
&=\vartheta_1(2u-\epsilon_1)\vartheta_1(2u)^2\vartheta_1(2u+\epsilon_1)\frac{\prod_{l=1}^{2N-8}\vartheta_1(u\pm m_l)}{\prod_{i=1}^N\vartheta_1(u\pm\alpha_i\pm \frac{\epsilon_1}{2})}\, .
\end{align}
\end{subequations}
It is worth noticing that $K(u, u^\prime; 0)=0$ by the virtue of the odd and even properties of $\vartheta_1$ and $\vartheta_1^\prime$. Therefore, in leading order $\mathcal O(\epsilon_2^{-1})$, one finds
\begin{align}
\begin{aligned}
Z_{\rm str}^{\rm{6d}}=\int\!\mathcal \mathcal D\lambda^\prime\ldots D\rho\,\exp &\bigg(\frac{1}{2\epsilon_2}\int\! \diff u \, \diff u^\prime\,\rho(u^\prime)K(u,u^\prime;\,\epsilon_1)\rho(u) \\
&+\frac{1}{\epsilon_2}\int\!\diff u\, \rho(u)\log\left(-q_\phi \mathcal Q_0(u)\right)+\mathcal O(1)\bigg)\,,
\end{aligned}
\label{eq:leading_order_of_instanton_partition_function}
\end{align}
where $\ldots$ denotes terms of $\lambda^\prime(u)$ that are irrelevant for the subsequent saddle point analysis.

\paragraph{Saddle point equation.}
After the preliminary considerations, one can proceed to study the saddle point of \eqref{eq:leading_order_of_instanton_partition_function}. 
For this, one varies \eqref{eq:leading_order_of_instanton_partition_function} with respect to $\rho(u)$, which leads to
\begin{align}
\int\! \diff u^\prime\, K(u,u^\prime; \epsilon_1)\rho(u^\prime)+\log\left(-q_\phi \mathcal Q_0(u)\right)=0\,.
\label{eq:saddle_point_equation_raw}
\end{align}
As in \cite{Chen:2020jla,Chen:2021ivd}, it is useful to introduce the following quantities:
\begin{align}
\omega(x)\equiv\exp\left(\int\!\diff u^\prime\,\rho(u^\prime)\frac{\vartheta_1^\prime(u^\prime+x)}{\vartheta_1(u^\prime+x)}\right) \qquad \mathrm{and} \qquad 
\Phi(x)\equiv\frac{\omega(x)}{\omega(-x)}.
\label{eq:defect_partition_function_in_path_integral_formalism}
\end{align}
The saddle point equation \eqref{eq:saddle_point_equation_raw} can be rewritten as
\begin{subequations}
\begin{align}
\log\left(\frac{\omega(u+\epsilon_1)}{\omega(-u-\epsilon_1)}\cdot\frac{\omega(-u+\epsilon_1)}{\omega(u-\epsilon_1)}\right)+\log\left(-q_\phi \mathcal Q_0(u)\right)=\log\left(-q_\phi \mathcal Q_0(u)\cdot\frac{\Phi(u+\epsilon_1)}{\Phi(u-\epsilon_1)}\right)=0\,,
\end{align}
or simply as 
\begin{align}
\Phi(u-\epsilon_1)+q_\phi \mathcal Q_0(u)\cdot\Phi(u+\epsilon_1)=0\,.
\end{align}
\end{subequations}
For later convenience, we shift $u\rightarrow u+\frac{\epsilon_1}{2}$, which leads to
\begin{align}
\left[Y+q_\phi Q(u)\cdot Y^{-1}\right]\Phi\left(u+\tfrac{\epsilon_1}{2}\right)=0\,,
\label{eq:saddle_point_equation}
\end{align}
with $Q(u)\equiv \mathcal Q_0(u+\epsilon_1/2)$ as defined in \eqref{eq:Q-fct} and the shift operator $Y$ as in \eqref{eq:def_shift_op}.
One finds that \eqref{eq:saddle_point_equation} is precisely the LHS of the proposed difference equation \eqref{eq:Dn_curve}. However, the saddle point equation is \emph{only} satisfied for certain fixed saddle point(s) ``$u$"; for generic defect parameter $x$, the intuition from \cite{Chen:2020jla,Chen:2021ivd} suggests to equate the difference equation to a codimension 4 defect or, say, a Wilson surface expectation value $\mathcal W^S(x)$, as follows:
\begin{align}
\left[Y+q_\phi Q(x)\cdot Y^{-1}\right]\Phi\left(x+\tfrac{\epsilon_1}{2}\right)=\mathcal W^S(x)\cdot \Phi\left(x+\tfrac{\epsilon_1}{2}\right)\,.
\end{align}
This eventually leads to the expected full difference equation \eqref{eq:Dn_curve}, as discussed now.
\paragraph{Defect partition function in path integral formalism.}
Now we show that $\Phi\left(x+\tfrac{\epsilon_1}{2}\right)$, as defined in \eqref{eq:defect_partition_function_in_path_integral_formalism}, can be identified with the partition function of the theory in the presence of a codimension 2 defect $\widetilde\Psi(x)$. Recall that the defect induces extra contributions to the $k$-string one-loop determinant,
\begin{align}
Z_{k,1-\text{loop}}^{\rm{4d}}(u, x)
=\prod_{I=1}^k\frac{\vartheta_1(\pm {u_I}+x+\epsilon_+)}{\vartheta_1(\pm {u_I}+x+\epsilon_-)}\,.
\end{align}
The above saddle point analysis can equally well be applied to the defect partition function. In this case, one starts with
\begin{align}
\begin{aligned}
Z_{\mathrm{str}}^{\mathrm{6d/4d}}&=\int\!\mathcal D\rho\,\mathcal D\lambda^\prime\,\exp\bigg(\frac{\Delta^2}{2}\int\! \diff u \, \diff u^\prime\,\rho(u^\prime)\log D(u, u^{\prime})\rho(u)\\
&\quad+\Delta\int\!\diff u\, \rho(u)\log\left(-q_\phi \mathcal Q(u)\cdot V_D(u, x)\right)+\ri\!\int\!\diff u\,\lambda^\prime(u)\cdot\rho(u)-\frac{\Delta}{4\pi \ri}\int\! \diff u\,\re^{-\frac{\ri}{\Delta}\lambda^\prime(u)}\bigg)\,,
\end{aligned}
\end{align}
and the defect contribution is defined as
\begin{align}
    V_D(u, x)\equiv\frac{\vartheta_1(\pm u+x+\epsilon_+)}{\vartheta_1(\pm u+x+\epsilon_-)} \,.
\end{align}
One notes in particular that
\begin{align}
\Delta\cdot\log V_D(u, x)=\left(\frac{\vartheta_1^\prime(u+x+\epsilon_1/2)}{\vartheta_1(u+x+\epsilon_1/2)}-\frac{\vartheta_1^\prime(u-x-\epsilon_1/2)}{\vartheta_1(u-x-\epsilon_1/2)}\right)+\mathcal O(\epsilon_2)\,.
\end{align}
Therefore, the defect does not contribute to the saddle point equation \eqref{eq:saddle_point_equation_raw}, but rather gives corrections at order $\mathcal{O}(1)$, analogous to the observations in \cite{Chen:2020jla,Chen:2021ivd}. Consequently, the normalized defect partition function is given by
\begin{align}
\widetilde{\Psi}(x)=\lim_{\epsilon_2\rightarrow 0}\frac{Z_{\mathrm{str}}^{\mathrm{6d/4d}}}{Z_{\mathrm{str}}^{\rm{6d}}}=\exp\left(\int\! \diff u\, \rho(u)\left(\frac{\vartheta_1^\prime(u+x+\epsilon_1/2)}{\vartheta_1(u+x+\epsilon_1/2)}-\frac{\vartheta_1^\prime(u-x-\epsilon_1/2)}{\vartheta_1(u-x-\epsilon_1/2)}\right)\right)=\Phi(x+\epsilon_1/2)
\end{align}
due to the  saddle point equation.

\section{6d \texorpdfstring{$\sorm(2N+1)$}{SO(2N+1)} theories}
\label{sec:BN}
In this section, we compute the partition functions of the 6d $\sorm(2N+1)$ gauge theories defined on a $-4$ curve in the presence of the codimension 2 and 4 defects.
We also present the difference equations satisfied by the defect partition functions and show that the difference equation realizes a quantization of the elliptic Seiberg-Witten curve for the 6d $\sorm(2N+1)$ gauge theory.
Since most of the arguments are analogous to the $\sorm(2N)$ case of Section \ref{sec:DN}, we are brief and highlight only the essentials.
\subsection{Self-dual strings and ADHM construction}
Let us first compute the partition function of the 6d $\sorm(2N+1)$ gauge theory with $2N-7$ flavors on $\mathbb{R}^4\times \mathbb{T}^2$ without defect. 
The form of the partition function is analogous to \eqref{eq:part_fct_explained}. First, the perturbative part is given by
\begin{align}
    Z^{\rm 6d}_{\rm pert} &= {\rm exp}\left(\frac{1}{\epsilon_1\epsilon_2}F_{\rm eff}\right)
    \cdot {\rm PE}\bigg[-\frac{1+p q}{(1-p)(1-q)(1-Q)} \left(\sum_{i<j}^N\left(A_i A_j^{\pm1}+A_i^{-1} A_j^{\pm1}Q\right)+\sum_{i=1}^N\left(A_i +A_i^{-1}Q\right) \right)\nonumber\\
&\qquad\qquad\qquad \qquad +\frac{\sqrt{pq}}{(1-p)(1-q)(1-Q)}\left(\sum_{i=1}^N(A_i+A_i^{-1}Q)+1\right)\sum_{l=1}^{2N-7}\left(M_l+M_l^{-1}\right)\bigg]\,, \label{Zpert_BN} \\
    F_{\rm eff} &= \frac{1}{6}\left(\sum_{i<j}(\alpha_i\pm\alpha_j)^3+\sum_i \alpha_i^3\right)-\frac{1}{12}\sum_{i=1}^{N}\sum_{j=1}^{2N-7}(\alpha_i\pm m_j)^3+\frac{1}{2}\phi_0\sum_{i=1}^{N}\alpha_i^2+\ldots,
\end{align}
where $A^{\pm}_i\equiv A_i+A_i^{-1}$, and the terms omitted in the last line are irrelevant for the discussion below.

To evaluate the self-dual string contributions, one considers the 2d $\Ncal=(0,4)$ quiver gauge theories for the self-dual strings in the 6d $\sorm(2N+1)$ gauge theory, which is reminiscent to the 2d theories in \eqref{eq:2d_quiver} for the self-dual strings in the 6d $\sorm(2N)$ gauge theory.
At $k$-strings, the worldsheet theory is an $\sprm(k)$ gauge theory coupled to the bulk gauge group $\sorm(2N+1)$ and the flavor group $\sprm(N_f)$ as follows:
\begin{align}
    \raisebox{-.5\height}{
    \begin{tikzpicture}
  \tikzstyle{gauge} = [circle, draw,inner sep=3pt];
  \tikzstyle{flavour} = [regular polygon,regular polygon sides=4,inner sep=3pt, draw];
  \node (g1) [gauge,label=below:{$\sprm(k)$}] {};
  \node (f1) [flavour,above of=g1, label=above:{$\sorm(2N+1)$}] {};
  \node (f2) [flavour,right of=g1, label=right:{$\sprm(N_f)$}] {};
  \draw (g1)--(f1);
  \draw[dashed] (g1)--(f2);
  \draw (g1) to [out=140,in=220,looseness=10] (g1);
  \node at (-1,0) {\small{asym}};
  \end{tikzpicture}
    } 
    \quad \text{with } N_f=2N-7.
    \label{eq:2d_quiver_BN}
\end{align}
The elliptic genus of this quiver theory then gives rise to $k$-string contribution to the partition function. We compute
\begin{align}\label{Z1loop_BN}
Z_{\text{str}}&=\sum_{k=0}^{\infty}q_{\phi}^k \oint[\rd u]Z_{k,1-\text{loop}}^{\rm{6d}}(u_I) \ , \nonumber  \\
Z_{k,1-\text{loop}}^{\rm{6d}}(u_I)&=\left(\frac{\vartheta _1\left(2 \epsilon _+\right)}{\vartheta _1\left(\epsilon_{1,2}\right)}\right)^k 
\cdot 
\prod _{I=1}^k \vartheta
   _1\left(\pm 2   u_I\right) \vartheta _1\left(2 \epsilon _+\pm 2 u_J\right)
   \cdot 
   \prod _{I<J}^k \frac{\vartheta _1\left(\pm
     u_I\pm   u_J\right) \vartheta _1\left(2 \epsilon _+\pm   u_I\pm
     u_J\right)}{\vartheta _1\left(\epsilon _1\pm   u_I\pm u_J\right) \vartheta _1\left(\epsilon _2\pm   u_I\pm u_J\right)}\nonumber \\
& \qquad \qquad \qquad \cdot 
\prod _{I=1}^k \frac{\prod _{l=1}^{2 N-7} \vartheta
   _1\left(\pm   u_I+m_l\right)}{\vartheta _1\left(\epsilon _+\pm
     u_I\right)\prod _{i=1}^N \vartheta _1\left(\epsilon _+\pm
     u_I\pm \alpha_i\right)},
\end{align}
where $u_I$, $\alpha_i$ and $m_l$ denote the holonomies for the $\sprm(k)$ gauge group, the bulk $\sorm(2N+1)$ gauge
group, and the $\sprm(N_f)$ flavor group, respectively. 
One may realize that the result  \eqref{Z1loop_BN} as well as \eqref{Zpert_BN} can also be obtained via Higgsing of the partition function of the $\sorm(2N+2)$ theory in \eqref{Z1loop_DN} by suitably tuning the parameters as 
\begin{align}\label{eq:Bn_higgs}
    \alpha_{N+1}=0,\quad m_{2N-6}=-\epsilon_+,
\end{align}
which is consistent with the Higgsing process of the associated brane webs in \eqref{eq:Higgs_D_to_B}.

As before, the elliptic genera are evaluated via the JK-residue prescription.
Choosing the analogous poles as in \eqref{eq:DN_Z1_poles}, the one-string elliptic genus is given by
\begin{align}\label{eq:BN_Z1}
Z_1^{\rm{6d}}
&= -\frac{1}{2} \sum_{i=1}^N \Bigg(\frac{\vartheta_1(2\epsilon_+ + 2\alpha_i) \vartheta_1(4\epsilon_+ + 2\alpha_i) \prod_{l=1}^{2N-7}\vartheta_1(\epsilon_+ + \alpha_i \pm m_l)} {\vartheta_1(\epsilon_{1,2})\vartheta_1(\alpha_i)\vartheta_1(\alpha_i+2\epsilon_+) \prod_{j\neq i}^N  \vartheta_1(\alpha_i \pm \alpha_j) \vartheta_1(2\epsilon_+ + \alpha_i \pm \alpha_j)} + (\alpha_i \to -\alpha_i) \Bigg) \,.
\end{align}
The two-string elliptic genus can be found in Appendix \ref{app:genera_details1}. 
We also computed the three-string elliptic genus by taking the results for the $\sorm(2N+2)$ given in Appendix \ref{app:genera_details2} and applying the Higgs mechanism with the choice of parameters as in \eqref{eq:Bn_higgs}. As the expressions are tedious but straightforward, we refrain from presenting them here.
\subsection{Codimension 2 defect}
Analogously to the $\sorm(2N)$ case, the minimal codimension 2 defect for a $\sorm(2N+1)$ theory can be introduced by Higgsing the $\sorm(2N+3)$ theory. This can be achieved by giving a position dependent VEV to a certain moment map operator. In the $\Omega$-deformed partition function, the defect Higgsing amounts to tuning the parameters of the UV $\sorm(2N+3)$ theory as follows:
\begin{align}
\begin{aligned}
    &m_{2N-6}+m_{2N-5}+2\epsilon_++\epsilon_2=0\,,\quad \alpha_{N+1}+m_{2N-5}+\epsilon_+=0\,, \\
    &\qquad \text{with} \qquad 
x \equiv −\alpha_{N +1} + \epsilon_2 \ .
\end{aligned}
\label{eq:codim_two_defect_BN}
\end{align}
After this tuning, the partition function of the $\sorm(2N+1)$ theory with codimension 2 defect has the same structure as in \eqref{eq:pt_fct_explained_defect}, but now the perturbative contribution from the defect is given by
\begin{align}
     Z^{\rm{4d}}_{\rm pert}(x)&=Z_{\rm eff}^{\rm4d}\cdot \Gamma_e(X^{-1}) \prod_{i=1}^N\frac{\Gamma_e(X^{-1}A_i)}{\Gamma_e(pXA_i)} \, ,\notag\\
    Z^{\rm{4d}}_{\rm eff}(x)&=\exp\left[-\frac{(x+\tfrac{\epsilon_1}{2})}{\epsilon_1}\left(\sum_{i=1}^N\alpha_i+\frac{\phi_0}{2}-\left(N+\tfrac{1}{2}\right)\frac{\tau}{6}\right)+ \ldots\right]
\end{align}
up to $A_i$ independent terms, and the defect contribution to the 1-loop integrand in the $k$-string elliptic genus is
\begin{align}
    Z_{k,1-\text{loop}}^{\rm{4d}}(u_I,x)=\prod_{I=1}^k\frac{\vartheta_1(-\epsilon_+-x\pm u_I)}{\vartheta_1(-\epsilon_--x\pm u_I)},
\end{align}
where $x$ denotes the chemical potential for the $\urm(1)$ defect symmetry, as introduced in \eqref{eq:codim_two_defect_BN}. 
Again, the defect can equally well be understood as originating from a free 4d $\mathcal{N}=1$ chiral multiplet coupled to the bulk $\sorm(2N+1)$ gauge symmetry. The chiral multiplet transforms in the fundamental representation of $\sorm(2N+1)$ and has a mass parameter $x$.

The elliptic genera can either be evaluated by the JK-residue prescription, taking into account additional poles like \eqref{eq:add_poles_codim2}, or by tuning the parameters in the elliptic genera of self-dual strings in the $\sorm(2N + 3)$ theory as in \eqref{eq:codim_two_defect_BN}.
At one-string, we compute
\begin{align}
    Z_1^{\rm{6d/4d}}=&-\frac{1}{2}\frac{\vartheta_1(2x+\epsilon_1)\vartheta_1(2x+2\epsilon_1)\vartheta_1(2x-2\epsilon_2)\vartheta_1(2x+2\epsilon_-)\vartheta_1(2\epsilon_+)\prod_l^{2N-7} \vartheta_1(x\pm m_l +\epsilon_-)}{\vartheta_1(\epsilon_1)\vartheta_1(x+\epsilon_1)\vartheta_1(x-\epsilon_2)\prod_{i=1}^N\vartheta_1(x\pm \alpha_i+\epsilon_1)\vartheta_1(x\pm \alpha_i-\epsilon_2)}\nonumber\\
    & -\frac{1}{2} \sum_{i=1}^N \Bigg(\frac{\vartheta_1(2\epsilon_+ + 2\alpha_i) \vartheta_1(4\epsilon_+ + 2\alpha_i) \prod_{l=1}^{2N-8}\vartheta_1(\epsilon_+ + \alpha_i \pm m_l)} {\vartheta_1(\epsilon_{1,2})\vartheta_1(\alpha_i)\vartheta_1(\alpha_i+2\epsilon_+) \prod_{j\neq i}^N  \vartheta_1(\alpha_i \pm \alpha_j) \vartheta_1(2\epsilon_+ + \alpha_i \pm \alpha_j)}\nonumber\\
    &\quad\quad\quad\quad\times \frac{\vartheta_1(x+\epsilon_+\pm(\alpha_i+\epsilon_+))}{\vartheta_1(x+\epsilon_-\pm(\alpha_i+\epsilon_+))} + (\alpha_i \to -\alpha_i) \Bigg) \,.
\end{align}
Elliptical genera of higher strings can be calculated in the same way.

\subsection{Codimension 4 defect}
\label{sec:codim4_SO2N-1}
We now study the codimension 4 defect partition function for the 6d $\sorm(2N+1)$ gauge theory, which is entirely analogous to the $\sorm(2N)$ case of Section \ref{sec:codim4_SO2N}.
The only difference is that the additional 2d fermion now transforms in the fundamental representation of the bulk $\sorm(2N+1)$ gauge group. 
We have to take this additional 2d field in the elliptic genus calculation into account, as in \eqref{eq:Ycal_SO2N}.

Alternatively, the same codimension 4 defect can be obtained from the Higgsing the $\sorm(2N + 5)$ theory by giving position dependent VEVs to two moment map components. On the level of the partition function, this Higgsing corresponds to setting the parameters in the UV $\sorm(2N + 5)$ gauge theory as follows:
\begin{alignat}{3}\label{Bn_cod4_Higgs}
\begin{aligned}
m_{2N-5}&=\alpha_{N+1}-\epsilon_+-\epsilon_2\,, & \qquad 
m_{2N-6}&=-\alpha_{N+1}-\epsilon_+\,, & \qquad 
\alpha_{N+1}&=x+2\epsilon_+\,, \\
m_{2N-3}&=\alpha_{N+2}-\epsilon_+-\epsilon_2\,,& \qquad 
m_{2N-4}&=-\alpha_{N+2}-\epsilon_+\,,& \qquad 
\alpha_{N+2}&=-x+2\epsilon_+ \, .
\end{aligned}
\end{alignat}
With this parameter tuning, we obtain the partition function of the $\sorm(2N+1)$ gauge theory in the presence of a codimension 4 defect.

Analogously to the $\sorm(2N)$ case, the partition function of the 6d/2d coupled system can be written as \eqref{eq:pt_fct_explained_codim4}. The perturbative contribution of the 2d fermion reads
\begin{align}\label{Bn_2d_pert}
  Z^{\rm{2d}}_{\rm pert} =  q_{\phi}^{-\frac{1}{2}}\vartheta_1(x)\prod_{i=1}^{N}\vartheta_1(x\pm \alpha_i) \ .
\end{align}
Utilizing the JK-residue prescription and taking the new poles \eqref{eq:poles_codim4_SO2N} from the defect into account, one can compute the elliptic genera of the self-dual strings in the coupled system.
At the $k=1$ sector, the result is
\begin{align}
    Z_1^{\rm{6d/2d}}=&\frac{1}{2}\Bigg(\frac{\vartheta_1(2x-\epsilon_{1})\vartheta_1(2x-\epsilon_{2})\vartheta_1(2x-3\epsilon_+\pm \epsilon_+)\vartheta_1(2\epsilon_+)\prod_l^{2N-7} \vartheta_1(-x\pm m_l +\epsilon_+)}{\vartheta_1(x)\vartheta_1(x-2\epsilon_+)\prod_{i=1}^N\vartheta_1(x\pm \alpha_i)\vartheta_1(x\pm \alpha_i-2\epsilon_+)}+(x\rightarrow -x)\Bigg)\nonumber\\
    & -\frac{1}{2} \sum_{i=1}^N \Bigg(\frac{\vartheta_1(2\epsilon_+ + 2\alpha_i) \vartheta_1(4\epsilon_+ + 2\alpha_i) \prod_{l=1}^{2N-7}\vartheta_1(\epsilon_+ + \alpha_i \pm m_l)} {\vartheta_1(\epsilon_{1,2})\vartheta_1(\alpha_i)\vartheta_1(\alpha_i+2\epsilon_+) \prod_{j\neq i}^N  \vartheta_1(\alpha_i \pm \alpha_j) \vartheta_1(2\epsilon_+ + \alpha_i \pm \alpha_j)}\nonumber\\
    &\quad\quad\quad\quad \times\frac{\vartheta_1(\pm x+\epsilon_+\pm\epsilon_-+\alpha_i)}{\vartheta_1(\pm x+\epsilon_+\pm\epsilon_++\alpha_i)} + (\alpha_i \to -\alpha_i) \Bigg) \,.
\end{align}
The same result can also be obtained from the Higgsing procedure by tuning the parameter in the elliptic genera of the $\sorm(2N+5)$ theory as described in \eqref{Bn_cod4_Higgs}.

The codimension 2 defect can also be interpreted as a generating function of $\frac{1}{2}$ BPS Wilson loop operators. We first note that,  by using \eqref{eq:theta_identity_1}, the 2d perturbative part of the 6d/2d partition function can be recast as
\begin{align}\label{Bn_2d_pert2}
\begin{aligned}
    Z^{\rm{2d}}_{\mathrm{pert}}&=q_{\phi}^{-\frac{1}{2}}\sum_{n=0}^{N}\mathcal{W}^{(n)}_{0}\vartheta_1(x)\theta_2^{N-n}(2x;2\tau)\theta_3^{n}(2x;2\tau)\,,  \\
    \mathcal{W}^{(n)}_{0}&=\frac{(-1)^{N-n}}{n!(N-n)!\eta(\tau)^{2N}}\left(\prod_{i=1}^{n}\theta_2(2\alpha_i;2\tau)\prod_{i=n +1}^{N}\theta_3(2\alpha_i;2\tau)+{\mathrm{Permutations\,\, of\,\,}\alpha_i}\right)\,.
    \end{aligned}
\end{align}
This decomposition separates the factors $\mathcal{W}^{(n)}_0$ depending on $\alpha_i$ from the other terms depending on $x$.
As in \eqref{eq:cod4splitting}, we conjecture that this splitting holds true for higher strings, which we checked up to three-strings, as
\begin{align}
\label{eq:cod4splitting_BN}
    Z^{\rm{6d/2d}}=\sum_{n=0}^{N}\mathcal{W}^{(n)}\vartheta_1(x)\theta_2^{N-n}(2x;2\tau)\theta_3^{n}(2x;2\tau) \;,
    \qquad \text{where} \qquad 
    \mathcal{W}^{(n)}=q_{\phi}^{-\frac{1}{2}}\sum_{k=0}^{\infty}q_{\phi}^k\mathcal{W}_{k}^{(n)}
\end{align}
is a $x$-independent function. Here we see that \eqref{eq:cod4splitting_BN} is, when we ignore the overall $\vartheta_1$-function, an even section of a degree $2N$ line bundle $L$ over the elliptic curve. As we saw before, such a line bundle has $N+1$ even sections which correspond to $ \theta_2^{N-n}(2x;2\tau)\theta_3^{n}(2x;2\tau)$ for $n=0,\ldots, N$. The multiplication with $\vartheta_1$ maps these into the space of odd degree $2N+1$ sections.

It turns out that the coefficients $\mathcal{W}^{(n)}$'s are the $\frac{1}{2}$-BPS Wilson loop partition functions in different representations of the $\sorm(2N+1)$ gauge group. We can, in fact, compute these Wilson loop partition functions using the blowup method \cite{Kim:2021gyj}. Building on this, we find that the blowup equations for the Wilson loop $\mathcal{W}^{(n)}$ of the 6d $\sorm(2N+1)$ theory are
\begin{align}
    \theta_3^{[a]}\bigg(&-({2N-3})\epsilon_+-2\epsilon_2+\sum_{i=1}^{2N-7}m_i;4\tau\bigg)
    \cdot \mathcal{W}_{k}^{(n)}(\alpha,m,\epsilon_1,\epsilon_2)=\\
    &\sum_{\lambda_{G}\in Q^{\vee}(G)}^{\frac{1}{2}||\lambda_G||^2+k_1+k_2=k}
    (-1)^{|\lambda_G|} \,
    \theta_3^{[a]}\left(4\lambda_{G}\cdot\alpha+\sum_{i=1}^{2N-7}m_i+4k_1\epsilon_1+4k_2\epsilon_2-(2N-3-2||\lambda_{G}||^2)\epsilon_+-2\epsilon_2;4\tau\right)\notag\\
    &\cdot 
    A_0(\alpha,m,\lambda_{G},\lambda_F)
    \cdot 
    \mathcal{W}_{k_1}^{(n)}\left(\alpha+\epsilon_2\lambda_{G},m-\tfrac{1}{2}\epsilon_2,\epsilon_1-\epsilon_2,\epsilon_2\right)
    \cdot 
    Z^{\rm{6d}}_{k_2}\left(\alpha+\epsilon_1\lambda_{G},m-\tfrac{1}{2}\epsilon_1,\epsilon_1,\epsilon_2-\epsilon_1\right), \notag
\end{align}
where $Q^{\vee}(G)$ is the co-root lattice of the gauge group $G=B_N$, and
\begin{align}
    \theta_3^{[a]}(z;4\tau)=\sum_{n=-\infty}^{\infty}\re^{2(n-a)^2\tau+(n-a)z},\quad\quad a=0,\frac{1}{4},\frac{1}{2},\frac{3}{4}.
\end{align}
Here we choose $\lambda_F=-\frac{1}{2}(1,\ldots,1)$. The detailed definition of $A_0(\alpha,m,\lambda_{G},\lambda_F)$, see also \eqref{eq:D-fct_blowup},
for generic $\lambda_{G}$ is quite elaborated and we refer to \cite[eq.\ (3.7)--(3.8)]{Gu:2020fem} for details.
One can use these blowup equations to solve $\mathcal{W}_{k}^{(n)}$. For example, at 1-string level, we compute the solutions
\begin{align}
    \mathcal{W}^{(n)}_{1}=-\sum_{i<j,i,j=1}^{N}&\sum_{r=\pm 1,s=\pm 1}\frac{\vartheta_1(r\alpha_i)^{-1}\vartheta_1(s\alpha_j)^{-1}\prod_{l=1}^{2N-7}\vartheta_1(r\alpha_i-m_l+\epsilon_+)\vartheta_1(s\alpha_j-m_l+\epsilon_+)}{\vartheta_1(r\alpha_i+s \alpha_j)\vartheta_1(r\alpha_i+s \alpha_j+\epsilon_1)\vartheta_1(r\alpha_i+s \alpha_j+\epsilon_1)\vartheta_1(r\alpha_i+s \alpha_j+2\epsilon_+)}\nonumber\\
    &\cdot \prod_{k\neq i,j}\prod_{p=\pm 1}{\vartheta_1(r\alpha_i+p \alpha_k)^{-1}\vartheta_1(s\alpha_j+p \alpha_k)}^{-1}
    \cdot 
    \mathcal{W}_{0}^{(n)}\left(\alpha_l+(r\delta_{l,i}+s\delta_{l,j})\epsilon_2\right)\nonumber\\
    &\cdot\frac{ 
    D_{a_1,a_2,a_3}\left(4(r\alpha_i+s\alpha_j+\epsilon_1+\epsilon_2),4\epsilon_1,4\epsilon_2;-(2N-3)\epsilon_+-2\epsilon_2+\sum_{l=1}^{2N-7}m_l\right) }{ 
    D_{a_1,a_2,a_3}\left(0,4\epsilon_1,4\epsilon_2;-(2N-3)\epsilon_+-2\epsilon_2+\sum_{l=1}^{2N-7}m_l\right)},
\end{align}
where $D_{a_1,a_2,a_3}(z_1,z_2,z_3;z)$ is defined in \eqref{eq:Da1a2a3}, with $a_{1,2,3}$ three arbitrary different values chosen from $\{ 0,\frac{1}{4},\frac{1}{2},\frac{3}{4} \}$.
\subsection{Quantum curve and difference equation}\label{sec:BN_curve}
In this section, we derive the quantum curve for the $\sorm(2N+1)$ theory via Higgsing and verify the proposal with the explicit defect partition functions we computed. Upon the Higgs mechanism from the $\sorm(2N+2)$ partition functions to the $\sorm(2N+1)$ partition functions by tuning the parameters as (\ref{eq:Bn_higgs}),
the curve \eqref{eq:Dn_curve} for the $\sorm(2N+2)$ theory reduces to
\begin{align}\label{eq:Bn_curve}
  &\left[Y+{\vartheta_1(2x)\vartheta_1(\epsilon_1+2x)^2\vartheta_1(2\epsilon_1+2x)\prod_{l=1}^{2N-7}\vartheta_1(x+\tfrac{\epsilon_1}{2}\pm m_l)}Y^{-1}-\chi(x)\right]{\Psi}(x) = 0 \ ,
\end{align}
with ${\Psi}(x)$ and ${\chi}(x)$ being the normalized codimension 2 and 4 defect partition functions, respectively, in the NS-limit of the  $\sorm(2N+1)$ theory as defined in \eqref{eq:norm_def_pf}. We claim that this equation quantizes the Seiberg-Witten curve of the $\sorm(2N+1)$ gauge theory on a torus coupled to $2N-7$ fundamentals.
Here, according to the conjecture \eqref{eq:cod4splitting_BN}, the normalized codimension 4 defect partition function $\chi(x)$ can be expanded in terms with a sequence of $x$ independent functions $H_n(\alpha)$
\begin{align}
    \chi(x)=\sum_{n=0}^{N}H_n(\alpha)\vartheta_1(x)\theta_2^{N-n}(2x;2\tau)\theta_3^{n}(2x;2\tau),
\end{align}
where $H_n=\lim_{\epsilon_2\rightarrow 0}\mathcal{W}^{(n)}/Z^{\rm{6d}}_{\rm{str}}$. We have verified that the quantum curve equation \eqref{eq:Bn_curve} is satisfied up to three-string order using the explicit elliptic genera in the presence of the defects.

\paragraph{Path integral derivation.}
One can also derive the curve in the saddle point approach. The process is analogous to the discussion of the $\sorm(2N)$ case in Section \ref{sec:path_integral_representation}. The difference here is that one needs to replace the function $\mathcal{Q}(u)$ in \eqref{eq:mathcalQ} by
\begin{align}
\mathcal{Q}(u) &=\vartheta_1(\pm 2u)\vartheta_1(\pm 2u+2\epsilon_+)\frac{\prod_{l=1}^{2N-7}\vartheta_1(\pm u+m_l)}{\vartheta_1(\epsilon_+\pm u)\prod_{i=1}^N\vartheta_1(\epsilon_+\pm u\pm\alpha_i)}\,.
\end{align}
Then the saddle point equation \eqref{eq:saddle_point_equation} becomes
\begin{align}
\left[Y+q_\phi Q(u)\cdot Y^{-1}\right]\Phi\left(u+\tfrac{\epsilon_1}{2}\right)=0\,,
\label{eq:saddle_pt_BN}
\end{align}
with the suitable modification \eqref{eq:Q-fct} given by
\begin{align}
    Q(u)&=\left.\mathcal Q\left(u+\tfrac{\epsilon_1}{2}\right)\right|_{\epsilon_2\rightarrow 0}\nonumber\\
    &=\vartheta_1(2u)\vartheta_1( 2u+\epsilon_1)^2\vartheta_1( 2u+2\epsilon_1)\frac{\prod_{l=1}^{2N-7}\vartheta_1( u+\frac{\epsilon_1}{2}\pm m_l)}{\vartheta_1( u)\vartheta_1( u+\epsilon_1)\prod_{i=1}^N\vartheta_1( u\pm\alpha_i)\vartheta_1( u+\epsilon_1\pm\alpha_i)}.
\end{align}
By a similar argument as in Section \ref{sec:path_integral_representation}, one can show that $\Phi(u+\tfrac{1}{2}\epsilon_1)$ equals the normalization of the elliptic genera $\widetilde\Psi(x)$ in presence of the codimension 2 defect at the saddle point $x=u$. For generic defect parameter $x$, inspired from \cite{Chen:2020jla,Chen:2021ivd}, the right-hand side of \eqref{eq:saddle_pt_BN} is supposed to be the expectation value $\mathcal{W}^S(x)$ of Wilson surface defect, which leads to 
\begin{align}
    \left[Y+q_\phi Q(x)\cdot Y^{-1}\right]\Phi\left(x+\tfrac{\epsilon_1}{2}\right)=\mathcal{W}^S(x)\cdot\Phi\left(x+\tfrac{\epsilon_1}{2}\right)\,.
\label{eq:saddle_pt_BN2}
\end{align}
This is again the expected difference equation \eqref{eq:Bn_curve}, proposed above.
\section{\texorpdfstring{6d $\sorm(2N)$ theories with $\mathbb{Z}_2$ twist}{6d SO(2N) theories with Z2 twist}}
\label{sec:DN-twist}
In this section, we return to the discussion in Section \ref{sec:braneweb} and compute the partition function of the 6d $\sorm(2N)$ gauge theories on a $-4$ curve with $\mathbb{Z}_2$ twisted circle compactification, and the partition functions in the presence of the codimension 2 and 4 defects. 
We then present the difference equations satisfied by the defect partition functions. We show that also in this case the difference equation realizes a quantization of the elliptic Seiberg-Witten curve for the $\mathbb{Z}_2$ twisted $\sorm(2N)$ gauge theory.  
\subsection{Higgsing from \texorpdfstring{$\mathrm{SO}(2N+1)$}{SO(2N+1)} theory}\label{sec:DN-twist5.1}
Consider a circle compactification of the 6d $\sorm(2N)$ gauge group with $\mathbb{Z}_2$ outer-automorphism twist. Then, the affine gauge algebra $D_N^{(1)}$ in the 6d theory on a circle reduces to the twisted affine algebra $D_N^{(2)}$, which results from the $\mathbb{Z}_2$ twist of $D_N^{(1)}$. For example, the fundamental representation $\mathbf{F}$ of the 6d $\sorm(2N)$ gauge algebra is decomposed under the $\mathbb{Z}_2$ twist into the fundamental representation $\mathbf{F}_0$ of the invariant sub-algebra $\sorm(2N-1)$ with Kaluza-Klein (KK) charge 0 and the trivial representation $\mathbf{1}_{\frac{1}{2}}$ with KK charge $\frac{1}{2}$, where the subscripts stand for the KK charges. See \cite{Kac:1990gs} for more details. 

As discussed in Section \ref{sec:braneweb} (see also \cite{Kim:2019dqn,Kim:2021cua}), the $\mathbb{Z}_2$ twisted compactification of the 6d $\sorm(2N)$ gauge theory on a $-4$ curve can be realized by a particular Higgsing of the 6d $\sorm(2N+1)$ gauge theory including KK momentum states. In the partition function computation, the Higgsing can be performed by tuning the parameters of the UV $\sorm(2N+1)$ theory as
\begin{align}\label{eq:DnZ2_higgs}
\quad \alpha_N=\frac{\tau}{2},\quad m_{2N-7}=\epsilon_+-\frac{\tau}{2} \ ,
\end{align}
which is consistent with the Higgsing process \eqref{eq:Higgs_twisted_D} in the brane webs  with the identification $\tau \equiv \tfrac{1}{R}$. Note that the $\epsilon_+ =\log\sqrt{pq}$ factor appears naturally, similar to the analysis around \eqref{eq:Higgs_cond_D_to_B}.
This Higgsing procedure gives rise to the partition function of the $\sorm(2N)$ gauge theory with $\mathbb{Z}_2$ twist which takes the form of
\begin{align}
Z^{\rm{6d}}=Z_{\text{pert}}^{\rm{6d}}\cdot Z_{\text{str}}^{\rm{6d}} \, ,
\qquad \text{with} \qquad 
Z_{\text{str}}^{\rm{6d}}=\sum_{k=0}^{\infty}q_{\phi}^k \oint[\rd u]Z_{k,1-\text{loop}}^{\rm{6d}'}(u_I),
\end{align}
with
\begin{align}
    Z^{\rm 6d}_{\rm pert} &= {\rm exp}\left(\frac{1}{\epsilon_1\epsilon_2} F_{\rm eff}\right)\cdot {\rm PE}\bigg[ -\frac{1+pq}{(1-p)(1-q)(1-Q)}\bigg(\sum_{i<j}^{N-1}(A_iA_j^{\pm1}+A_i^{-1}A_j^{\pm1}Q)+\sum_{i=1}^{N-1}(A_i+A_i^{-1}Q) \nonumber \\
    & \qquad \qquad\qquad \qquad\qquad \qquad \qquad \qquad \qquad \qquad + \sum_{i=1}^{N-1}(A_i+A_i^{-1}+1)Q^{1/2}\bigg) \\
    &\qquad \qquad \qquad \qquad + \frac{\sqrt{pq}}{(1-p)(1-q)(1-Q)}\left(\sum_{i=1}^{N-1}(A_i+A_i^{-1}Q)+1+Q^{1/2}\right)\sum_{l=1}^{2N-8}(M_l+M_l^{-1})\bigg] \nonumber \\
    F_{\rm eff} &= \frac{1}{6}\left(\sum_{i<j}^{N-1}(\alpha_i\pm\alpha_j)^3+\sum_i^{N-1}\alpha_i^3\right)-\frac{1}{12}\sum_{i=1}^{N-1}\sum_{l=1}^{2N-8}(\alpha_i\pm m_l)^3 + \frac{1}{2}\phi_0\sum_{i=1}^{N-1}\alpha_i^2  +\ldots
\end{align}
where $[\rd u]=\frac{1}{2^k k!}\prod_{I=1}^k\frac{\rd u_I}{2\pi \ri}$, $q_{\phi}$ is the tensor fugacity and $Z_{k,{\rm 1-loop}}^{\rm 6d'}$ is the one-loop determinant in the $k$-th string elliptic genus given by
\begin{align}\label{eq:oneloopprime}
Z_{k,1-\text{loop}}^{\rm{6d}'}(u_I)=
\left(\frac{\vartheta _1\left(2 \epsilon _+\right)}{\vartheta _1\left(\epsilon_{1,2}\right)}\right)^k 
&\cdot \prod _{I=1}^k \vartheta
   _1\left(\pm 2   u_I\right) \vartheta _1\left(2 \epsilon _+\pm 2 u_I\right)
   \cdot \prod _{I<J}^k \frac{\vartheta _1\left(\pm
     u_I\pm   u_J\right) \vartheta _1\left(2 \epsilon _+\pm   u_I\pm
     u_J\right)}{\vartheta _1\left(\epsilon _1\pm   u_I\pm u_J\right) \vartheta _1\left(\epsilon _2\pm   u_I\pm u_J\right)}\nonumber \\
&\cdot \prod _{I=1}^k \frac{\prod _{l=1}^{2 N-8} \vartheta
   _1\left(\pm   u_I+m_l\right)}{\vartheta _1\left(\epsilon _+\pm
     u_I\right)\vartheta _1\left(\epsilon _+\pm
     u_I+\frac{\tau}{2}\right)\prod _{i=1}^{N-1} \vartheta _1\left(\epsilon _+\pm
     u_I\pm \alpha_i\right)} \,.
\end{align}
One observes that the expression \eqref{eq:oneloopprime} is not completely modular. Utilizing the identity
\begin{align}\label{eq:5.4}
    \theta_1\left( z+\frac{\tau}{2}\right)=\ri \ \re^{-\frac{1}{8} \tau-\frac{1}{2}z} \ \theta_4(z),
\end{align}
there is a factor $ \re^{\frac{1}{4}k\tau+k\epsilon_+}$ appearing at the $k$-string elliptic genus, which should be absorbed by the tensor fugacity $q_{\phi}$. This suggests that the tensor fugacity should be changed to 
\begin{align}\label{eq:qphiDNZ2}
\tilde{q}_{\phi}=\re^{\frac{1}{4}\tau+\epsilon_+}q_{\phi},
\end{align} 
when considering a $\mathbb{Z}_2$ twist for the $\sorm(2N)$ theories. In fact, if one does not absorb the $\re^{\epsilon_+}$, the refined BPS expansion of the free energy $F=\log Z$ is not consistent, which gives another reason for such a modification. In conclusion, the instanton string partition function of the $\sorm(2N)$ theory with $\mathbb{Z}_2$ twist is
\begin{align}
    Z_{\text{str}}^{\rm{6d}}&=\sum_{k=0}^{\infty}\tilde{q}_{\phi}^k \oint[\rd u]Z_{k,1-\text{loop}}^{\rm{6d}}(u_I),
\end{align}
with
\begin{align}
Z_{k,1-\text{loop}}^{\rm{6d}}(u_I)=
\left(-\frac{\vartheta _1\left(2 \epsilon _+\right)}{\vartheta _1\left(\epsilon_{1,2}\right)}\right)^k 
&\cdot \prod _{I=1}^k \vartheta_1\left(\pm 2   u_I\right) \vartheta _1\left(2 \epsilon _+\pm 2 u_I\right)
   \cdot \prod _{I<J}^k \frac{\vartheta _1\left(\pm
     u_I\pm   u_J\right) \vartheta _1\left(2 \epsilon _+\pm   u_I\pm
     u_J\right)}{\vartheta _1\left(\epsilon _1\pm   u_I\pm u_J\right) \vartheta _1\left(\epsilon _2\pm   u_I\pm u_J\right)}\nonumber \\
&\cdot \left(\prod _{I=1}^k \frac{\prod _{l=1}^{2 N-8} \vartheta_1\left(\pm   u_I+m_l\right)}{\vartheta _1\left(\epsilon _+\pm
     u_I\right)\vartheta _4\left(\epsilon _+\pm
     u_I\right)\prod _{i=1}^{N-1} \vartheta _1\left(\epsilon _+\pm
     u_I\pm \alpha_i\right)}\right)\, .
\end{align}
We can evaluate the contour integral for the elliptic genus by employing the JK-residue prescription.
At one-string level, the JK-poles giving non-trivial contributions are
\begin{align}\label{eq:DNZ2twist_Z1_poles}
\epsilon_+\pm\alpha_i+u_1=0,\quad\quad i=1,\ldots,N \ .
\end{align}
By taking the relevant JK-residues, we compute the one-string elliptic genus as
\begin{align}\label{eq:DNZ2twist_Z1}
Z_1^{\rm{6d}}
= -\frac{1}{2\vartheta_1(\epsilon_{1,2})} \sum_{i=1}^{N-1} \Bigg(&\frac{\vartheta_1(2\epsilon_+ + 2\alpha_i) \vartheta_1(4\epsilon_+ + 2\alpha_i) \prod_{l=1}^{2N-8}\vartheta_1(\epsilon_+ + \alpha_i \pm m_l)} {\vartheta_1(\alpha_i)\vartheta_1(\alpha_i+2\epsilon_+)\vartheta_4(\alpha_i)\vartheta_4(\alpha_i+2\epsilon_+) \prod_{j\neq i}^{N-1}  \vartheta_1(\alpha_i \pm \alpha_j) \vartheta_1(2\epsilon_+ + \alpha_i \pm \alpha_j)} \nonumber\\
&+ (\alpha_i \to -\alpha_i) \Bigg).
\end{align}
We also computed the two- and three-string elliptic genera, where the two-string genus is listed in Appendix \ref{app:genera_details1}. The expression of three-string elliptic genera are cumbersome; thus, we refrain from presenting them  here.
\subsection{Codimension 2 defect}
We now introduce the codimension 2 defect in the $\mathbb{Z}_2$ twisted compactification of the $\sorm(2N)$ theory. For this we use an RG-flow from the $\sorm(2N+2)$ theory with $\Z_2$ twist by Higgsing it with a position dependent VEV to a moment map operator. This RG-flow is realized in the partition function by setting the parameters of the UV theory as
\begin{align}
\begin{aligned}
    &m_{2N-7}+m_{2N-6}+2\epsilon_++\epsilon_2=0\,,\quad \alpha_{N+1}+m_{2N-6}+\epsilon_+=0\,,\\
    &\qquad \text{with} \qquad 
    x\equiv -\alpha_{N+1}+\epsilon_2 \,.
\end{aligned}
\label{eq:codim_two_defect_DnZ2}
\end{align}
After the Higgsing \eqref{eq:codim_two_defect_DnZ2}, one obtains a defect partition function of the $\sorm(2N)$ gauge theory with $\Z_2$ twist given by
\begin{align}
Z^{\rm{6d/4d}}&=Z_{\text{pert}}^{\rm{6d}}
\cdot Z_{\text{pert}}^{\rm{4d}}
\cdot Z_{\text{str}}^{\text{6d/4d}},\quad Z_{\text{str}}^{\text{6d/4d}}=\sum_{k=0}^{\infty}\tilde{q}_\phi^k \oint[\rd{u}]Z_{k,1-\text{loop}}^{\rm{6d}}( u_I)\cdot\prod_{I=1}^k\frac{\vartheta_1(-\epsilon_+-x\pm  u_I)}{\vartheta_1(-\epsilon_--x\pm  u_I)} \ , \nonumber \\
Z^{\rm 4d}_{\rm pert} &= Z^{\rm 4d}_{\rm eff}\cdot \Gamma_e(X^{-1})\Gamma_e(Q^{1/2}X^{-1})\prod_{i=1}^{N-1}\frac{\Gamma_e(X^{-1}A_i)}{\Gamma_e(pXA_i)} \ , \nonumber \\
Z^{\rm{4d}}_{\rm eff}(x)&=\exp\left[-\frac{(x+\epsilon_1/2)}{\epsilon_1}\left(\sum_{i=1}^N\alpha_i+\frac{\phi_0}{2}-(N-1)\frac{\tau}{6}\right) + \ldots\right]
\end{align}
and $x$ denotes the chemical potential for the $\urm(1)$ defect symmetry, as introduced in \eqref{eq:codim_two_defect_DnZ2}. $Z_{\text{pert}}^{\rm{4d}}$ is the perturbative contribution from the 4-dimensional defect, which can be understood as originating from a 4d free chiral multiplet coupled to the bulk $\sorm(2N-1)$ gauge group, with mass $x$.
Before the compactification, this 4d chiral multiplet transforms in the fundamental representation of the 6d $\sorm(2N)$ gauge group. Under the $\Z_2$ twist, the fundamental representation decomposes into a fundamental representation  of the $\sorm(2N-1)$ invariant sub-group carrying zero KK-charge and a singlet with KK-charge $\frac{1}{2}$. The 4d perturbative contribution reflects this decomposition. The extra terms in the integrand of the self-dual string contribution comes from the self-dual string modes coupled to the 4d chiral field.

Again, the elliptic genera of the self-dual string can be evaluated by the JK-prescription. We note that the 4d defect contribution provides the following additional poles: 
\begin{align}
  -\epsilon_- - x + u_I = 0 \ .
\end{align}
At one-string, we calculate the elliptic genus as
\begin{align}
    Z_1^{\rm{6d/4d}}=&-\frac{1}{2}\frac{\vartheta_1(2x+\epsilon_1)\vartheta_1(2x+2\epsilon_1)\vartheta_1(2x-2\epsilon_2)\vartheta_1(2x+2\epsilon_-)\vartheta_1(2\epsilon_+)\prod_l^{2N-8} \vartheta_1(x\pm m_l +\epsilon_-)}{\vartheta_1(\epsilon_1)\vartheta_1(x+\epsilon_1)\vartheta_1(x-\epsilon_2)\vartheta_4(x+\epsilon_1)\vartheta_4(x-\epsilon_2)\prod_{i=1}^{N-1}\vartheta_1(x\pm \alpha_i+\epsilon_1)\vartheta_1(x\pm \alpha_i-\epsilon_2)}\nonumber\\
    & -\frac{1}{2} \sum_{i=1}^{N-1} \Bigg(\frac{\vartheta_1(2\epsilon_+ + 2\alpha_i) \vartheta_1(4\epsilon_+ + 2\alpha_i) \prod_{l=1}^{2N-8}\vartheta_1(\epsilon_+ + \alpha_i \pm m_l)} {\vartheta_1(\epsilon_{1,2})\vartheta_1(\alpha_i)\vartheta_1(\alpha_i+2\epsilon_+) \vartheta_4(\alpha_i)\vartheta_4(\alpha_i+2\epsilon_+) \prod_{j\neq i}^{N-1} \vartheta_1(\alpha_i \pm \alpha_j) \vartheta_1(2\epsilon_+ + \alpha_i \pm \alpha_j)}\nonumber\\
    &\quad\quad\quad\quad\cdot \frac{\vartheta_1(x+\epsilon_+\pm(\alpha_i+\epsilon_+))}{\vartheta_1(x+\epsilon_-\pm(\alpha_i+\epsilon_+))} + (\alpha_i \to -\alpha_i) \Bigg) \,.
\end{align}
The same result can be obtained by tuning the parameters as \eqref{eq:codim_two_defect_DnZ2} in the elliptic genera of $\sorm(2N + 2)$ theory with $\mathbb{Z}_2$ twist.
One can compute higher string contributions in the similar manner.
\subsection{Codimension 4 defect}
Consider the codimension 4 defect of the type we discussed in previous sections. There are several ways to introduce such a defect into the $\sorm(2N)$ gauge theory with $\Z_2$ twist. One way is to Higgs the $\sorm(2N+1)$ gauge theory in the presence of the codimension 4 defect by following the prescription in Section \ref{sec:DN-twist5.1}. Using this, we compute the partition function of the codimension 4 defect and provide a natural interpretation of it.

After the Higgsing from the $\sorm(2N+1)$ theory with a codimension 2 defect, we obtain the defect partition function of the $\sorm(2N)$ theory with $\Z_2$ twist as follows:
\begin{align}
Z^{\rm{6d/2d}}&=Z^{\rm{6d}}_{\rm pert} 
\cdot Z^{\rm{2d}}_{\rm pert} 
\cdot Z^{\rm{6d/2d}}_{\rm str} \notag \\
&=Z^{\rm{6d}}_{\rm{pert}}
\cdot Z^{\rm{2d}}_{\rm{pert}}
\cdot \sum_{k=0}^{\infty}\tilde{q}_\phi^k \oint[\rd u]Z_{k,1-\text{loop}}^{\rm{6d}}( u_I)
\cdot \prod_{I=1}^k\frac{\vartheta_1(-\epsilon_-\pm x\pm u_I)}{\vartheta_1(-\epsilon_+\pm x\pm u_I)} \ , \nonumber \\
Z^{\rm{2d}}_{\rm{pert}}&=\tilde{q}_{\phi}^{-\frac{1}{2}}\vartheta_1(x)\vartheta_4(x)\prod_{i=1}^{N-1}\vartheta_1(x\pm \alpha_i) \ ,
\end{align}
where $x$ is the chemical potential for the $U(1)$ defect symmetry.
Here, $Z^{\rm{2d}}_{\rm{pert}}$ is the perturbative contribution from the codimension 4 defect. This implies that the codimension 4 defect comes from coupling a 2d free fermion transforming in the fundamental representation of the 6d $\sorm(2N)$ gauge group with $\Z_2$ twist: the factors of $\vartheta_1$ are KK-charge 0 modes and the factor of $\vartheta_4$ is the KK-charge $\frac{1}{2}$ mode, respectively, in the decomposition of the $\sorm(2N)$ fundamental representation under $\Z_2$ twist.

The self-dual string part $Z_{\rm str}^{\rm 6d/2d}$ receives extra contributions in the integrand from the defect. These defect contributions provide additional JK-poles
\begin{align}
  -\epsilon_+ \pm x + u_I = 0 \ ,
\end{align}
which we take into account when we evaluate the contour integral using the JK-prescription.
We compute the elliptic genus for a single string and obtain
\begin{align}
    Z_1^{\rm{6d/2d}}=&\frac{1}{2}\Bigg(\frac{\vartheta_1(2x-\epsilon_{1})\vartheta_1(2x-\epsilon_{2})\vartheta_1(2x-3\epsilon_+\pm \epsilon_+)\vartheta_1(2\epsilon_+)\prod_l^{2N-8} \vartheta_1(-x\pm m_l +\epsilon_+)}{\vartheta_1(x)\vartheta_1(x-2\epsilon_+)\vartheta_4(x)\vartheta_4(x-2\epsilon_+)\prod_{i=1}^{N-1}\vartheta_1(x\pm \alpha_i)\vartheta_1(x\pm \alpha_i-2\epsilon_+)}+(x\rightarrow -x)\Bigg)\nonumber\\
    & -\frac{1}{2} \sum_{i=1}^{N-1} \Bigg(\frac{\vartheta_1(2\epsilon_+ + 2\alpha_i) \vartheta_1(4\epsilon_+ + 2\alpha_i) \prod_{l=1}^{2N-8}\vartheta_1(\epsilon_+ + \alpha_i \pm m_l)} {\vartheta_1(\epsilon_{1,2})\vartheta_1(\alpha_i)\vartheta_1(\alpha_i+2\epsilon_+)\vartheta_4(\alpha_i)\vartheta_4(\alpha_i+2\epsilon_+) \prod_{j\neq i}^{N-1}  \vartheta_1(\alpha_i \pm \alpha_j) \vartheta_1(2\epsilon_+ + \alpha_i \pm \alpha_j)}\nonumber\\
    &\quad\quad\quad\quad \cdot \frac{\vartheta_1(\pm x+\epsilon_+\pm\epsilon_-+\alpha_i)}{\vartheta_1(\pm x+\epsilon_+\pm\epsilon_++\alpha_i)} + (\alpha_i \to -\alpha_i) \Bigg) \,.
\end{align}
Again, the elliptic genera for higher strings can be calculated similarly using the JK-prescription.

The codimension 4 defect partition function can also be interpreted as a generating function of the $\frac{1}{2}$ BPS Wilson loop partition functions.
First, note that we can recast the 2d perturbative part of the 6d/2d partition function, by using the identity in \eqref{eq:theta_identity_1}, as 
\begin{align}\label{DnZ2_2d_pert2}
    Z^{\rm{2d}}_{\mathrm{pert}}&=\tilde{q}_{\phi}^{-\frac{1}{2}}\sum_{n=0}^{N-1}\mathcal{W}_{0}^{(n)}\vartheta_1(x)\vartheta_4(x)\theta_2^{N-n-1}(2x;2\tau)\theta_3^{n}(2x;2\tau)\,, \nonumber \\
    \mathcal{W}_{0}^{(n)}&=\frac{(-1)^{N-n-1}}{n!(N-n-1)!\eta(\tau)^{2N-2}}\left(\prod_{j=1}^{n}\theta_2(2\alpha_j;2\tau)\prod_{j=n +1}^{N-1}\theta_3(2\alpha_j;2\tau)+{\mathrm{Permutations\,\, of\,\,}\alpha_j}\right)\,,
\end{align}
which basically separates the functions of $\alpha_i$ from the other functions depending only on $x$.
It turns out that this separation still holds for higher string orders, i.e.
\begin{align}\label{eq:cod4splitting_DNZ2}
    Z^{\rm{6d/2d}}(\alpha,x)&=\sum_{n=0}^{N-1}\mathcal{W}^{(n)}(\alpha)\ \vartheta_1(x)\ \vartheta_4(x)\ \theta_2^{N-n-1}(2x;2\tau)\ \theta_3^{n}(2x;2\tau) \,, \nonumber \\
    \mathcal{W}^{(n)}(\alpha)&=\tilde{q}_{\phi}^{-\frac{1}{2}}\sum_{k=0}^{\infty}\tilde{q}_{\phi}^k\mathcal{W}_{k}^{(n)}(\alpha)\ ,
\end{align}
which we checked up to three-string order. We conjecture that this relation holds for all self-dual string sectors and the function $\mathcal{W}^{(n)}$ captures the $\frac{1}{2}$ BPS Wilson loop partition functions in the 6d $\sorm(2N)$ gauge theory on a circle with $\Z_2$ twist.
We notice that the 6d/2d partition function is mathematically a linear combination of 
\begin{equation}
    h^0(L)_{+} = \frac{1}{2}h^0(L) + 1 = \frac{1}{2}(2N-2) +1 = N
\end{equation}
even sections of a degree $2N-2$ line bundle $L$ over the elliptic curve. The multiplication with $\vartheta_1(x) \vartheta_4(x)$ maps these into the space of odd sections of degree $2N$.

The Wilson loop partition functions can also be computed via the blowup method. In the case at hand, the blowup equations for the Wilson loops $\mathcal{W}^{(n)}$ are
\begin{align}
    \theta_3^{[a]}\bigg(&-({2N-4})\epsilon_+-2\epsilon_2+\sum_{i=1}^{2N-8}m_i;4\tau\bigg) \
    \mathcal{W}_{k}^{(n)}(\alpha,m,\epsilon_1,\epsilon_2)=\\
    &\sum_{\lambda_{G}\in Q^{\vee}(B_{N-1})}^{\frac{1}{2}||\lambda_G||^2+k_1+k_2=k}
    (-1)^{|\lambda_G|} \ \theta_3^{[a]}\left(4\lambda_{G}\cdot\alpha+\sum_{i=1}^{2N-8}m_i+4k_1\epsilon_1+4k_2\epsilon_2-(2N-3-2||\lambda_{G}||^2)\epsilon_+-2\epsilon_2;4\tau \right)\notag\\
    &\cdot A_0(\alpha,m,\lambda_{G},\lambda_F)
    \cdot \mathcal{W}_{k_1}^{(n)}\left(\alpha+\epsilon_2\lambda_{G},m-\tfrac{1}{2}\epsilon_2,\epsilon_1-\epsilon_2,\epsilon_2 \right)
    \cdot Z^{\rm{6d}}_{k_2} \left(\alpha+\epsilon_1\lambda_{G},m-\tfrac{1}{2}\epsilon_1,\epsilon_1,\epsilon_2-\epsilon_1\right), \notag
\end{align}
where $\lambda_F=-\frac{1}{2}(1,\ldots,1)$ and $Q^{\vee}(B_{N-1})$ is the co-root lattice of the group $B_{N-1}$, and
\begin{align}
    \theta_3^{[a]}(z;4\tau)=\sum_{n=-\infty}^{\infty}\re^{2(n-a)^2\tau+(n-a)z},\quad\quad a=\frac{1}{8},\frac{3}{8},\frac{5}{8},\frac{7}{8}.
\end{align}
In contrast to the untwisted cases in Sections \ref{sec:codim4_SO2N} and \ref{sec:codim4_SO2N-1}, the magnetic flux $2a$ is half-integral.
Again, the reader is referred to \cite[eq.\ (3.7)--(3.8)]{Gu:2020fem} for the definition of $A_0(\alpha,m,\lambda_{G},\lambda_F)$ with generic $\lambda_{G}$.

One can solve the blowup equations to compute $\mathcal{W}_{k}^{(n)}$.
At one-string, we find the solutions
\begin{align}
    \mathcal{W}_{1}^{(n)}=\sum_{i<j,i,j=1}^{N-1}&\sum_{r=\pm 1,s=\pm 1}\frac{\prod_{l=1}^{2N-8}\vartheta_1(r\alpha_i-m_l+\epsilon_+)\vartheta_1(s\alpha_j-m_l+\epsilon_+)}{\vartheta_1(r\alpha_i+s \alpha_j)\vartheta_1(r\alpha_i+s \alpha_j+\epsilon_1)\vartheta_1(r\alpha_i+s \alpha_j+\epsilon_1)\vartheta_1(r\alpha_i+s \alpha_j+2\epsilon_+)}\nonumber\\
    &\cdot \vartheta_1(r\alpha_i)^{-1}\vartheta_1(s\alpha_j)^{-1}\vartheta_4(r\alpha_i)^{-1}\vartheta_4(s\alpha_j)^{-1}\nonumber\\
    &\cdot\prod_{k\neq i,j}\prod_{p=\pm 1}{\vartheta_1(r\alpha_i+p \alpha_k)^{-1}\vartheta_1(s\alpha_j+p \alpha_k)}^{-1}
    \cdot \mathcal{W}_{n}^{(0)}\left(\alpha_l+(r\delta_{l,i}+s\delta_{l,j})\epsilon_2\right)\nonumber\\
    &\cdot\frac{D_{a_1,a_2,a_3}(4(r\alpha_i+s\alpha_j+\epsilon_1+\epsilon_2),4\epsilon_1,4\epsilon_2;-(2N-4)\epsilon_+-2\epsilon_2+\sum_{l=1}^{2N-8}m_l)}{D_{a_1,a_2,a_3}(0,4\epsilon_1,4\epsilon_2;-(2N-4)\epsilon_+-2\epsilon_2+\sum_{l=1}^{2N-8}m_l)},
\end{align}
where $D_{a_1,a_2,a_3}(z_1,z_2,z_3;z)$ is defined in \eqref{eq:Da1a2a3}, with $a_{1,2,3}$ three arbitrary different values chosen from $\{ \frac{1}{8},\frac{3}{8},\frac{5}{8},\frac{7}{8} \}$. We checked that these solutions agree with the results from the elliptic genus at $k=1$, which we computed using the contour integral expressions above.
\subsection{Quantum curve and difference equation}
In the previous subsections, we have mainly used RG-flows of the 6d $\sorm(2N+1)$ and $\sorm(2N+2)$ gauge theories and their partition functions in order to compute the partition functions of the $\Z_2$ twisted compactifications of the $\sorm(2N)$ gauge theory with or without defects. The same idea can be used to derive the quantum Seiberg-Witten curve of the $\sorm(2N)$ theory with $\Z_2$ twist. We can simply take the quantum curve for the $\sorm(2N+1)$ gauge theory given in \eqref{eq:Bn_curve} and tune the parameters as \eqref{eq:DnZ2_higgs}. This then Higgses the $\sorm(2N+1)$ curve and gives rise to the curve for the $\sorm(2N)$ theory with $\Z_2$ twist given by
\begin{align}
\label{eq:DnZ2_curve}
  \left[Y+{\vartheta_1(2x)\vartheta_1(\epsilon_1+2x)^2\vartheta_1(2\epsilon_1+2x)\prod_{l=1}^{2N-8}\vartheta_1(x+\tfrac{\epsilon_1}{2}\pm m_l)}Y^{-1}-\chi(x)\right]{\Psi}(x) = 0 \ , \end{align}
where ${\Psi}(x)$ and ${\chi}(x)$ are the codimension 2 and 4 defect partition functions in the NS-limit normalized by the bare partition function, analogous to \eqref{eq:norm_def_pf}.
Using the decomposition in \eqref{eq:cod4splitting_DNZ2}, the normalized partition function $\chi(x)$ can be expanded in terms of the BPS Wilson loop expectation values as
\begin{align}
    \chi(x)=\sum_{n=0}^{N-1}H_n\vartheta_1(x)\vartheta_4(x)\theta_2^{N-n-1}(2x;2\tau)\theta_3^{n}(2x;2\tau),
\end{align}
where $H_n=\lim_{\epsilon_2\rightarrow 0} \mathcal{W}^{(n)}/Z^{\rm{6d}}$ is the Wilson loop expectation value in the NS-limit which is independent of the defect mass parameter $x$. We have verified that the quantum curve equation \eqref{eq:DnZ2_curve} holds up to three-string order by using the explicit defect partition functions.
\section{Conclusion}
\label{sec:conc}
In this paper we studied the codimension 2 and 4 defects in the 6d $\Ncal=(1,0)$ SCFTs with $\sorm(N_c)$ gauge groups defined on a $-4$ curve and their circle compactifications with $\Z_2$ twist. On the tensor branch, these defects can be introduced by coupling the bulk gauge fields to a 4d $\Ncal=1$ fundamental chiral multiplet and a 2d fundamental fermion, respectively. It turns out that the interplay between the 6d bulk theory and the defect operators provides a quantization of the Seiberg-Witten curve for the 6d theory compactified on a circle with/without twist. 
More specifically, we have shown that the BPS partition function of the 6d theory on a $\Omega$-deformed $\R^4$ times a torus in the presence of the codimension 2 defect wrapping the torus becomes a formal wave function for a certain difference operator in the Nekrasov-Shatashvili limit $\epsilon_2\rightarrow 0$. Crucially, once this difference operator acts on the wave function it introduces the codimension 4 defect. 
The difference equation is in fact a quantum SW-curve in the 6d theory that reduces to the classical SW-curve given in \cite{Haghighat:2018dwe} in the flat $\R^4$ limit $\epsilon_1,\epsilon_2\rightarrow0$.

An immediate generalization of this work would be to study the quantizations of SW-curves in the 6d $\Ncal=(1,0)$ SCFTs realized on a linear tensor chain of $-4$ and $-1$ curves. The BPS defects defined in this paper and those studied in \cite{Chen:2021ivd} for the E-string theory on $-1$ curve will also play a significant role in the study of the linear tensor chain. A preliminary investigation suggests that the classical SW-curve for a linear tensor chain in \cite{Haghighat:2018dwe} can be promoted to a difference equation acting on the partition functions of these defects and it provides a natural quantization for the classical SW-curve. This will be reported in upcoming papers \cite{Chen:preperation}. A generalization along these lines to the little string theories will also be interesting.

The Seiberg-Witten curves of supersymmetric 5d gauge theories are known to correspond to the spectral curves of associated relativistic integrable systems \cite{Nekrasov:1996cz}. 
 For example, the SW-curve of the 5d $\Ncal=2$ $\surm(N)$ gauge theory arising from the 6d $\Ncal=(2,0)$ $A_{N-1}$ theory on a circle is the spectral curve of the  $N$-body elliptic Ruijsenaars-Schneider (RS) system \cite{Nekrasov:1996cz}. In this case the Nekrasov partition function in the presence of the Gukov-Witten monodromy defect in the NS limit becomes a formal eigenfunction of the elliptic RS system and the eigenvalues are identified with the expectation values of Wilson loops \cite{Bullimore:2014awa}, which also provides a quantization of the SW-curve. It is then natural to wonder if we can establish this correspondence with integrable systems for any 6d SCFTs compactified on a circle. When this is the case, it will be tempting to interpret the partition function of the 6d SCFT coupled to some codimension 2 defects as an eigenfunction of the Hamitonians in the related elliptic integrable system. There are some tantalizing hints for such correspondence for particular 5d KK theories admitting dual 5d gauge theory descriptions: novel mutually commuting difference operators acting on codimension 2 defects in the 6d $\surm(2)$ gauge theory with four flavors \cite{Gaiotto:2015usa} and in the twisted compactifications of the 6d minimal $\surm(3)$ and $\sorm(8)$ gauge theories \cite{Razamat:2018zel} have been constructed, and they may be translated into Hamiltonians of some many-body integrable systems. It would be interesting to investigate  these difference operators as well as the related defect operators to identify corresponding integrable systems.
 
 Yet another direction to proceed is \cite{Jeong:2021bbh} where the authors analyze gauge theory defect partition functions from the point of view of Knizhnik-Zamolodchikov (KZ) equations. That such a viewpoint is justified can be deduced from the fact that the theories studied in \cite{Jeong:2021bbh} admit 4d/2d correspondences where the dual 2d picture is that of a four-punctured sphere on which the 6d (2,0) theory is compactified. Gauge theory defect operators can then be thought of as point-like insertions of 2d CFT operators. The corresponding conformal blocks which are at the same time defect partition functions then satisfy KZ-equations. In our case, we would expect an elliptic generalization of this picture leading to elliptic KZ-equations (see \cite{ellKZ} for a review).

\paragraph{Acknowledgments.}
It is our pleasure to acknowledge useful discussions with Sung-Soo Kim, Kaiwen Sun, and Futoshi Yagi.
The work of J.C., B.H., and M.S. is supported by the National Thousand-Young-Talents Program of China. H.K. is supported  by Samsung Science and Technology Foundation under Project Number SSTF-BA2002-05 and by the National Research Foundation of Korea (NRF) Grant 2018R1D1A1B07042934. K.L. is supported  by KIAS Individual Grant PG006904 and by the National Research Foundation of Korea Grant NRF-2017R1D1A1B06034369. X.W. is supported by KIAS Individual Grant QP079201.   
M.S. is further supported by the National Natural Science Foundation of China (grant no. 11950410497), and the China Postdoctoral Science Foundation (grant no. 2019M650616).

%
%
\appendix
\section{Elliptic functions}
\label{app:functions}
In this paper, we use the following standard theta functions
\begin{align}
\theta_1(z;\tau)&=-i Q^{\frac{1}{8}}X^{\frac{1}{2}}\prod_{n=1}^{\infty}\left(1-Q^n\right)\left(1-XQ^n\right)\left(1-X^{-1}Q^{n-1}\right)\,,\\
\theta_2(z;\tau)&=Q^{\frac{1}{8}}X^{\frac{1}{2}}\prod_{n=1}^{\infty}\left(1-Q^n\right)\left(1+XQ^n\right)\left(1+X^{-1}Q^{n-1}\right) \, ,\\
\theta_3(z;\tau)&=\prod_{n=1}^{\infty}\left(1-Q^{n}\right)\left(1+XQ^{n+\frac{1}{2}}\right)\left(1+X^{-1}Q^{{n+\frac{1}{2}}}\right)\, ,\\
\theta_4(z;\tau)&=\prod_{n=1}^{\infty}\left(1-Q^{n}\right)\left(1-XQ^{n+\frac{1}{2}}\right)\left(1-X^{-1}Q^{{n+\frac{1}{2}}}\right) \,,    
\end{align}
with $Q=\re^{\tau},\, X=\re^z$. We also use the notation for a  variant of the theta functions
\begin{align}
    \vartheta_i(z)=\frac{\ri \,\theta_i(z;\tau)}{\eta(\tau)},
\end{align}
where $\eta(\tau)$ is the Dedekind eta function, which is defined as 
\begin{align}
    {\eta(\tau)}=Q^{\frac{1}{24}}\prod_{n=1}^{\infty}\left(1-Q^n\right).
\end{align}
\section{Field theory Higgsing}
\label{app:Higgs}
In this appendix, the Higgsing of $\sorm(2N)$ is detailed from the field theory perspective. Consider $\sorm(2N)$ with $N_f=2N-8$ fundamental hypermultiplets. It is a well-know fact that the $\urm(N_f)$ flavor symmetry is enhanced to $\sprm(N_f)\equiv \mathrm{USp}(2N_f)$, due to existence of the $\sorm$-invariant tensor $\delta_{ab}$, with $a,b\in \{1,2,\ldots, 2N\}$. The hypermultiplets can be rearranged into half-hypermultiplets as follows
\begin{subequations}
\begin{alignat}{2}
 &\text{hyper}\quad  &   H_a^f &=\left( Y_{a}^f, (\widetilde{Y}_f^a )^\dagger \right) \qquad  f \in \{ 1,2,\ldots, N_f\} \\
 &\text{half-hyper}\quad &  X_a^I &=
 \begin{cases} Y_a^f &, I =f\\
 \delta_{ab} \widetilde{Y}_f^b  &, I=f+N_f
\end{cases}
\end{alignat}
\end{subequations}
such that $\{X_a^I \}_{a=1,\ldots,2N}^{I=1,\ldots,2N_f}$ transforms properly in the fundamental representation $[1,0,\ldots,0]_D$ of $\sorm(2N)$ and in the fundamental representation $[1,0,\ldots,0]_C$ of $\sprm(N_f)$. The basic gauge invariant operator is the following meson
\begin{align}
 \mathcal{M}^{(IJ)} = X_a^I \delta^{ab} X_{b}^J
  = \begin{pmatrix}
     Y_a^f \delta^{ab} Y_b^l & Y_a^f \widetilde{Y}_l^a \\
      \widetilde{Y}_f^a Y_a^l  &  \widetilde{Y}_f^a \delta_{ab} \widetilde{Y}_l^b
    \end{pmatrix}
\end{align}
which is symmetric in the $I,J$ indices. Hence, $\mathcal{M}$ transforms in the adjoint representation $[2,0,\ldots,0]_C$ of $\sprm(N_f)$, and is known as moment map of the global symmetry. 
The different operators have the following fugacities associated to them:
\begin{align}
 Y_a^f \delta^{ab} Y_b^l \sim pq e^{-m_f - m_l }\,, \quad
 Y_a^f \widetilde{Y}_l^a \sim pq e^{-m_f+ m_l }\,, \quad 
 \widetilde{Y}_f^a Y_a^l \sim pq e^{m_f - m_l }\,, \quad
 \widetilde{Y}_f^a \delta_{ab} \widetilde{Y}_l^b \sim pq e^{m_f+ m_l }\,,
\end{align}
where the $p q$ term are due to the $(\tfrac{1}{2},\tfrac{1}{2})$ charges of each hypermultiplet under the $\sorm(4)\cong \surm(2) \times \surm(2)$ rotation symmetry of $\R^4$.

Next, assign non-trivial VEVs to the meson operator. There are two cases to consider.

\paragraph{Higgsing to SO(odd).}
To begin with, assign a VEV to a diagonal matrix element $\langle\mathcal{M}^{JJ} \rangle \neq 0$. Without loss of generality, one may choose $J=2N_f$ such that
\begin{align}
 \langle\mathcal{M}^{2N_f,2N_f} \rangle =  \widetilde{Y}_{N_f}^a \delta_{ab} \widetilde{Y}_{N_f}^b  =  \widetilde{Y}_{N_f}^N  \widetilde{Y}_{N_f}^N 
\end{align}
where the last equality holds after a suitable gauge transformation. Such a VEV breaks the $\sprm(N_f)$ flavor symmetry to $\sprm(N_f-1)$, because only one hypermultiplet needs to acquire a VEV. As a consequence, the theory after Higgsing is $\sorm(2N-1)$ with $N_f-1$ flavors.

In terms of fugacities, the Higgsing is realized by
\begin{align}
  \langle\mathcal{M}^{2N_f,2N_f} \rangle \neq 0:  \qquad 
  pq \cdot e^{2 m_{N_f}} = 1 \quad \Rightarrow \quad \sqrt{pq} \cdot e^{ m_{N_f}} = 1 
  \label{eq:Higgs_cond_D_to_B}
\end{align}
and the rank reduction of the gauge symmetry can be realized by $\alpha_N=0$.
\paragraph{Higgsing to SO(even).}
Next, assign a VEV to an off-diagonal matrix element $\langle\mathcal{M}^{IJ} \rangle \neq 0$. Without loss of generality, one may choose $I=2N_f-1$ and $J=2N_f$ such that
\begin{align}
 \langle\mathcal{M}^{2N_f-1,2N_f} \rangle =  \widetilde{Y}_{N_f-1}^a \delta_{ab} \widetilde{Y}_{N_f}^b  =  \widetilde{Y}_{N_f-1}^N  \widetilde{Y}_{N_f}^N 
\end{align}
where the last equality holds after a suitable gauge transformation. Such a VEV breaks the $\sprm(N_f)$ flavor symmetry to $\sprm(N_f-2)$, because two hypermultiplets need to acquire a VEV. As a consequence, the theory after Higgsing is $\sorm(2N-2)$ with $N_f-2$ flavors.

In terms of fugacities, the Higgsing is realized by
\begin{align}
  \langle\mathcal{M}^{2N_f-1,2N_f} \rangle \neq 0:  \qquad 
  pq \cdot e^{m_{N_f-1} +m_{N_f}} = 1  \,.
  \label{eq:Higgs_cond_D_to_D}
\end{align}
Equivalently, one could also consider $ \langle\mathcal{M}^{2N_f-1,2N_f} \rangle $ which implies $  pq \cdot e^{-m_{N_f-1} -m_{N_f}} = 1$. Both version are useful for the main text.

\section{Finite sector under NS-limit}
\label{app:NS_finitie_sector}
For simplicity, we consider the $\sorm(8)$ theory. Since we focus on the NS-finite sector, the poles are taken only from the contributions of the defect and anti-symmetric fields in the 2d ADHM construction.
\subsection{Codimension 2 defect partition function \texorpdfstring{$\widetilde{\Psi}_{\mathrm{finite}}$}{Z} }
\paragraph{1-instanton.} The 1-instanton partition function with codimension 2 defect is given by
\begin{align}
Z_1=\int\!\!\frac{\rd u}{2\pi \ri}\, \frac{\vartheta_1(2\epsilon_+)}{\vartheta_1(\epsilon_{1,2})}\frac{\vartheta_1(\pm2{u})\vartheta_1(\pm2{u}+2\epsilon_+)}{\prod_{i=1}^4\vartheta_1(\epsilon_+\pm {u} \pm \alpha_i)}
\cdot\frac{\vartheta_1(\pm{u}+x+\epsilon_+)}{\vartheta_1(\pm{u}+x+\epsilon_-)}\,.
\end{align}
It is straightforward to verify that the contribution from poles of $\theta(\epsilon_+\pm{u}\pm\alpha_i)$ is divergent in the $\epsilon_2\rightarrow 0$ limit. Therefore, we only pick the pole of
\begin{align}
\vartheta_1({u}+x+\epsilon_-)=0\,,
\end{align}
from the defect contribution. This yields the residue
\begin{align}
\widetilde{\Psi}_1(x)=\frac{\vartheta_1(2x)\vartheta_1(2x+\epsilon_1)^2\vartheta_1(2x+2\epsilon_1)}{\prod_{i=1}^{4}\vartheta_1(x\pm\alpha_i)\vartheta_1(x\pm\alpha_i+\epsilon_1)}\equiv Q(x)\,,
\end{align}
in the NS-limit.

\paragraph{2-instanton.} The 2-instanton partition function with codimension 2 defect is given by
\begin{align}
Z_2=\int \!\!\frac{\rd{u}_1}{2\pi \ri}\frac{\rd{u}_2}{2\pi \ri} \left(\frac{\vartheta_1(2\epsilon_+)}{\vartheta_1(\epsilon_{1,2})}\right)^2
&\frac{\vartheta_1(\pm2{u}_{1,2})\vartheta_1(\pm2{u}_{1,2}+2\epsilon_+)\vartheta_1(\pm{u}_1\pm{u}_2)\vartheta_1(\pm{u}_1\pm{u}_2+2\epsilon_+)}{\vartheta_1(\pm{u}_1\pm{u}_2+\epsilon_{1,2})\prod_{i=1}^4\vartheta_1(\epsilon_+\pm {u}_{1,2} \pm \alpha_i)}\notag\\
&\cdot\frac{\vartheta_1(\pm{u}_{1,2}+x+\epsilon_+)}{\vartheta_1(\pm{u}_{1,2}+x+\epsilon_-)} \,.
\end{align}
For the JK-residue prescription, we have chosen the auxiliary vector $\nu$ to be inside the cone spanned by $\{u_1,u_1+u_2\}$. This choice defines which poles are to be taken into account.
The NS-finite sector corresponds to selecting poles from:
\begin{align}
\label{eq:cones_2string}
\begin{aligned}
\mathrm{a}: &\quad \vartheta_1(+{u}_1+x+\epsilon_-)=0\,, \quad \vartheta_1(-{u}_1+{u}_2+\epsilon_{1,2})=0\,,\\
\mathrm{b}: &\quad \vartheta_1(+{u}_2+x+\epsilon_-)=0\,, \quad \vartheta_1(+{u}_1-{u}_2+\epsilon_{1,2})=0\,,\\
\mathrm{c}: &\quad \vartheta_1(-{u}_2+x+\epsilon_-)=0\,, \quad \vartheta_1(+{u}_1+{u}_2+\epsilon_{1,2})=0\,,\\
\mathrm{d}: &\quad \vartheta_1(+{u}_1+x+\epsilon_-)=0\,, \quad \vartheta_1(+{u}_1+{u}_2+\epsilon_{1,2})=0\,.
\end{aligned}
\end{align}
The first three sectors correspond to the cones spanned by $\{{u}_1,\, -{u}_1+{u}_2\}$, $\{{u}_2,\, {u}_1-{u}_2\}$ and $\{-{u}_2,\, {u}_1+{u}_2\}$, which are related to each other by the Weyl group of $\sprm(2)$, and thus have same contributions. We find
\begin{align}
\widetilde{\Psi}_2|_{\rm a}=\widetilde{\Psi}_2|_{\rm b}=\widetilde{\Psi}_2|_{\rm c}=Q(x)Q(x+\epsilon_1)\,,
\end{align}
with $Q$ as defined in \eqref{eq:Q-fct}.
Meanwhile the sector ``d", corresponding to the cone spanned by $\{{u}_1,\, {u}_1+{u}_2\}$, gives
\begin{align}
\widetilde{\Psi}_2|_{\rm d}=-Q(x)^2+Q(x)Q(x-\epsilon_1)\,.
\end{align}
Overall, the residue contributions are
\begin{align}
\widetilde{\Psi}_{2}=\widetilde{\Psi}_2 |_{\mathrm{d}}+3\widetilde{\Psi}_2 |_{\mathrm{a}}+=-Q(x)^2+Q(x)Q(x-\epsilon_1)+3Q(x)Q(x+\epsilon_1)\,.
\label{eq:SO_codim_two_finite_sector_2}
\end{align}
\paragraph{3-instanton.} 
The 3-instanton partition function with codimension 2 defect is given by
\begin{align}
Z_3=\int \!\!\prod_{I=1}^3\frac{\rd{u_I}}{2\pi \ri} \left(\frac{\vartheta_1(2\epsilon_+)}{\vartheta_1(\epsilon_{1,2})}\right)^3
&\prod_{I=1}^3\frac{\vartheta_1(\pm2{u}_{I})\vartheta_1(\pm2{u}_{I}+2\epsilon_+)}{\prod_{i=1}^4\vartheta_1(\epsilon_+\pm {u}_{I} \pm \alpha_i)}
\cdot\prod_{I<J}^3\frac{\vartheta_1(\pm{u_I}\pm{u_J})\vartheta_1(\pm{u_I}\pm{u_J}+2\epsilon_+)}
{\vartheta_1(\pm{u_I}\pm{u_J}+\epsilon_{1,2})}\notag\\
&\cdot\prod_{I=1}^3\frac{\vartheta_1(\pm{u}_{I}+x+\epsilon_+)}{\vartheta_1(\pm{u}_{I}+x+\epsilon_-)}
\end{align}
We chose the auxiliary vector $\nu={u}_1+\frac{1}{2}{u}_2+\frac{1}{3}{u}_3$ in the plane spanned by $\{{u}_1,\,{u}_2,\,{u}_3\}$. The cones that contain $\nu$ are as follows:
\begin{align}
\mathrm{A}: \quad &({u}_1,\,+{u}_2+{u}_3,\,+{u}_1+{u}_2),\ \ ({u}_1,\,+{u}_1+{u}_3,\,+{u}_1+{u}_2),\ \ ({u}_1,\,+{u}_2+{u}_3,\,+{u}_1-{u}_3),\notag\\
&({u}_1,\,+{u}_2+{u}_3,\,-{u}_1+{u}_2),\ \ ({u}_1,\,+{u}_1+{u}_3,\,-{u}_1+{u}_2),\ \ ({u}_1,\,+{u}_2+{u}_3,\,-{u}_1-{u}_3),\notag\\
&({u}_1,\,+{u}_2-{u}_3,\,+{u}_1+{u}_3),\ \ ({u}_1,\,+{u}_2-{u}_3,\,-{u}_1+{u}_3),\ \ ({u}_1,\,-{u}_2+{u}_3,\,+{u}_1+{u}_2),\notag\\
&({u}_1,\,-{u}_1+{u}_3,\,+{u}_1+{u}_2),\ \ ({u}_1,\,-{u}_2+{u}_3,\,-{u}_1+{u}_2),\ \ ({u}_1,\,-{u}_1+{u}_3,\,-{u}_1+{u}_2),\notag\\
\mathrm{B}: \quad &({u}_2,\,+{u}_2+{u}_3,\,+{u}_1-{u}_2),\ \ ({u}_2,\,+{u}_1+{u}_3,\,+{u}_1-{u}_2),\ \ ({u}_2,\,-{u}_2+{u}_3,\,+{u}_1-{u}_3),\notag\\
&({u}_2,\,-{u}_2+{u}_3,\,+{u}_1-{u}_2),\ \ ({u}_2,\,-{u}_1+{u}_3,\,+{u}_1-{u}_2),\ \ ({u}_2,\,-{u}_2-{u}_3,\,+{u}_1+{u}_3),\notag\\
\mathrm{C}: \quad &({u}_3,\,+{u}_2+{u}_3,\,+{u}_1-{u}_3),\ \ ({u}_3,\,+{u}_1-{u}_3,\,+{u}_1+{u}_2),\ \ ({u}_3,\,+{u}_2-{u}_3,\,+{u}_1-{u}_3),\notag\\
&({u}_3,\,+{u}_2-{u}_3,\,+{u}_1-{u}_2),\ \ ({u}_3,\,+{u}_1-{u}_3,\,-{u}_1+{u}_2),\ \ ({u}_3,\,-{u}_2-{u}_3,\,+{u}_1+{u}_2),\notag\\
\mathrm{D}:  \quad &(-{u}_2,\,-{u}_2+{u}_3,\,+{u}_1+{u}_2),\ \ (-{u}_2,\,+{u}_1+{u}_3,\,+{u}_1+{u}_2),\ \ (-{u}_2,\,+{u}_2+{u}_3,\,+{u}_1-{u}_3),\notag\\
&(-{u}_2,\,+{u}_2+{u}_3,\,+{u}_1+{u}_2),\ \ (-{u}_2,\,-{u}_1+{u}_3,\,+{u}_1+{u}_2),\ \ (-{u}_2,\,+{u}_2-{u}_3,\,+{u}_1+{u}_3),\notag\\
\mathrm{E}: \quad &(-{u}_3,\,+{u}_2-{u}_3,\,+{u}_1+{u}_3),\ \ (-{u}_3,\,+{u}_1+{u}_3,\,+{u}_1+{u}_2),\ \ (-{u}_3,\,+{u}_2+{u}_3,\,+{u}_1+{u}_3),\notag\\
&(-{u}_3,\,+{u}_2+{u}_3,\,+{u}_1-{u}_2),\ \ (-{u}_3,\,+{u}_1+{u}_3,\,-{u}_1+{u}_2),\ \ (-{u}_3,\,-{u}_2+{u}_3,\,+{u}_1+{u}_2).
\label{eq:JK-cones}
\end{align}
Many of these cones are related by the Weyl group of $\sprm(3)$, for example, all sectors B, C, D and E can be obtained by acting with Weyl group elements on the sector A. We evaluated the residues of the poles from
\begin{align}
\vartheta_1(\alpha+x+\epsilon_-)=0,\quad \vartheta_1(\beta+\epsilon_{1,2})=0,\quad \vartheta_1(\gamma+\epsilon_{1,2})=0\,,
\end{align}
with $(\alpha,\,\beta,\,\gamma)$ running over the cones listed in \eqref{eq:JK-cones}. One finds
\begin{align}
&\widetilde{\Psi}_3|_A=3Q(x-2\epsilon_1)Q(x-\epsilon_1)Q(x)-Q(x-\epsilon_1)^2Q(x)-4Q(x-\epsilon_1)Q(x)^2-2Q(x)^2Q(x+\epsilon)+3Q(x)^3\notag\\
&\quad\quad-Q(x)Q(x+\epsilon_1)^2+2Q(x-\epsilon_1)Q(x)Q(x+\epsilon_1)+3Q(x)Q(x+\epsilon_1)Q(x+2\epsilon_1)\notag\\
&\widetilde{\Psi}_3|_{B,\,C,\,D,\,E}=-Q(x)Q(x+\epsilon_1)^2+Q(x-\epsilon_1)Q(x)Q(x+\epsilon_1)+3Q(x)Q(x+\epsilon_1)Q(x+2\epsilon_1)\,,
\end{align}
and
\begin{align}
\widetilde{\Psi}_3=\widetilde{\Psi}_3|_A+4\widetilde{\Psi}_3|_B\,.
\end{align}
\subsection{Codimension 4 defect partition function \texorpdfstring{$\tilde\chi_{\mathrm{finite}}$}{Z} }

\paragraph{1-instanton.}  Now we turn to the codimension 4 defect partition function. The 1-instanton partition function with codimension 4 defect is given by
\begin{align}
W_1=\int\!\!\frac{\rd{u}}{2\pi \ri}\, \frac{\vartheta_1(2\epsilon_+)}{\vartheta_1(\epsilon_{1,2})}\frac{\vartheta_1(\pm2{u})\vartheta_1(\pm2{u}+2\epsilon_+)}{\prod_{i=1}^4\vartheta_1(\epsilon_+\pm {u} \pm \alpha_i)}
\cdot\frac{\vartheta_1(-\epsilon_-\pm x\pm{u})}{\vartheta_1(-\epsilon_+\pm x\pm{u})}\,.
\end{align}
The relevant poles originate from
\begin{align}
\vartheta_1(-\epsilon_+\pm x+{u})=0\,,
\end{align}
and we find in the NS-limit
\begin{align}
\tilde\chi_1=Q(x)+Q(x-\epsilon_1)\,,
\label{eq:SO_codim_four_finite_sector}
\end{align}
with $Q$ as defined in \eqref{eq:Q-fct}.
\paragraph{2-instanton.} The 2-instanton partition function with codimension 4 defect is given by
\begin{align}
W_2=\int \!\!\frac{\rd{u}_1}{2\pi \ri}\frac{\rd{u}_2}{2\pi \ri} \left(\frac{\vartheta_1(2\epsilon_+)}{\vartheta_1(\epsilon_{1,2})}\right)^2
&\frac{\vartheta_1(\pm2{u}_{1,2})\vartheta_1(\pm2{u}_{1,2}+2\epsilon_+)\vartheta_1(\pm{u}_1\pm{u}_2)\vartheta_1(\pm{u}_1\pm{u}_2+2\epsilon_+)}{\vartheta_1(\pm{u}_1\pm{u}_2+\epsilon_{1,2})\prod_{i=1}^4\vartheta_1(\epsilon_+\pm {u}_{1,2} \pm \alpha_i)}\notag\\
&\cdot\frac{\vartheta_1(-\epsilon_-\pm x\pm{u}_{1,2})}{\vartheta_1(-\epsilon_+\pm x\pm{u}_{1,2})}\,.
\end{align}
We follow the same JK-residue prescription as before, see \eqref{eq:cones_2string}. One finds that all contributions from the cones labeled by ``a", ``b" and  ``c" vanish for the codimension 4 defect. The only nonzero contributions come from the cone spanned by $\{{u}_1,\, {u}_1+{u}_2\}$, i.e.
\begin{align}
\mathrm{d}: \quad \vartheta_1(-\epsilon_+\pm x+{u}_1)=0\,, \quad \mathrm{and} \quad \vartheta_1({u}_1+{u}_2+\epsilon_{1,2})=0\,.
\end{align}
Collecting all the contributions from these poles and taking the NS-limit, we find 
\begin{align}
\tilde\chi_2=-Q(x)^2+Q(x-2\epsilon_1)Q(x-\epsilon_1)-Q(x-\epsilon_1)^2+Q(x)Q(x+\epsilon_1)\,.
\label{eq:SO_codim_four_defect_inst2}
\end{align}
\paragraph{3-instanton.}
The 3-instanton partition function with codimension 4 defect is given by
\begin{align}
W_3=\int \!\!\prod_{I=1}^3\frac{\rd{u_I}}{2\pi \ri} \left(\frac{\vartheta_1(2\epsilon_+)}{\vartheta_1(\epsilon_{1,2})}\right)^3
&\prod_{I=1}^3\frac{\vartheta_1(\pm2{u}_{I})\vartheta_1(\pm2{u}_{I}+2\epsilon_+)}{\prod_{i=1}^4\vartheta_1(\epsilon_+\pm {u}_{I} \pm \alpha_i)}
\cdot\prod_{I<J}^3\frac{\vartheta_1(\pm{u_I}\pm{u_J})\vartheta_1(\pm{u_I}\pm{u_J}+2\epsilon_+)}
{\vartheta_1(\pm{u_I}\pm{u_J}+\epsilon_{1,2})}\notag\\
&\cdot\prod_{I=1}^3\frac{\vartheta_1(-\epsilon_-\pm x\pm{u_I})}{\vartheta_1(-\epsilon_+\pm x\pm{u_I})}\,.
\end{align}
The relevant poles are given by
\begin{align}
\vartheta_1(\alpha\pm x-\epsilon_+)=0,\quad \vartheta_1(\beta+\epsilon_{1,2})=0,\quad \vartheta_1(\gamma+\epsilon_{1,2})=0\,,
\end{align}
with $(\alpha,\,\beta,\,\gamma)$ in \eqref{eq:JK-cones}. One observes that the contributions from cones $B$, $C$, $D$, and $E$ are zero. Therefore, we end up with
\begin{align}
\begin{aligned}
\tilde\chi_3&=\tilde\chi_3|_A=3Q(x-3\epsilon_1)Q(x-2\epsilon_1)Q(x-\epsilon_1)+3Q(x)Q(x+\epsilon_1)Q(x+2\epsilon_1)\\
&-Q(x-2\epsilon_1)^2Q(x-\epsilon_1)-Q(x)Q(x+\epsilon_1)^2-Q(x)^2Q(x-\epsilon_1)-Q(x-\epsilon_1)^2Q(x)\\
&-4Q(x-2\epsilon_1)Q(x-\epsilon_1)^2-4Q(x+\epsilon_1)Q(x)^2+3Q(x)^3+3Q(x-\epsilon_1)^3\,.
\end{aligned}
\end{align}
%
%
\section{Elliptic genera}
\label{app:ell_genera}
In this appendix, we collect the results for two-string and three-string elliptic genera. For the $\sorm(2N)$ case without defects, we list the three-string result. 
\subsection{Two-string elliptic genera}
\label{app:genera_details1}
\paragraph{$\boldsymbol{\sorm(2N)}$ case.}
The two-string elliptic genus of $\sorm(2N)$ theory receives contributions from three types of poles:
\begin{align}\label{eq:DN_Z2_poles}
\begin{aligned}
&1)\quad 1\times (\epsilon_+\pm\alpha_i+u_1=0,\,\epsilon_{1,2}+u_2+u_1=0)\,, \\
&2)\quad 3\times (\epsilon_+\pm\alpha_i+u_1=0,\,\epsilon_{1,2}+u_2-u_1=0)\,, \\
&3)\quad 2\times (\epsilon_+\pm\alpha_i+u_1=0,\,\epsilon_{+}\pm\alpha_j+u_2=0)\,,\quad \, (i<j)\, , 
\end{aligned}
\end{align}
where $n$ in $n\times \cdots$ denotes the multiplicity of the JK-residue contribution from the JK-pole in $\cdots$. From these poles, we have the residues \footnote{Here, the notation $\vartheta_1(2\epsilon_+ + 2\alpha_{i,j})\equiv \vartheta_1(2\epsilon_+ + 2\alpha_{i})\vartheta_1(2\epsilon_+ + 2\alpha_{j})$ is adopted in \eqref{eq:Dn_Z2_Res3} to simplify the expression.}
\begin{subequations}
\label{eq:2-str_ell_genus_SO2N}
\begin{align}
\mathrm{Res}_1
&= \frac{1}{8} \sum_{i=1}^N \bigg[ \frac{\vartheta_1(3\epsilon_+ + \epsilon_-) \vartheta_1(2\epsilon_- - 2\alpha_i) \vartheta_1(2\epsilon_1 - 2\alpha_i) \vartheta_1(4\epsilon_+ + 2\alpha_i) \vartheta_1(\epsilon_2 + 2\alpha_i)}{\vartheta_1(\epsilon_1) \vartheta_1(\epsilon_2)^2 \vartheta_1(2\epsilon_1) \vartheta_1(2\epsilon_-) \vartheta_1(2\alpha_i)} \nonumber \\
& \qquad \quad \cdot \frac{ \vartheta_1(3\epsilon_+ - \epsilon_- + 2\alpha_i) \prod_{l=1}^{2N-8}\vartheta_1(\epsilon_+ + \alpha_i \pm m_{l}) \vartheta_1(\epsilon_- - \alpha_i \pm m_{l})}{ \prod_{j \neq i}^N\vartheta_1(\alpha_i \pm \alpha_j) \vartheta_1(\epsilon_1 - \alpha_i \pm \alpha_j)\vartheta_1(\epsilon_2 + \alpha_i \pm \alpha_j) \vartheta_1(2\epsilon_+ + \alpha_i \pm \alpha_j)}  \\
& \qquad \qquad  + (\alpha_i \to -\alpha_i) \bigg] + (\epsilon_1 \leftrightarrow \epsilon_2)\ ,\nonumber\\
\mathrm{Res}_{2}
&= -\frac{1}{8} \sum_{i=1}^N \bigg[ \frac{\vartheta_1(3\epsilon_+ + \epsilon_- + 2\alpha_i) \vartheta_1(4\epsilon_+ + 2\epsilon_- + 2\alpha_i) \vartheta_1(5\epsilon_+ + \epsilon_- + 2\alpha_i)}{\vartheta_1(\epsilon_{1,2}) \vartheta_1(2\epsilon_1) \vartheta_1(2\epsilon_-)} \nonumber \\
& \qquad \quad \cdot \frac{\vartheta_1(6\epsilon_+ \!+ 2\epsilon_- + 2\alpha_i) \prod_{l=1}^{2N-8}\vartheta_1(\epsilon_+\! + \alpha_i \pm m_{l}) \vartheta_1(2\epsilon_+ +\epsilon_-\! + \alpha_i \pm m_{l})}{\prod_{j\neq i}^N\vartheta_1(\alpha_i \pm \alpha_j) \vartheta_1(\epsilon_1\! + \alpha_i \pm \alpha_j) \vartheta_1(2\epsilon_+\! + \alpha_i \pm \alpha_j) \vartheta_1(3\epsilon_+ \!+ \epsilon_-\! + \alpha_i \pm \alpha_j)}  \\
& \qquad \qquad + (\alpha_i \to -\alpha_i) \bigg] + (\epsilon_1 \leftrightarrow \epsilon_2)\ ,\nonumber\\
\mathrm{Res}_{3}
&= \frac{1}{8} \sum_{i < j}^N \bigg[ \frac{\vartheta_1(2\epsilon_+ + 2\alpha_{i,j}) \vartheta_1(4\epsilon_+ + 2\alpha_{i,j}) \vartheta_1(4\epsilon_+ + \alpha_i + \alpha_j)}{\vartheta_1(\epsilon_{1,2})^2 \vartheta_1(\alpha_i + \alpha_j) \vartheta_1(\epsilon_{1,2} + \alpha_i + \alpha_j) \vartheta_1(\epsilon_{1,2} \pm (\alpha_i - \alpha_j))} \nonumber \\
& \qquad \quad \cdot \frac{\prod_{l=1}^{2N-8}\vartheta_1(\epsilon_+ + \alpha_{i,j} \pm m_{l})}{ \vartheta_1(3\epsilon_+ \pm \epsilon_- + \alpha_i + \alpha_j) \prod_{k\neq i, j}^N \vartheta_1(\alpha_{i,j} \pm \alpha_k) \vartheta_1(2\epsilon_+ + \alpha_{i,j} \pm \alpha_k)}  \label{eq:Dn_Z2_Res3}\\
& \quad \qquad + (\alpha_i \to -\alpha_i) + (\alpha_j \to -\alpha_j) + (\alpha_i \to -\alpha_i, \: \alpha_j \to -\alpha_j) \bigg].\nonumber
\end{align}
\end{subequations}
The two-string elliptic genus is then given by
\begin{align}\label{eq:DN_Z2}
Z_2^{\rm{6d}} = \mathrm{Res}_1+ 3 \times \mathrm{Res}_2 + 2\times \mathrm{Res}_3\ .
\end{align}

\paragraph{$\boldsymbol{\sorm(2N+1)}$ case.}
The two-string elliptic genus of the $\sorm(2N+1)$ theory receives contributions from four types of poles:
\begin{align}\label{eq:BN_Z2_poles}
\begin{aligned}
&1)\quad 1\times (\epsilon_+\pm\alpha_i+u_1=0,\,\epsilon_{1,2}+u_2+u_1=0)\,, \\
&2)\quad 3\times (\epsilon_+\pm\alpha_i+u_1=0,\,\epsilon_{1,2}+u_2-u_1=0)\,, \\
&3)\quad 2\times (\epsilon_+\pm\alpha_i+u_1=0,\,\epsilon_{+}\pm\alpha_j+u_2=0)\,,\quad \, (i<j)\, ,  \\
&4)\quad 2\times (u_1=-\epsilon_+,\,u_2=-\epsilon_++\epsilon_1)\, ,
\end{aligned}
\end{align}
 where $n$ in $n\times \cdots$ denotes the multiplicity of the JK-residue contribution from the JK-pole in $\cdots$. The last type of poles are degenerate poles, which come from the intersection of three hyperplanes in the JK-residue description. From these poles, we computes the following residues:
\begin{subequations}
\begin{align}
\mathrm{Res}_1
&= \frac{1}{8} \sum_{i=1}^N \bigg[ \frac{\vartheta_1(3\epsilon_+ + \epsilon_-) \vartheta_1(2\epsilon_- - 2\alpha_i) \vartheta_1(2\epsilon_1 - 2\alpha_i) \vartheta_1(4\epsilon_+ + 2\alpha_i) \vartheta_1(\epsilon_2 + 2\alpha_i)\vartheta_1(3\epsilon_+ - \epsilon_- + 2\alpha_i)}{\vartheta_1(\epsilon_1) \vartheta_1(\epsilon_2)^2 \vartheta_1(2\epsilon_1) \vartheta_1(2\epsilon_-) \vartheta_1(2\alpha_i)\vartheta_1(\alpha_i)\vartheta_1(\alpha_i-\epsilon_1)\vartheta_1(\alpha_i+\epsilon_2)\vartheta_1(\alpha_i+2\epsilon_+)} \nonumber \\
& \qquad \quad\quad \cdot \frac{  \prod_{l=1}^{2N-7}\vartheta_1(\epsilon_+ + \alpha_i \pm m_{l}) \vartheta_1(\epsilon_- - \alpha_i \pm m_{l})}{ \prod_{j \neq i}^N\vartheta_1(\alpha_i \pm \alpha_j) \vartheta_1(\epsilon_1 - \alpha_i \pm \alpha_j)\vartheta_1(\epsilon_2 + \alpha_i \pm \alpha_j) \vartheta_1(2\epsilon_+ + \alpha_i \pm \alpha_j)} \nonumber \\
& \qquad \qquad  + (\alpha_i \to -\alpha_i) \bigg] + (\epsilon_1 \leftrightarrow \epsilon_2)\ ,
\\
\mathrm{Res}_{2}
&= -\frac{1}{8} \sum_{i=1}^N \bigg[ \frac{\vartheta_1(3\epsilon_+ + \epsilon_- + 2\alpha_i) \vartheta_1(4\epsilon_+ + 2\epsilon_- + 2\alpha_i) \vartheta_1(5\epsilon_+ + \epsilon_- + 2\alpha_i)\vartheta_1(6\epsilon_+ \!+ 2\epsilon_- + 2\alpha_i)}{\vartheta_1(\epsilon_{1,2}) \vartheta_1(2\epsilon_1) \vartheta_1(2\epsilon_-)\vartheta_1(\alpha_i)\vartheta_1(\alpha_i+\epsilon_1)\vartheta_1(\alpha_i+2\epsilon_+)\vartheta_1(\alpha_i+\epsilon_1+2\epsilon_+)} \nonumber \\
& \qquad \quad \quad \cdot  \frac{ \prod_{l=1}^{2N-7}\vartheta_1(\epsilon_+\! + \alpha_i \pm m_{l}) \vartheta_1(2\epsilon_+ +\epsilon_-\! + \alpha_i \pm m_{l})}{\prod_{j\neq i}^N\vartheta_1(\alpha_i \pm \alpha_j) \vartheta_1(\epsilon_1\! + \alpha_i \pm \alpha_j) \vartheta_1(2\epsilon_+\! + \alpha_i \pm \alpha_j) \vartheta_1(3\epsilon_+ \!+ \epsilon_-\! + \alpha_i \pm \alpha_j)} \nonumber \\
& \qquad \qquad + (\alpha_i \to -\alpha_i) \bigg] + (\epsilon_1 \leftrightarrow \epsilon_2)\ ,
\\
\mathrm{Res}_{3}
&= \frac{1}{8} \sum_{i < j}^N \bigg[ \frac{\vartheta_1(2\epsilon_+ + 2\alpha_{i,j}) \vartheta_1(4\epsilon_+ + 2\alpha_{i,j}) \vartheta_1(4\epsilon_+ + \alpha_i + \alpha_j)}{\vartheta_1(\epsilon_{1,2})^2 \vartheta_1(\alpha_i + \alpha_j) \vartheta_1(\epsilon_{1,2} + \alpha_i + \alpha_j) \vartheta_1(\epsilon_{1,2} \pm (\alpha_i - \alpha_j))\vartheta_1(\alpha_i)\vartheta_1(\alpha_j)} \nonumber \\
& \qquad \quad\quad \cdot \frac{\vartheta_1(\alpha_i+2\epsilon_+)^{-1}\vartheta_1(\alpha_j+2\epsilon_+)^{-1}\prod_{l=1}^{2N-7}\vartheta_1(\epsilon_+ + \alpha_{i,j} \pm m_{l})}{ \vartheta_1(3\epsilon_+ \pm \epsilon_- + \alpha_i + \alpha_j) \prod_{k\neq i, j}^N \vartheta_1(\alpha_{i,j} \pm \alpha_k) \vartheta_1(2\epsilon_+ + \alpha_{i,j} \pm \alpha_k)} \nonumber \\
& \quad \qquad + (\alpha_i \to -\alpha_i) + (\alpha_j \to -\alpha_j) + (\alpha_i \to -\alpha_i, \: \alpha_j \to -\alpha_j) \bigg]\ ,
\\
\mathrm{Res}_{4}
&= -\frac{1}{8} \frac{\vartheta_1(2\epsilon_+)\vartheta_1(4\epsilon_+)\vartheta_1(2\epsilon_++\epsilon_1)\vartheta_1(2\epsilon_++\epsilon_2)\prod_{l=1}^{2N-7}\vartheta_1(\pm m_l+\epsilon_+)\vartheta_1(\pm m_l+\epsilon_-)}{\prod_{i=1}^N{\vartheta_1(\alpha_i)^2\vartheta_1(\alpha_i\pm \epsilon_1)\vartheta_1(\alpha_i\pm \epsilon_2)\vartheta_1(\alpha_i\pm 2\epsilon_+)}}\ .
\end{align}
\end{subequations}
The two-string elliptic genus is then given by
\begin{align}\label{eq:BN_Z2}
Z_2^{\rm{6d}} = \mathrm{Res}_1+ 3 \cdot \mathrm{Res}_2 + 2\cdot \mathrm{Res}_3+2\cdot \mathrm{Res}_4\ .
\end{align}

\paragraph{$\boldsymbol{\sorm(2N)}$ with $\boldsymbol{\mathbb{Z}_2}$ twist.}
The two-string elliptic genus of the $\sorm(2N)$ theory with $\mathbb{Z}_2$ twist receives contributions from five types of poles:
\begin{align}
\label{eq:DNZ2twist_Z2_poles}
\begin{aligned}
&1)\quad 1\times (\epsilon_+\pm\alpha_i+u_1=0,\,\epsilon_{1,2}+u_2+u_1=0)\,,\\
&2)\quad 3\times (\epsilon_+\pm\alpha_i+u_1=0,\,\epsilon_{1,2}+u_2-u_1=0)\,, \\
&3)\quad 2\times (\epsilon_+\pm\alpha_i+u_1=0,\,\epsilon_{+}\pm\alpha_j+u_2=0)\,,\quad \, (i<j)\, , \\
&4) \quad 2\times ( \epsilon_++\frac{\tau}{2}+u_1=0, \,\epsilon_{1}+u_2+u_1=0)\,,  \\
&5) \quad 2\times (u_1=-\epsilon_+,\, u_2=-\epsilon_++\epsilon_{1})\,,  
\end{aligned}
\end{align}
where $n$ in $n\times \cdots$ denotes the multiplicity of the JK-residue contribution from the JK-pole in $\cdots$. The last poles are degenerate poles, which come from the intersection of the hyperplanes $u_1+\epsilon_+=0$, $\epsilon_1+u_1+u_2=0$, $\epsilon_2+u_1-u_2=0$ and $u_1+\epsilon_+=0$, $\epsilon_2+u_1+u_2=0$, $\epsilon_1+u_1-u_2=0$.
From these poles, we obtain the residues
\begin{subequations}
\begin{align}
\mathrm{Res}_1
&= \frac{1}{8} \sum_{i=1}^{N-1} \bigg[ \frac{\vartheta_1(3\epsilon_+ + \epsilon_-) \vartheta_1(2\epsilon_- - 2\alpha_i) \vartheta_1(2\epsilon_1 - 2\alpha_i) \vartheta_1(4\epsilon_+ + 2\alpha_i) \vartheta_1(\epsilon_2 + 2\alpha_i)}{\vartheta_1(\epsilon_1) \vartheta_1(\epsilon_2)^2 \vartheta_1(2\epsilon_1) \vartheta_1(2\epsilon_-) \vartheta_1(2\alpha_i)}  \\
& \qquad \quad \cdot\frac{ \vartheta_1(3\epsilon_+ - \epsilon_- + 2\alpha_i) \prod_{l=1}^{2N-8}\vartheta_1(\epsilon_+ + \alpha_i \pm m_{l}) \vartheta_1(\epsilon_- - \alpha_i \pm m_{l})}{ \prod_{j \neq i}^{N-1}\vartheta_1(\alpha_i \pm \alpha_j) \vartheta_1(\epsilon_1 - \alpha_i \pm \alpha_j)\vartheta_1(\epsilon_2 + \alpha_i \pm \alpha_j) \vartheta_1(2\epsilon_+ + \alpha_i \pm \alpha_j)} \notag  \\
& \qquad \quad \cdot \frac{1}{\prod_{I=1,4}\vartheta_I(\alpha_i)\vartheta_I(\alpha_i-\epsilon_1)\vartheta_I(\alpha_i+\epsilon_2)\vartheta_I(\alpha_i+2\epsilon_+)} + (\alpha_i \to -\alpha_i) \bigg] + (\epsilon_1 \leftrightarrow \epsilon_2)\ ,
\notag \\
\mathrm{Res}_{2}
&= -\frac{1}{8} \sum_{i=1}^{N-1} \bigg[ \frac{\vartheta_1(3\epsilon_+ + \epsilon_- + 2\alpha_i) \vartheta_1(4\epsilon_+ + 2\epsilon_- + 2\alpha_i) \vartheta_1(5\epsilon_+ + \epsilon_- + 2\alpha_i)}{\vartheta_1(\epsilon_{1,2}) \vartheta_1(2\epsilon_1) \vartheta_1(2\epsilon_-)}  \\
& \qquad \quad \cdot \frac{\vartheta_1(6\epsilon_+ \!+ 2\epsilon_- + 2\alpha_i) \prod_{l=1}^{2N-8}\vartheta_1(\epsilon_+\! + \alpha_i \pm m_{l}) \vartheta_1(2\epsilon_+ +\epsilon_-\! + \alpha_i \pm m_{l})}{\prod_{j\neq i}^{N-1}\vartheta_1(\alpha_i \pm \alpha_j) \vartheta_1(\epsilon_1\! + \alpha_i \pm \alpha_j) \vartheta_1(2\epsilon_+\! + \alpha_i \pm \alpha_j) \vartheta_1(3\epsilon_+ \!+ \epsilon_-\! + \alpha_i \pm \alpha_j)} \notag \\
& \qquad \quad \cdot \frac{1}{\prod_{I=1,4}\vartheta_I(\alpha_i)\vartheta_I(\alpha_i+\epsilon_1)\vartheta_I(\alpha_i+2\epsilon_+)\vartheta_I(\alpha_i+3\epsilon_++\epsilon_-)} + (\alpha_i \to -\alpha_i) \bigg] + (\epsilon_1 \leftrightarrow \epsilon_2)\ ,
\notag \\
\mathrm{Res}_{3}
&= \frac{1}{8} \sum_{i < j}^{N-1} \bigg[ \frac{\vartheta_1(2\epsilon_+ + 2\alpha_{i,j}) \vartheta_1(4\epsilon_+ + 2\alpha_{i,j}) \vartheta_1(4\epsilon_+ + \alpha_i + \alpha_j)}{\vartheta_1(\epsilon_{1,2})^2 \vartheta_1(\alpha_i + \alpha_j) \vartheta_1(\epsilon_{1,2} + \alpha_i + \alpha_j) \vartheta_1(\epsilon_{1,2} \pm (\alpha_i - \alpha_j))\vartheta_1(3\epsilon_+ \pm \epsilon_- + \alpha_i + \alpha_j)} \nonumber \\
& \quad \quad \cdot\frac{\prod_{l=1}^{2N-8}\vartheta_1(\epsilon_+ + \alpha_{i,j} \pm m_{l})}{  \prod_{I=1,4}\vartheta_I(\alpha_{i,j})\vartheta_I(2\epsilon_++\alpha_{i,j})\prod_{k\neq i, j}^{N-1} \vartheta_1(\alpha_{i,j} \pm \alpha_k) \vartheta_1(2\epsilon_+ + \alpha_{i,j} \pm \alpha_k)} \nonumber \\
& \quad \qquad + (\alpha_i \to -\alpha_i) + (\alpha_j \to -\alpha_j) + (\alpha_i \to -\alpha_i, \: \alpha_j \to -\alpha_j) \bigg]\ ,
\\
\mathrm{Res}_4+\mathrm{Res}_5
&= -\frac{1}{8} \sum_{I=1,4}  \frac{\vartheta_1(3\epsilon_+ \pm \epsilon_-) \vartheta_1(2\epsilon_+) \vartheta_1(4\epsilon_+)\prod_{l=1}^{2N-8}\vartheta_I(\epsilon_+\pm m_{l}) \vartheta_I(\epsilon_-  \pm m_{l})}{\vartheta_4(0)\vartheta_4(\epsilon_{1,2}) \vartheta_4(2\epsilon_+) \prod_{i=1}^{N-1}\vartheta_I(\alpha_i )^2 \vartheta_I(\epsilon_{1,2}\pm\alpha_i )\vartheta_I(2\epsilon_+\pm \alpha_i )} \ .
\end{align}
\end{subequations}
The two-string elliptic genus is
\begin{align}\label{eq:DNZ2twist_Z2}
Z_2^{\rm{6d}} = \mathrm{Res}_1+ 3 \times \mathrm{Res}_2 + 2\times \mathrm{Res}_3+2\times (\mathrm{Res}_4+\mathrm{Res}_5)\ .
\end{align}

\subsection{Three-string elliptic genera}
\label{app:genera_details2}
The three-string elliptic genus of the $\sorm(2N)$ theory receives contributions from ten types of poles:
\begin{align}
\begin{aligned}
&1)\quad 12\times (-\alpha_i+\epsilon_++u_1=0,\,\epsilon_1-u_1+u_2=0,\epsilon_2-u_1+u_3=0)\,, \\
&2)\quad 15\times (-\alpha_i+\epsilon_++u_1=0,\,\epsilon_{1,2}-u_1+u_2=0,\epsilon_{1,2}-u_2+u_3=0)\,, \\
&3)\quad 6\times (-\alpha_i+\epsilon_++u_1=0,\,\epsilon_{1,2}+u_1+u_2=0,\,\epsilon_{1,2}-u_1+u_3=0)\,, \\
&4)\quad 6\times (-\alpha_i+\epsilon_++u_1=0,\,\alpha_i+\epsilon_++u_2=0,\,\epsilon_2+u_2-u_3=0,\,\epsilon_1+u_1+u_3=0)\,, \\
&5)\quad 3\times (-\alpha_i+\epsilon_++u_1=0,\,\epsilon_{1,2}+u_1-u_3=0,\,\epsilon_{1,2}+u_2+u_3=0)\,, \\
&6)\quad 3\times (-\alpha_i+\epsilon_++u_1=0,\,\epsilon_{1,2}+u_1+u_2=0,\,\epsilon_{2,1}+u_2+u_3=0)\,, \\
&7)\quad 3\times (-\alpha_i+\epsilon_++u_1=0,\,\epsilon_{1,2}-u_1+u_2=0,\,\epsilon_{2,1}+u_2+u_3=0)\,, \\
&8)\quad 9\times (-\alpha_i+\epsilon_++u_1=0,\,-\alpha_j+\epsilon_++u_2=0,\,\epsilon_{1,2}-u_1+u_3=0)\,, \quad (i\neq j),\\
&9)\quad 3\times (-\alpha_i+\epsilon_++u_1=0,\,-\alpha_j+\epsilon_++u_2=0,\,\epsilon_{1,2}+u_1+u_3=0)\,, \quad (i\neq j),\\
&10)\quad 6\times (-\alpha_i+\epsilon_++u_1=0,\,-\alpha_j+\epsilon_++u_2=0,\,-\alpha_k+\epsilon_++u_3=0)\,, \quad (i<j<k),\\
\end{aligned}
\end{align}
where $n$ in $n\times \cdots$ denotes the multiplicity of the JK-residue contribution from the JK-pole(s) in $\cdots$. From these poles, we have the residues
\begin{subequations}
\begin{align}
\mathrm{Res}_1
&= \frac{1}{48} \sum_{i=1}^N \bigg[ \frac{\vartheta_1(\pm 2 \epsilon _--6 \epsilon _++2 \alpha_i) \vartheta_1(\pm \epsilon _+-5 \epsilon _++2 \alpha_i) \vartheta_1(\pm 2 \epsilon _--4 \epsilon _++2 \alpha_i)}{ \vartheta_1(3 \epsilon _-\pm \epsilon _+) \vartheta_1(\epsilon _1)^2 \vartheta_1(\epsilon _2)^2}\nonumber \\
& \qquad \quad \cdot \frac{\prod _{l=1}^{2 N-8} \vartheta_1(\pm \epsilon _--2 \epsilon _+\pm m_l+\alpha_i)  \vartheta_1(-\epsilon _+\pm m_l+\alpha_i)}{\prod _{j\neq i}^{N} \vartheta_1(\alpha_i\pm \alpha_j) \vartheta_1(\pm \epsilon _--3 \epsilon _++\alpha_i\pm \alpha_j) \vartheta_1(-2 \epsilon _++\alpha_i\pm \alpha_j) \vartheta_1(\pm \epsilon _--\epsilon _++\alpha_i\pm \alpha_j)} \nonumber\\
& \qquad \qquad  + (\alpha_i \to -\alpha_i) \bigg] \ ,\\
\mathrm{Res}_{2}
&= -\frac{1}{48} \sum_{i=1}^N \bigg[\frac{ \vartheta_1(2 \alpha_i-5 \epsilon _1-2 \epsilon _2\pm \epsilon _1) \vartheta_1(2 \alpha_i-5 \epsilon _1-2 \epsilon _2) \vartheta_1(2 \alpha_i-4 \epsilon _1-\epsilon _2) }{ \vartheta_1(2 \epsilon _-+\epsilon _+\pm \epsilon _+) \vartheta_1(3 \epsilon _-+2 \epsilon _+\pm \epsilon _+) \vartheta_1(\epsilon _1) \vartheta_1(\epsilon _2)} \nonumber \\
& \qquad \quad \cdot \frac{\prod _{l=1}^{2 N-8} \vartheta_1(-\epsilon _+\pm m_l+\alpha_i) \vartheta_1(-2 \epsilon _--3 \epsilon _+\pm m_l+\alpha_i) \vartheta_1(-\epsilon _--2 \epsilon _+\pm m_l+\alpha_i)}{\prod _{j\neq i}^{N} \vartheta_1(\alpha_i\pm \alpha_j) \vartheta_1(-2 \epsilon _1+\alpha_i\pm \alpha_j) \vartheta_1(-\epsilon _1+\alpha_i\pm \alpha_j) \vartheta_1(-2 \epsilon _++\alpha_i\pm \alpha_j)} \nonumber\\
& \qquad \quad \cdot \frac{\vartheta_1(2 \alpha_i-3 \epsilon _1-\epsilon _2)\vartheta_1(2 \alpha_i-5 \epsilon _1-\epsilon _2)}{\prod _{j\neq i}^{N} \vartheta_1(-2 \epsilon _--4 \epsilon _++\alpha_i\pm \alpha_j) \vartheta_1(-\epsilon _--3 \epsilon _++\alpha_i\pm \alpha_j)} \nonumber \\
& \qquad \qquad + (\alpha_i \to -\alpha_i) \bigg] + (\epsilon_1 \leftrightarrow \epsilon_2)\ ,\\
\mathrm{Res}_{3}
&= \frac{1}{48} \sum_{i=1}^N \bigg[\frac{\vartheta_1(2 \alpha_i+2 \epsilon _1) \vartheta_1(2 \alpha_i-3 \epsilon _1-2 \epsilon _2) \vartheta_1(2 \alpha_i-3 \epsilon _1-2 \epsilon _2\pm \epsilon _1) \vartheta_1(2 \alpha_i-3 \epsilon _1-\epsilon _2)}{ \vartheta_1(2 \alpha_i-\epsilon _1) \vartheta_1(2 \epsilon _-) \vartheta_1(3 \epsilon _-+\epsilon _+) \vartheta_1(3 \epsilon_1) \vartheta_1(\epsilon _1)^2 \vartheta_1(\epsilon _2)^2} \nonumber \\
& \qquad \quad \cdot \frac{\prod _{l=1}^{2 N-8} \vartheta_1(\epsilon _-\pm m_l+\alpha_i) \vartheta_1(-\epsilon _+\pm m_l+\alpha_i) \vartheta_1(-\epsilon _--2 \epsilon _+\pm m_l+\alpha_i)}{\prod _{j\neq i}^{N} \vartheta_1(\alpha_i\pm \alpha_j) \vartheta_1(-\epsilon _--3 \epsilon _++\alpha_i\pm \alpha_j) \vartheta_1(-2 \epsilon _++\alpha_i\pm \alpha_j) } \nonumber \\
& \qquad \quad \cdot  \frac{ \vartheta_1(2 \alpha_i\pm \epsilon _1-\epsilon _2) \vartheta_1(3 \epsilon _1+\epsilon _2)}{\prod _{j\neq i}^{N}\vartheta_1(\pm \epsilon_1+\alpha_i\pm \alpha_j) \vartheta_1(-\epsilon_2+\alpha_i\pm \alpha_j)}+ (\alpha_i \to -\alpha_i) \bigg] + (\epsilon_1 \leftrightarrow \epsilon_2)\ , \\
\mathrm{Res}_{4}
&= -\frac{1}{48} \sum_{i=1}^N \bigg[\frac{\vartheta_1(2 \epsilon _+) \vartheta_1(4 \epsilon _+) \vartheta_1(\pm 4 \epsilon _++2 \alpha_i) \vartheta_1(\pm 2 \epsilon _++2 \alpha_i) \vartheta_1(\epsilon _-\pm \epsilon _++2 \alpha_i)}{ \vartheta_1(2 \alpha_i-\epsilon _1) \vartheta_1(2 \alpha_i+\epsilon _2) \vartheta_1(2 \epsilon _1) \vartheta_1(2 \epsilon _2) \vartheta_1(2 \epsilon _-)^2 \vartheta_1(\epsilon _1)^2 \vartheta_1(\epsilon _2)^2} \nonumber \\
& \qquad \quad \cdot \frac{ \vartheta_1(\epsilon _-\pm 3 \epsilon _++2 \alpha_i)\prod _{l=1}^{2 N-8} \vartheta_1(\epsilon _-\pm m_l+\alpha_i) \vartheta_1(\pm \epsilon _+\pm m_l+\alpha_i)}{\prod _{j\neq i}^{N} \vartheta_1(\alpha_i\pm \alpha_j)^2 \vartheta_1(\pm 2 \epsilon _++\alpha_i\pm \alpha_j) \vartheta_1(\epsilon _-\pm \epsilon _++\alpha_i\pm \alpha_j)}+ (\alpha_i \to -\alpha_i) \bigg] \ , \\
\mathrm{Res}_{5}
&= \frac{1}{48} \sum_{i=1}^N \bigg[\frac{\vartheta_1(\epsilon _-+3 \epsilon _+) \vartheta_1(2 \epsilon _-+4 \epsilon _+) \vartheta_1(2 \epsilon _-+2 \alpha_i) \vartheta_1(-4 \epsilon _++2 \alpha_i) \vartheta_1(2 \alpha_i+3 \epsilon _1)}{ \vartheta_1(2 \epsilon _-)^2 \vartheta_1(3 \epsilon _-+\epsilon _+) \vartheta_1(2 \epsilon _1) \vartheta_1(3 \epsilon _1) \vartheta_1(\epsilon _1) \vartheta_1(\epsilon _2)^2 \vartheta_1(2 \alpha_i) \vartheta_1(\epsilon _1+2 \alpha_i)} \nonumber \\
& \qquad \quad \cdot \frac{\prod _{l=1}^{2N-8} \vartheta_1(\epsilon _-\pm m_l+\alpha_i) \vartheta_1(-\epsilon _+\pm m_l+\alpha_i) \vartheta_1(2 \epsilon _-+\epsilon _+\pm m_l+\alpha_i)}{\prod _{j\neq i}^{N} \vartheta_1(\alpha_i\pm \alpha_j) \vartheta_1(2 \epsilon _-+\alpha_i\pm \alpha_j) \vartheta_1(-2 \epsilon _++\alpha_i\pm \alpha_j) \vartheta_1(-\epsilon _2+\alpha_i\pm \alpha_j)} \nonumber \\
&\qquad \quad \cdot  \frac{ \vartheta_1(2 \alpha_i+4 \epsilon _1) \vartheta_1(2 \alpha_i-2 \epsilon _2) \vartheta_1(2 \alpha_i-\epsilon _1-2 \epsilon _2) \vartheta_1(2 \alpha_i+2 \epsilon _1-\epsilon _2) \vartheta_1(2 \alpha_i+3 \epsilon _1-\epsilon _2)}{\prod _{j\neq i}^{N} \vartheta_1(\epsilon _1+\alpha_i\pm \alpha_j) \vartheta_1(2 \epsilon _1+\alpha_i\pm \alpha_j)}\nonumber\\
&\qquad \qquad  + (\alpha_i \to -\alpha_i) \bigg] + (\epsilon_1 \leftrightarrow \epsilon_2)\ , \\
\mathrm{Res}_{6}
&= \frac{1}{48} \sum_{i=1}^N \bigg[\frac{\vartheta_1(\pm \epsilon _-+3 \epsilon _+) \vartheta_1(2 \epsilon _-+2 \alpha_i) \vartheta_1(-4 \epsilon _++2 \alpha_i) \vartheta_1(-4 \epsilon _2+2 \alpha_i) \vartheta_1(3 \epsilon _-\pm \epsilon _++2 \alpha_i) }{ \vartheta_1(2 \epsilon _-) \vartheta_1(3 \epsilon _-\pm \epsilon _+) \vartheta_1(2 \epsilon _2) \vartheta_1(\epsilon _1)^2 \vartheta_1(\epsilon _2)^2 \vartheta_1(2 \alpha_i) \vartheta_1(2 \epsilon _-+2 \alpha_i)}\nonumber \\
& \qquad \quad \cdot \frac{\prod _{l=1}^{2N-8} \vartheta_1(\epsilon _-\pm m_l+\alpha_i) \vartheta_1(-\epsilon _+\pm m_l+\alpha_i) \vartheta_1(2 \epsilon _--\epsilon _+\pm m_l+\alpha_i)}{\prod _{j\neq i}^{N} \vartheta_1(\alpha_i\pm \alpha_j) \vartheta_1(2 \epsilon _-+\alpha_i\pm \alpha_j) \vartheta_1(-2 \epsilon _++\alpha_i\pm \alpha_j) \vartheta_1(-2 \epsilon _2+\alpha_i\pm \alpha_j) }\nonumber \\
&\qquad \quad \cdot \frac{\vartheta_1(2 \epsilon _1+2 \alpha_i) \vartheta_1(2 \alpha_i-\epsilon _1-3 \epsilon _2) \vartheta_1(2 \alpha_i+\epsilon _1-3 \epsilon _2)}{\prod _{j\neq i}^{N}\vartheta_1(-\epsilon _2+\alpha_i\pm \alpha_j) \vartheta_1(\epsilon _1+\alpha_i\pm \alpha_j)}+ (\alpha_i \to -\alpha_i) \bigg] + (\epsilon_1 \leftrightarrow \epsilon_2)\ , \\
\mathrm{Res}_{7}
&= \frac{1}{48} \sum_{i=1}^N \bigg[\frac{\vartheta_1(-\epsilon _-+3 \epsilon _+) \vartheta_1(-4 \epsilon _-+2 \alpha_i) \vartheta_1(2 \alpha_i-3 \epsilon _1) \vartheta_1(2 \alpha_i-4 \epsilon _1-2 \epsilon _2) }{ \vartheta_1(2 \epsilon _-) \vartheta_1(3 \epsilon _-\pm \epsilon _+) \vartheta_1(2 \epsilon _1) \vartheta_1(\epsilon _1)^2 \vartheta_1(\epsilon _2) \vartheta_1(-\epsilon _1+2 \alpha_i)} \nonumber \\
& \qquad \quad \cdot \frac{\prod _{l=1}^{2N-8} \vartheta_1(-2 \epsilon _--\epsilon _+\pm m_l+\alpha_i) \vartheta_1(-\epsilon _+\pm m_l+\alpha_i) \vartheta_1(-\epsilon _--2 \epsilon _+\pm m_l+\alpha_i)}{\prod _{j\neq i}^{N} \vartheta_1(\alpha_i\pm \alpha_j) \vartheta_1(-2 \epsilon _-+\alpha_i\pm \alpha_j) \vartheta_1(-\epsilon _--3 \epsilon _++\alpha_i\pm \alpha_j) \vartheta_1(-2 \epsilon _++\alpha_i\pm \alpha_j) }\nonumber \\
&\qquad \quad \cdot \frac{\vartheta_1(2 \alpha_i-3 \epsilon _1-2 \epsilon _2) \vartheta_1(2 \alpha_i-4 \epsilon _1-\epsilon _2) \vartheta_1(2 \alpha_i-3 \epsilon _1\pm \epsilon _2)}{\prod _{j\neq i}^{N}\vartheta_1(-2 \epsilon _1+\alpha_i\pm \alpha_j) \vartheta_1(-\epsilon _1+\alpha_i\pm \alpha_j)}+ (\alpha_i \to -\alpha_i) \bigg]  \nonumber \\
&\qquad \qquad   +(\epsilon_1 \leftrightarrow \epsilon_2)\ , \\
\mathrm{Res}_{8}
&= -\frac{1}{48} \sum_{i\neq j} \bigg[\frac{ \vartheta_1(-2 \epsilon _--5 \epsilon _+\pm \epsilon _++2 \alpha_i) \vartheta_1(-\epsilon _--4 \epsilon _+\pm \epsilon _++2 \alpha_i) \vartheta_1(-\epsilon _--5 \epsilon _++\alpha_i+\alpha_j) }{ \vartheta_1(2 \epsilon _-) \vartheta_1(2 \epsilon _1) \vartheta_1(\epsilon _1)^2 \vartheta_1(\epsilon _2)^2 \vartheta_1(\alpha_i\pm \alpha_j) \vartheta_1(-2 \epsilon _-+\alpha_i-\alpha_j) } \nonumber \\
& \qquad \quad \cdot\frac{\vartheta_1(-4 \epsilon _++2 \alpha_j) \vartheta_1(-2 \epsilon _++2 \alpha_j)\vartheta_1^{-1}(-2 \epsilon _--4 \epsilon _++\alpha_i+\alpha_j) \vartheta_1^{-1}(\epsilon _--3 \epsilon _++\alpha_i+\alpha_j)}{\vartheta_1(-2 \epsilon _++\alpha_i\pm \alpha_j) \vartheta_1(-2 \epsilon _1+\alpha_i\pm \alpha_j) \vartheta_1(-\epsilon _2+\alpha_i\pm \alpha_j) \vartheta_1(\epsilon _1\pm \alpha_i-\alpha_j) }\nonumber \\
& \qquad \quad \cdot \frac{\prod _{l=1}^{2N-8} \vartheta_1(-\epsilon _--2 \epsilon _+\pm m_l+\alpha_i) \vartheta_1(-\epsilon _+\pm m_l+\alpha_i) \vartheta_1(-\epsilon _+\pm m_l+\alpha_j)}{\prod _{k\neq i,j}^{N} \vartheta_1(\alpha_i\pm \alpha_k) \vartheta_1(-2 \epsilon _++\alpha_i\pm \alpha_k)\vartheta_1(\alpha_j\pm \alpha_k) \vartheta_1(-2 \epsilon _++\alpha_j\pm \alpha_k) }\nonumber \\
& \qquad \quad \cdot {\prod _{k\neq i,j}^{N}\vartheta_1^{-1}(-\epsilon _--3 \epsilon _++\alpha_i\pm \alpha_k) \vartheta_1^{-1}(-\epsilon _1+\alpha_i\pm \alpha_k)} + (\alpha_i \to -\alpha_i) + (\alpha_j \to -\alpha_j)  \nonumber \\
&\qquad \qquad  + (\alpha_i \to -\alpha_i, \: \alpha_j \to -\alpha_j)\bigg] + (\epsilon_1 \leftrightarrow \epsilon_2)\ , \\
\mathrm{Res}_{9}
&= -\frac{1}{48} \sum_{i\neq j} \bigg[\frac{\vartheta_1(-\epsilon _-+3 \epsilon _+) \prod_{l=i,j}\vartheta_1(-2 \epsilon _++2 \alpha_l) \vartheta_1(-4 \epsilon _++2 \alpha_l)}{ \vartheta_1(2 \epsilon _-) \vartheta_1(2 \epsilon _2) \vartheta_1(\epsilon _1)^3 \vartheta_1(\epsilon _2)^2 \vartheta_1(2 \alpha_i) \vartheta_1(-2 \epsilon _++2 \alpha_i) \vartheta_1(\alpha_i-\alpha_j)} \nonumber \\
&\qquad \quad \cdot\frac{\vartheta_1(-2 \epsilon _-+2 \alpha_i) \vartheta_1(-\epsilon _--3 \epsilon _++2 \alpha_i) \vartheta_1(-\epsilon _1+2 \alpha_i) \vartheta_1(2 \epsilon _2+2 \alpha_i) }{ \vartheta_1(\alpha_i+\alpha_j)^2 \vartheta_1(-2 \epsilon _-+\alpha_i\pm \alpha_j) \vartheta_1(\pm \epsilon _1+\alpha_i-\alpha_j) \vartheta_1(2 \epsilon _+\pm \alpha_i-\alpha_j) }\nonumber \\
&\qquad \quad \cdot \frac{\vartheta_1(-2 \epsilon _1+\alpha_i-\alpha_j) \vartheta_1(-\epsilon _-+3 \epsilon _++\alpha_i-\alpha_j) \vartheta_1(-4 \epsilon _++\alpha_i+\alpha_j)}{\vartheta_1(-\epsilon _2+\alpha_i\pm \alpha_j) \vartheta_1(2 \epsilon _2+\alpha_i-\alpha_j) \vartheta_1(-2 \epsilon _1+\alpha_i\pm \alpha_j) \vartheta_1(\epsilon _--3 \epsilon _++\alpha_i+\alpha_j)}\nonumber \\
& \qquad \quad \cdot \frac{\prod _{l=1}^{2N-8} \vartheta_1(-\epsilon _-\pm m_l+\alpha_i) \vartheta_1(-\epsilon _+\pm m_l+\alpha_i) \vartheta_1(-\epsilon _+\pm m_l+\alpha_j)}{\prod _{k\neq i,j}^{N} \vartheta_1(\alpha_i\pm \alpha_k) \vartheta_1(\alpha_j\pm \alpha_k) \vartheta_1(-2 \epsilon _++\alpha_i\pm \alpha_k) \vartheta_1(-\epsilon_{1}+\alpha_i\pm \alpha_k)  }\nonumber \\
& \qquad \quad \cdot \prod _{k\neq i,j}^{N}\vartheta_1^{-1}(\epsilon_{2}+\alpha_i\pm \alpha_k)\vartheta_1^{-1}(-2 \epsilon _++\alpha_j\pm \alpha_k)+(\alpha_i \to -\alpha_i) + (\alpha_j \to -\alpha_j)  \nonumber \\
&\qquad \qquad  +  (\alpha_i \to -\alpha_i, \: \alpha_j \to -\alpha_j) \bigg] + (\epsilon_1 \leftrightarrow \epsilon_2)\ , \\
\mathrm{Res}_{10}
&= -\frac{1}{48\vartheta_1(\epsilon _1)^3 \vartheta_1(\epsilon _2)^3} \sum_{{i<j<k}} \bigg[ \prod_{l={i,j,k}}\vartheta_1(-4 \epsilon _++2 \alpha_l) \vartheta_1(-2 \epsilon _++2 \alpha_l) \prod _{s=1}^{2 N-8} \vartheta_1(-\epsilon _+\pm m_s+\alpha_l)\nonumber \\
& \qquad\qquad \quad \cdot \prod_{l_2< l_1}\frac{\vartheta_1^{-1}(\pm \epsilon _--\epsilon _++\alpha_{l_1}+\alpha_{l_2})\vartheta_1(-4 \epsilon _++\alpha_{l_1}+\alpha_{l_2})}{\vartheta_1(\pm \epsilon _-\pm \epsilon _++\alpha_{l_1}-\alpha_{l_2}) \vartheta_1(\alpha_{l_1}+\alpha_{l_2}) \vartheta_1(\pm \epsilon _--3 \epsilon _++\alpha_{l_1}+\alpha_{l_2}) }\nonumber \\
& \qquad\qquad \quad \cdot\prod_{\substack{l_1=i,j,k,\\l_2\neq i,j,k}}\vartheta_1^{-1}(\alpha_{l_1}\pm \alpha_{l_2}) \vartheta_1^{-1}(-2 \epsilon _++\alpha_{l_1}\pm \alpha_{l_2})+(\alpha_i \to -\alpha_i)+(\alpha_j \to -\alpha_j)\nonumber \\
&\qquad\qquad \qquad  + (\alpha_k \to -\alpha_k) + (\alpha_i \to -\alpha_i, \: \alpha_j \to -\alpha_j)+ (\alpha_i \to -\alpha_i, \: \alpha_k \to -\alpha_k) \nonumber \\
&\qquad\qquad \qquad  + (\alpha_j \to -\alpha_j, \: \alpha_k \to -\alpha_k)+ (\alpha_i \to -\alpha_i, \: \alpha_j \to -\alpha_j, \: \alpha_k \to -\alpha_k)\bigg]\ .
\end{align}
\end{subequations}
The three-string elliptic genus is then given by
\begin{align}\label{eq:DN_Z3}
Z_3^{\rm{6d}} = 12\times\mathrm{Res}_1+ 15 \times \mathrm{Res}_2 + 6\times \mathrm{Res}_3+&6\times \mathrm{Res}_4+3\times \mathrm{Res}_5+3\times \mathrm{Res}_6\nonumber\\
&+3\times \mathrm{Res}_7+9\times \mathrm{Res}_8+3\times \mathrm{Res}_9+6\times \mathrm{Res}_{10}\ .
\end{align}
%
%
 \bibliographystyle{JHEP}     
 {\small{\bibliography{references}}}
\end{document}